\documentclass[longauth]{aa} 
\usepackage{graphicx}
\usepackage{wasysym}
\usepackage{txfonts}
\usepackage[colorlinks]{hyperref}
\usepackage{color}
\usepackage[normalem]{ulem}
\usepackage{natbib}
\usepackage{multirow}
\usepackage{textcomp}
\usepackage{rotating}
\usepackage{amsmath}

\newcommand{\specialcell}[2][c]{\begin{tabular}[#1]{@{}c@{}}#2\end{tabular}}

\begin{document} 

\title{Metis: the Solar Orbiter visible light and ultraviolet coronal imager\thanks{Metis website: \url{http://metis.oato.inaf.it}}}

\author{Ester Antonucci\inst{\ref{inaf_oato}}\thanks{\email{ester.antonucci@inaf.it}} \and Marco Romoli\inst{\ref{unifi},\ref{inaf_ass}} \and Vincenzo Andretta\inst{\ref{inaf_oac}} \and Silvano Fineschi\inst{\ref{inaf_oato}} \and Petr Heinzel\inst{\ref{astr_inst}} \and J. Daniel Moses\inst{\ref{nasa}} \and Giampiero Naletto\inst{\ref{unipd},\ref{ifn_cnr}} \and Gianalfredo Nicolini\inst{\ref{inaf_oato}} \and Daniele Spadaro\inst{\ref{inaf_oact}} \and Luca Teriaca\inst{\ref{mps}} \and 
Arkadiusz Berlicki\inst{\ref{uniwrocl},\ref{astr_inst}} \and Gerardo Capobianco\inst{\ref{inaf_oato}} \and Giuseppe Crescenzio\inst{\ref{inaf_oato}} \and Vania Da Deppo\inst{\ref{ifn_cnr},\ref{inaf_ass}} \and Mauro Focardi\inst{\ref{inaf_oafi}} \and Fabio Frassetto\inst{\ref{ifn_cnr}} \and Klaus Heerlein\inst{\ref{mps}} \and Federico Landini\inst{\ref{inaf_oato}} \and Enrico Magli\inst{\ref{polito}} \and Andrea Marco Malvezzi\inst{\ref{unipv}} \and Giuseppe Massone\inst{\ref{inaf_oato}} \and Radek Melich\inst{\ref{turnov}} \and Piergiorgio Nicolosi\inst{\ref{unipd}} \and Giancarlo Noci\inst{\ref{unifi}} \and Maurizio Pancrazzi\inst{\ref{unifi},\ref{inaf_ass}} \and Maria G. Pelizzo\inst{\ref{ifn_cnr}} \and Luca Poletto\inst{\ref{ifn_cnr}} \and Clementina Sasso\inst{\ref{inaf_oac}} \and Udo Sch\"uhle\inst{\ref{mps}} \and Sami K. Solanki\inst{\ref{mps}} \and Leonard Strachan\inst{\ref{nrl}} \and Roberto Susino\inst{\ref{inaf_oato}} \and Giuseppe Tondello\inst{\ref{unipd}} \and Michela Uslenghi\inst{\ref{inaf_iasf}} \and Joachim Woch\inst{\ref{mps}} \and
Lucia Abbo\inst{\ref{inaf_oato}} \and Alessandro Bemporad\inst{\ref{inaf_oato}} \and Marta Casti\inst{\ref{altec},\ref{inaf_oato}} \and Sergio Dolei\inst{\ref{inaf_oact}} \and Catia Grimani\inst{\ref{uniurb},\ref{infn_fi}} \and Mauro Messerotti\inst{\ref{inaf_oatr}} \and Marco Ricci\inst{\ref{polito}} \and Thomas Straus\inst{\ref{inaf_oac}} \and Daniele Telloni\inst{\ref{inaf_oato}} \and Paola Zuppella\inst{\ref{ifn_cnr}} \and
Frederic Auch\`ere\inst{\ref{iasp}} \and Roberto Bruno\inst{\ref{inaf_iaps}} \and Angela Ciaravella\inst{\ref{inaf_oapa}} \and Alain J. Corso\inst{\ref{ifn_cnr}} \and Miguel Alvarez Copano\inst{\ref{mps}} \and Regina Aznar Cuadrado\inst{\ref{mps}} \and Raffaella D'Amicis\inst{\ref{inaf_iaps}} \and Reiner Enge\inst{\ref{mps}} \and Alessio Gravina\inst{\ref{sanitas}} \and Sonja Jej\v{c}i\v{c}\inst{\ref{uniljub}} \and Philippe Lamy\inst{\ref{lab_astr}} \and Alessandro Lanzafame\inst{\ref{unict},\ref{inaf_ass}} \and Thimo Meierdierks\inst{\ref{mps}} \and Ioanna Papagiannaki\inst{\ref{mps}} \and Hardi Peter\inst{\ref{mps}} \and German Fernandez Rico\inst{\ref{mps}} \and Mewael Giday Sertsu\inst{\ref{helm}} \and Jan Staub\inst{\ref{mps}} \and Kanaris Tsinganos\inst{\ref{uniat}} \and Marco Velli\inst{\ref{unical}} \and Rita Ventura\inst{\ref{inaf_oact}} \and Enrico Verroi\inst{\ref{trento}} \and Jean-Claude Vial\inst{\ref{iasp}} \and Sebastien Vives\inst{\ref{lab_astr}} \and Antonio Volpicelli\inst{\ref{inaf_oato}} \and Stephan Werner\inst{\ref{mps}} \and Andreas Zerr\inst{\ref{mps}} \and 
Barbara Negri\inst{\ref{asi}} \and Marco Castronuovo\inst{\ref{asi}} \and Alessandro Gabrielli\inst{\ref{asi}} \and Roberto Bertacin\inst{\ref{asi}} \and Rita Carpentiero\inst{\ref{asi}} \and Silvia Natalucci\inst{\ref{asi}} \and
Filippo Marliani\inst{\ref{estec}} \and Marco Cesa\inst{\ref{estec}} \and Philippe Laget\inst{\ref{estec}} \and
Danilo Morea\inst{\ref{ohb}} \and Stefano Pieraccini\inst{\ref{ohb}} \and Paolo Radaelli\inst{\ref{ohb}} \and Paolo Sandri\inst{\ref{ohb}} \and Paolo Sarra\inst{\ref{ohb}} \and Stefano Cesare\inst{\ref{thales}} \and Felice Del Forno\inst{\ref{thales}} \and Ernesto Massa\inst{\ref{thales}} \and Mauro Montabone\inst{\ref{thales}} \and Sergio Mottini\inst{\ref{thales}} \and Daniele Quattropani\inst{\ref{thales}} \and Tiziano Schillaci\inst{\ref{thales}} \and Roberto Boccardo\inst{\ref{thales}} \and Rosario Brando\inst{\ref{thales}} \and Arianna Pandi\inst{\ref{thales}} \and Cristian Baietto\inst{\ref{thales}} \and Riccardo Bertone\inst{\ref{thales}} \and Alberto Alvarez-Herrero\inst{\ref{inta}} \and Pilar Garc\'ia Parejo\inst{\ref{inta}} \and Mar\'ia Cebollero\inst{\ref{inta}} \and Mauro Amoruso\inst{\ref{sitael}}  \and Vito Centonze\inst{\ref{sitael}}}

\titlerunning{Metis -- the Solar Orbiter coronal imager}
\authorrunning{Antonucci et al.}

\institute{INAF -- Astrophysical Observatory of Torino, Italy\label{inaf_oato} \and University of Florence, Italy\label{unifi} \and INAF associated scientist\label{inaf_ass} \and INAF -- Astronomical Observatory of Capodimonte, Naples, Italy\label{inaf_oac}\and Astronomical Institute of the Czech Academy of Sciences, Czech Republic\label{astr_inst} \and NASA HQ, Washington DC, US\label{nasa} \and University of Padua, Italy\label{unipd} \and IFN CNR Padova, Italy\label{ifn_cnr} \and INAF -- Astrophysical Observatory of Catania, Italy\label{inaf_oact} \and Max-Planck-Institut f\"ur Sonnensystemforschung (MPS), G\"ottingen, Germany\label{mps} \and University of Wroclaw, Astronomical Institute, Poland\label{uniwrocl} \and INAF -- Astrophysical Observatory of Arcetri, Florence, Italy\label{inaf_oafi} \and Politecnico di Torino, Italy\label{polito} \and University of Pavia, Italy\label{unipv} \and Turnov OPToElectronic Centre, Czech Republic\label{turnov} \and Naval Research Laboratory, Washington DC, US\label{nrl} \and INAF -- Institute for Space Astrophysics and Cosmic Physics, Milan, Italy\label{inaf_iasf} \and ALTEC, Torino, Italy\label{altec} \and University of Urbino ``Carlo Bo'', Italy\label{uniurb} \and INFN, Florence, Italy\label{infn_fi} \and INAF -- Astrophysical Observatory of Trieste, Italy\label{inaf_oatr} \and Institut d'Astrophysique Spatiale, Paris, France\label{iasp} \and INAF -- Institute for Space Astrophysics and Planetology, Rome, Italy\label{inaf_iaps} \and INAF -- Astronomical Observatory of Palermo, Italy\label{inaf_oapa} \and Sanitas, EG, Milano, Italy\label{sanitas} \and Faculty of Mathematics and Physics, University of Ljubljana, Slovenia\label{uniljub} \and Laboratoire d'Astrophysique de Marseille, France\label{lab_astr} \and University of Catania, Italy\label{unict} \and Helmholtz-Zentrum, Berlin, Germany \label{helm} \and University of Athens, Greece\label{uniat} \and University of California, Los Angeles, California, US\label{unical} \and INFN -- Trento Institute for Fundamental Physics and Applications, Trento, Italy\label{trento} \and Agenzia Spaziale Italiana, Roma, Italy\label{asi} \and ESTEC/ESA, Noordwijk, Netherlands\label{estec} \and OHB Italia, Milano, Italy\label{ohb} \and Thales Alenia Space Italia, Torino, Italy\label{thales} \and Instituto Nacional de T\'ecnica Aeroespacial, Madrid, Spain\label{inta} \and SITAEL, Bari, Italy\label{sitael}}

\date{}

\abstract{}{Metis is the first solar coronagraph designed for a space mission and is capable of performing simultaneous imaging of the off-limb solar corona in both visible and UV light. The observations obtained with Metis aboard
the Solar Orbiter ESA-NASA observatory will enable us to diagnose, with unprecedented temporal coverage and spatial resolution, the structures and dynamics of the full corona in a square field of view (FoV) of  $\pm 2.9$\degr\ in width, with an inner circular FoV at 1.6\degr, thus spanning the solar atmosphere from 1.7~$R_\odot$ to about 9~$R_\odot$, owing to the eccentricity of the spacecraft orbit. Due to the uniqueness of the Solar Orbiter mission profile, Metis will be able to observe the solar corona from a close (0.28 AU, at the closest perihelion) vantage point, achieving increasing out-of-ecliptic views with the increase of the orbit inclination over time. Moreover, observations near perihelion, during the phase of lower rotational velocity of the solar surface relative to the spacecraft, allow longer-term studies of the off-limb coronal features, thus finally disentangling their intrinsic evolution from effects due to solar rotation.}
{Thanks to a novel occultation design and a combination of a UV interference coating of the mirrors and a spectral bandpass filter, Metis images the solar corona simultaneously in the visible light band, between 580 and 640~nm, and in the UV \ion{H}{i}~Lyman-$\alpha$ line at 121.6~nm.
The visible light channel also includes a broadband polarimeter able to observe the linearly polarised component of the K corona. The coronal images in both the UV \ion{H}{i}~Lyman-$\alpha$   and polarised visible light are obtained at high spatial resolution with a spatial scale down to about 2000~km and 15000~km at perihelion, in the cases of the visible and UV light, respectively. A temporal resolution down to 1 second can be achieved when observing coronal fluctuations in visible light.}
{The Metis measurements, obtained from different latitudes, will allow for complete characterisation of the main physical parameters and dynamics of the electron and neutral hydrogen/proton plasma components of the corona in the region where the solar wind undergoes the acceleration process and where the onset and initial propagation of coronal mass ejections (CMEs) take place. The near-Sun multi-wavelength coronal imaging performed with Metis, combined with the unique opportunities offered by the Solar Orbiter mission, can effectively address crucial issues of solar physics such as: the origin and heating/acceleration of the fast and slow solar wind streams; the origin, acceleration, and transport of the solar energetic particles; and the transient ejection of coronal mass and its evolution in the inner heliosphere, thus significantly improving our understanding of the region connecting the Sun to the heliosphere and of the processes generating and driving the solar wind and coronal mass ejections.}
{This paper presents the scientific objectives and requirements, the
overall optical design of the Metis instrument, the thermo-mechanical design, and the processing and power unit; reports on the results of the campaigns dedicated to integration, alignment, and tests, and to the characterisation of the instrument performance; describes the operation concept, data handling, and software tools; and, finally, the diagnostic techniques to be applied to the data, as well as a brief description of the expected scientific products. The performance of the instrument measured during calibrations ensures that the scientific objectives of Metis can be pursued with success.}{}


\maketitle

\section{Introduction}
\label{sec:intro}

In the last two solar cycles, namely 23 and 24, the coronagraphs of the Solar and Heliospheric Observatory, SOHO, launched in 1995 have enormously enhanced our capability to study the solar atmosphere. The Large Angle Spectroscopic Coronagraph, LASCO \citep{brueckner1995}, formed by a suite of three visible light coronagraphs, has significantly extended the spatial coverage of coronal imaging out to 30~solar radii ($R_\odot$), with a dynamic range able to reveal  faint coronal structures and weak coronal mass ejections (CMEs). Previous space-borne coronagraphs, on the other hand, imaged the corona  out to at most 10 $R_\odot$. Concomitantly, the Ultraviolet Coronagraph Spectrometer, UVCS \citep{kohl1995a}, allowed us to gain access to the unexplored domain of the ultraviolet spectroscopy of the outer  corona, thus permitting complete diagnostics out to several solar radii of the physical properties and dynamics of the plasma of the solar atmosphere \citep{kohl1997,raymond1997a,noci1997,antonucci1998,antonucci1999,cranmer1999}.

The idea of using the Doppler dimming of the resonantly scattered UV emission to measure the outward expansion of the hot corona, put forward by Giancarlo Noci in the seventies as quoted by \cite{withbroe1982a}, was the basis of the development of the UVCS, one of the most innovative instruments of the SOHO mission. The UVCS, and its prototype on Spartan 201 \citep{kohl1995b}, made the first observations of the continuous outward  flow of the corona, which generates the solar wind, and allowed for the identification of the signatures of energy deposition in the region where the wind accelerates. 
The UVCS instrument, being primarily dedicated to UV spectroscopy, had excellent spectroscopic capabilities although within a limited instantaneous field of view (FoV). 

Metis\footnote{According to the Greek mythology, Metis, Zeus first spouse and Athena mother, was the Titaness of all wisdom and knowledge.}, the coronagraph included in the payload of Solar Orbiter, is based on the UVCS heritage and diagnostic methods. The aim of Metis is to extend the knowledge acquired with the SOHO UV coronagraph-spectrometer by focusing on the imaging of the full off-limb corona, with the purpose of tracing its dynamics and evolution on a global-scale and at high temporal resolution. Metis is designed to simultaneously detect both the polarised visible light and the \ion{H}{i}~Lyman-$\alpha$ line emission, thus allowing the use of the Doppler dimming technique to derive the outflow speed of the coronal wind.  The coronagraph will image the full off-limb corona and inner heliosphere in broadband (580--640~nm) polarised visible light (VL) and the narrow-band ($121.6 \pm 10$~nm) UV \ion{H}{i}~Lyman-$\alpha$ line, with unprecedented temporal coverage and spatial resolution. For instance, a spatial scale of about 2000~km is achieved in visible light when the spacecraft is at the closest perihelion at 0.28~AU, and a temporal resolution down to 1 second can be achieved when observing coronal fluctuations in visible light. The FoV of the coronagraph spans over a wide range of heliocentric distances owing to the eccentric orbit of the spacecraft, thus permitting the study of the solar atmosphere from 1.7~$R_\odot$ to about 9~$R_\odot$, during the nominal observation windows along the orbit; although only part of this range will be accessible at a given heliodistance. 

The polarised visible light observations of the corona outline the shape and evolution of coronal structures, modelled by the solar magnetic field. 
These observations provide a direct way of measuring  the electron density, a quantity which is required to relate the dimming of the \ion{H}{i}~Lyman-$\alpha$ line to the expansion speed of the coronal plasma. 

The \ion{H}{i}~Lyman-$\alpha$ line in the corona is primarily produced through resonance scattering of the chromospheric \mbox{\ion{H}{i}~Lyman-$\alpha$} photons by the tiny fraction of residual neutral hydrogen still present in the hot coronal plasma as first deduced by  \cite{gabriel1971}. This line is Doppler dimmed depending on the outflow velocity of the neutral hydrogen component of the expanding corona.  This is because in the frame of reference of the expanding solar wind, the relative wavelength shift of incident chromospheric photons and coronal absorbing profiles causes a dimming of the resonant emission relative to that of a static corona \citep{withbroe1982a, withbroe1982b}.  The intensity reduction of the coronal  \ion{H}{i}~Lyman-$\alpha$ line, increasing with the outflow speed, is detectable in the velocity range  from about 80~to  500~km~s$^{-1}$. The dimming effect of the  \ion{H}{i}~Lyman-$\alpha$ line intensity is negligible for velocities lower than 80~km~s$^{-1}$; whereas,  for speeds higher than 500~km~s$^{-1}$, this line is almost completely dimmed. Therefore, the dimming effect is sensitive to the typical outflow velocities of neutral hydrogen atoms measured in the outer corona \citep[e.g.,][and references therein]{dolei2018}. Moreover,  Doppler dimming is the only technique available for investigating the coronal wind dynamics on the basis of optical line emission, when no tracers, such as for instance coronal blobs \citep{sheeley1997}, can be identified.

Since protons in the outer solar corona are still coupled to neutral hydrogen through different processes, such as photo-ionisation, collisional ionisation, radiative recombination, and charge-exchange \citep{olsen1994,allen1998,allen2000}, up to a heliodistance that depends on the physical conditions of the coronal region, Metis measurements allow for complete characterisation of the main properties and dynamics of the two most important constituents of the coronal and solar wind plasma, namely the electron component and the proton component. 

The objective of Metis is to achieve a better understanding of energy deposition and heating processes occurring in the extended corona and causing its expansion. The key issues are the study of energy deposition processes taking place in the polar regions, where the fast solar wind is generated and accelerated, and the identification of the sources of the slow solar wind at lower latitudes, in order to assess their contribution in different phases of the solar cycle. Related to these issues, Metis observations will advance our comprehension of the influence on the wind velocity of the magnetic field topology which channels coronal outflows along the open field lines, as well as our understanding of the origin of the continuous fluctuations of visible and UV light emission found in the corona and of their possible role in accelerating the solar wind on its way toward interplanetary space. Concerning the topic of coronal activity, the main goals are to trace the evolution of the global corona in order to understand how CMEs are triggered and to comprehend how protons, as well as other particles, are accelerated to almost relativistic velocities at the shock front formed during their outward propagation.

The coronagraph is designed to exploit the unique mission profile of Solar Orbiter \citep[e.g.,][]{marsch2002,mueller2019} and to take advantage of close-up observations of the solar corona from different latitudes \citep{antonucci2001,antonucci2005a}. In particular, the observations performed out of the ecliptic will provide a completely new perspective of the solar corona and its longitudinal structure and evolution. The quasi-heliosynchronous conditions achieved along the orbit when approaching perihelion are important to disentangle effects due to the intrinsic evolution of the coronal features and those due to solar rotation. This is essential to understanding both the nature of coronal fluctuations and the evolution and re-configuration of the corona, primarily prior to and during the early phase of coronal eruptions. Metis observations of the dynamic corona, combined with those obtained with the complementary instruments of Solar Orbiter, thus allow for a thorough investigation of the connections of the heliosphere with its roots on the Sun. 

\section{Scientific objectives}
\label{sec:sci_objectives}

\begin{table*}[h]
\centering
\caption{Key scientific questions of the Solar Orbiter mission addressed by Metis.}\label{tab:questions}
\begin{tabular}{p{0.45\textwidth}|p{0.45\textwidth}}
\hline
\hline
\multirow{2}{*}{Solar Orbiter top-level questions} &  \multirow{2}{*}{Metis contribution} \\
& \\
\hline
What drives the solar wind and where does the coronal magnetic field originate? &
Investigation of the region where the solar wind is accelerated to near its asymptotic value.\\
& \\
How do solar transients drive heliospheric variability? &
Investigation of the region where the first, most dramatic phase of the propagation of CMEs occurs.\\
& \\
How do solar eruptions produce the energetic particle radiation that fills the heliosphere? &
Identification of the path of shock fronts accelerating particles in the solar corona.\\
& \\
How does the solar dynamo work and drive connections between the Sun and the heliosphere? &
Study of the overall magnetic configuration by identifying  the closed and open magnetic field regions in the corona.\\
\hline
\hline
\end{tabular}
\end{table*}

The Metis instrument versatility combined with the characteristics of the Solar Orbiter mission, addresses all the four key scientific questions of Solar Orbiter as summarised in Table~\ref{tab:questions} by providing unique contributions to investigating the following scientific issues:
\begin{itemize}
\item Energy deposition and outflows in the expanding corona.
\item Role of magnetic field lines in channelling the coronal wind.
\item Coronal fluctuations and their role in the solar wind acceleration.
\item Coronal mass ejection onset and early propagation.
\item Eruption of prominences and their propagation in the corona.
\item Global evolution of the streamer belt.
\item Acceleration of the solar energetic particles.
\end{itemize}

Solar Orbiter remote sensing
and \emph{in situ} instruments, as well as observations obtained by simultaneous space missions
dedicated to heliophysics, will complement the Metis observations and facilitate our understanding of the scientific questions being addressed, as discussed in the papers devoted to the Solar Orbiter joint science \citep{auchere2019,horbury2019a}.

\subsection{Energy deposition and outflows in the expanding corona} 

In the corona, the emission of the \ion{H}{i}~Lyman-$\alpha$ line is primarily due to chromospheric photons resonantly scattered by the coronal neutral hydrogen atoms.
The radiative emissivity depends on the velocity, with respect to the chromospheric source of radiation, of the coronal volume element absorbing and re-emitting the radiation, owing to the Doppler effect. If the radiation emitted by the source consists of an emission line,  in the frame of the absorbing hydrogen atoms the relative velocity causes its centre to move away from the centre of the absorbing profile. The consequence is a reduced absorption, and therefore a reduced re-emission. 
Therefore, by measuring this Doppler dimming effect, it is possible to identify the coronal regions where the wind flows and derive the outflow velocity of its neutral hydrogen component. Hydrogen atoms are coupled to the protons that constitute the main component of the solar wind, at least out to 3~$R_\odot$ in the case of the fast wind in coronal holes, and approximately out to 5~$R_\odot$--6~$R_\odot$ in the case of the denser slow wind \citep{olsen1994,allen1998,allen2000}. The coronal electron density, needed to apply the Doppler dimming technique to analyse the \ion{H}{i}~Lyman-$\alpha$ emission, can be derived from the VL polarised emission \citep{vandeHulst:50}. Section~\ref{ops:sci:analysis} describes in detail these diagnostic techniques for Metis. 

The global maps of the outflow velocity of the neutral hydrogen/proton component, which will be obtained with Metis observations, provide the wind velocity gradient and thus enable us to identify where and at which rate the energy deposition occurs in the corona. The height where energy is deposited more efficiently is derived by determining the maximum coronal expansion velocity gradient for the major component of the solar wind \citep{telloni2007}. The deposition rate can be estimated from the measurements of the coronal density, outflow velocity, and wind acceleration. Moreover, the first two quantities allow us to determine the mass and energy flux in the coronal wind. These measurements of the coronal outward expansion directly provide\ the information necessary to discriminate fast and slow streams of the solar wind.

\subsection{Role of magnetic field lines in channeling the coronal wind} 

Metis can outline the magnetic features of the solar atmosphere on the basis of polarised visible light images, and trace the regions where the coronal magnetic field is predominantly open by mapping the coronal plasma outflows in the various wind regimes, thus allowing us to infer the most important features of the solar magnetic topology in terms of connection of the Sun with the heliosphere. 

These observations are also crucial to the assessment of the role of magnetic topology in regulating the solar wind speed and properties and thus to test alternative hypotheses on the origin and sources of the fast and slow wind. UVCS observations of the slow coronal wind \citep{marocchi2001,antonucci2005b,antonucci2006} support the hypothesis that the solar wind speed and the divergence rate of the coronal magnetic field are inversely correlated \citep{noci1997}, and that the expansion factor of flux tubes controls the proton flux, and in turn the Coulomb drag of ions, into the corona. During solar minimum, the slow wind flows predominantly at the edges of coronal holes, where the magnetic flux tubes exhibit a rapid non-monotonic expansion of their cross-sections. Field lines at the coronal hole boundaries diverge much more rapidly than in the core of coronal holes out to about 3~$R_\odot$, and then converge again to attain an almost constant expansion factor \citep{antonucci2012}. These UVCS observations obtained in solar minimum conditions, suggest that both slow and fast wind can originate from the same coronal source and differentiate further out in the corona since they are channelled in flux tubes with different expansion factors.

During near-perihelion phases, the spacecraft orbital motion relative to the solar surface is significantly slower than at 1~AU. This condition favours the investigation of the intrinsic evolution of the coronal magnetic topology, and thus its influence on the wind parameters can be directly assessed. In particular, global maps of the outflow velocity combined with those of the electron density allow us to better resolve the structure, evolution, and dynamics of the streamer borders, where the slow wind is guided by over-expanded magnetic fields \citep[see above discussion, and][]{raymond1997b}. Similarly, it is possible to observe at high resolution the open field lines separating sub-streamers, likely to be the site of the very slow wind component, which further out flows close to the heliospheric current sheet \citep{noci2007}. 

These data are also expected to represent a powerful tool to test the role of transient openings of closed magnetic fields and intermittent reconnection events - such as those generating the coronal plasma blobs at the cusp of helmet streamers \citep{sheeley1997} - in the origin of the slow wind, and assess their contribution, especially when the Sun is active. On the other hand, high-resolution observations of the thin pseudo-streamers, and the corresponding plasma sheet separating same-polarity magnetic field lines, will be performed in order to test the hypothesis that these structures are sources of fast wind flows in the corona \citep{wang2007}. 

\subsection {Coronal fluctuations and their role in the solar wind acceleration} 

The Metis coronagraph can also significantly contribute to the understanding of the nature of coronal density fluctuations and assess their role in transferring energy from the inner corona to the regions where the solar wind plasma accelerates. Coronal fluctuations were observed with UVCS in the outer layers of the solar atmosphere: in polar coronal holes at high latitude - at 1.9~$R_\odot$ \citep{ofman1997,ofman2000}  and at 2.1~$R_\odot$ \citep{bemporad2008} - and in the slow wind, at mid and low latitudes - at 1.7~$R_\odot$  \citep{telloni2009a}. Recently, in a systematic study based on STEREO-A and STEREO-B COR-1 white-light images mapping the outer corona from 1.4 to 4~$R_\odot$, \citet{telloni2013,telloni2014} revealed quasi-periodic, intermittent plasma density fluctuations characterised by a wide range of temporal and spatial scales and observed along the magnetic fields. The investigation of how the observed fluctuations connect to the dynamics of the photosphere on the one hand, and evolve from the corona into the interplanetary space on the other, is still in its infancy. Nevertheless, it is possible to state that the density fluctuations of the fast wind maintain their characteristics in the transition to the heliosphere. On the contrary, a quick dynamic evolution is observed in the slow wind, where large-scale coronal fluctuations with strong phase correlations evolve toward a nearly incompressible turbulent cascade at 0.3~AU \citep{telloni2009b}.

A thorough investigation of coronal fluctuations in different structures and solar wind regimes during the various phases of the solar activity cycle could provide the key to recognising  the sources of fluctuations and fine structures observed in the interplanetary wind. The observations of the solar corona from a vantage point near perihelion, during the periods of slower prograde motion of the Solar Orbiter spacecraft with respect to the solar surface, offer a unique opportunity to isolate the coronal fluctuations arising from the rotation of magnetic flux tubes separated by tangential discontinuities from those to be ascribed to wave propagation and/or inhomogeneities carried by the wind. 
The investigation of the possible connection of these fluctuations, occurring on scales resolvable by the instrument, both with the \ion{H}{i} outflow velocity and with the morphology and evolution of the magnetic coronal structures, will provide valuable information for understanding the transfer of energy to the outflowing solar wind plasma. 

\subsection{Coronal mass ejection onset and early propagation}

A crucial objective of the Metis investigation is that of identifying the mechanism driving the eruption of coronal mass. Coronal mass ejection events commonly result from an abrupt disruption of the balance between the upward pressure exerted by strongly sheared magnetic fields and the downward pressure, either of magnetic nature or due to the weight of an overlying mass. Limitations inherent to coronal observations have until now prevented the identification of the eruption mechanism leading to a CME.

At perihelia, Metis imaging of the solar corona extends down to 1.7~$R_\odot$. Under these conditions, Metis observations, combined with the overlapping ones obtained with the full disk imager of  the Extreme Ultraviolet Imager (EUI) of Solar Orbiter \citep{Rochus2019}, provide an unprecedented coverage of the region where the eruption and early propagation of the  CME occur. Moreover, at perihelion, there is a unique opportunity to observe the evolution of the global corona over a time-scale longer than the 2-3 days passage at the limb of coronal features when observed with coronagraphs located at 1~AU or at the Lagrangian points. This extends the temporal coverage, thus increasing the probability to trace the entire evolution of the event from the pre-eruption to the eruption phase, as well as possibly to the reconfiguration of  the solar corona in response to a CME. This aspect is crucial in order to identify the mechanism/s driving the eruption, to ascertain whether the main source of the flux injection into the heliosphere resides in the corona, and to study the restructuring of the global solar atmosphere following a mass ejection.

The out-of ecliptic vantage point later in the mission offers the opportunity to observe for the first time the corona from mid-latitudes and to investigate the longitudinal structure and distribution of the propagating CME plasma. These measurements will provide the geometry of the \ion{H}{i} and electron global corona and its evolution, providing information on the timing, mass content, and overall dynamics of CMEs. They are also crucial to measuring the directionality of the plasma erupted from the Sun, in order to infer its geo-effectiveness and to predict the impact on the near-Earth environment, when allowed by the orbital phase.

\subsection {Eruption of prominences and their propagation in the corona}

When the CME activation is associable from an observational point of view with an eruptive prominence, this event can be followed in its development \citep{heinzel2016,jejcic2017,susino2018} whenever the prominence temperature is relatively low, thus implying a not-too-high ionisation ratio. During the entire eruptive phase, it will then be possible to evaluate the amount of material moving outwards, and that returning to the solar surface. This relies upon diagnostic capabilities which concern optically thick illuminated plasmas in the initiation phase, up to optically thin illuminated plasmas in the corona. Such diagnostic tools can even be implemented by using the integrated intensity \citep{heinzel2015,patsourakos2007}. Such a diagnostic capability will be enhanced with the combination of the \ion{H}{i}~Lyman-$\alpha$ emission measured by Metis, and the \ion{He}{ii} line intensity obtained with EUI \citep[see, e.g., the joint modelling of the two lines in a cylindrical geometry by][]{labrosse2012}.

\subsection{Global evolution of the streamer belt}

Presently, the physics of the streamer belt is derived from limb observations of the corona on only relatively short timescales, due to the solar rotation effect. However, understanding the global corona is indeed related to the capability of observing its readjustment in response to different phenomena, varying from moderate episodic heating to disruptive CMEs, on still-unexplored temporal scales significantly longer than the mere passage of coronal structures at the limb. Imaging the corona close to the limb, when the spacecraft is near perihelion and the temporal coverage of coronal features is extended, allows us to monitor the longer-term evolution and thus address the physics and evolution of the large-scale corona in response to the evolution of the magnetic field of the Sun, which is changing from a quasi-dipolar configuration at sunspot minimum to a much more complex topology at sunspot maximum \citep[e.g.,][]{solanki2006}.

Out-of-ecliptic observations, in addition, permit investigations of the longitudinal extent of coronal features, and thus the transition of the large-scale coronal structures, projected on the solar equatorial plane, from a slightly warped streamer belt at solar minimum to a more complex configuration when approaching solar maximum, for example, to a quadrupolar structure as predicted by extrapolating the photospheric magnetic field \citep{wang2014}. In addition to the electron density distribution in the solar corona, these observations permit us to derive the mass and energy flux carried away by the solar wind in the solar equatorial plane.

Coronal observations from different latitudes also allow a novel approach to the study of the degree of rigidity of the coronal rotation with heliodistance \citep{lewis1999,giordano2008,mancuso2012}, by tracing the motion of persistent coronal features on the plane of the sky as this moves toward the solar equatorial plane when the spacecraft latitude increases.

\subsection{Acceleration of the solar energetic particles}

Understanding the origin of the solar energetic particles (SEPs) and the large variations of their physical properties from one event to another can be achieved by distinguishing flare-accelerated particles from those associated with CMEs, and in the latter case by determining the key parameters of the shocks and their evolution as the mass ejection propagates outward. The timing of type II bursts, indicating radio emission at the local coronal plasma density due to shock-accelerated electron beams, allows for the unambiguous identification of the shock produced by the passage of a CME \citep{mancuso2009}. Global maps of coronal electron density derived from the polarised visible light images and of the intensity and outflow velocity of the hydrogen corona, both in the pre-CME ambient conditions and during CME propagation, can be combined with radio observations. The joint study of such parameters is essential to identifying the SEP events produced by CMEs, distinguishing them from the events due to the acceleration processes occurring during magnetic reconnection in flares. In addition, this allows us to characterise their associated shocks as they cross the corona, thus increasing our capability of investigating the mechanisms producing solar energetic particles.

\subsection {Synergy with the Parker Solar Probe mission} 

The Solar Orbiter mission is projected to overlap with the NASA Parker Solar Probe (PSP) space mission \citep{fox2016, velli2019}, successfully  launched on August 12, 2018, which is carrying \emph{in situ} instrumentation very close to the Sun in the ecliptic plane. Metis has the capability, when allowed by the relative geometry of the two spacecraft, to image the coronal regions crossed by the Parker Solar Probe during its transit close to the Sun and to characterise the properties of the plasma environment sampled locally by the probe in its journey around the Sun. Thus, for the first time, plasma of the extended solar atmosphere will be simultaneously probed \emph{in situ} and observed by a remote-sensing instrument. This is particularly important in the case of CMEs, whose propagation and evolution can be followed out to the orbit of the Parker Solar Probe ($\sim 9\ R_\odot$, minimum perihelion), when Solar Orbiter is at 0.8~AU and the Metis FoV ranges from 4.2~$R_\odot$ to 12~$R_\odot$. 

\subsection{Joint science with BepiColombo, Proba-3, ASO-S, and Aditya-L1 space missions}

BepiColombo is Europe's first mission to Mercury \citep{benkhoff2010}. Developed by ESA and JAXA (Japan Aerospace Exploration Agency), it was launched in October 20, 2018, and will arrive at Mercury, the smallest and least explored terrestrial planet in the solar system, in late 2025, to gather data during its one-year nominal mission, with a possible one-year extension. Hence, the science phase of the BepiColombo mission will overlap with the last two orbits of the nominal mission phase of Solar Orbiter and hopefully with the first two of the extended mission phase, when the inclination of the orbit with respect to the plane of the ecliptic will be at least 25\degr and higher.
The scientific objectives of BepiColombo include the study of the interaction of the magnetic field of Mercury with the solar wind, as well as the investigation of the influence of the solar activity on the planet magnetosphere.
During the out-of-ecliptic phase of the Solar Orbiter mission, Metis will be able to determine the distribution in longitude of the solar wind plasma streams and CMEs, and single out those potentially affecting the magnetosphere and the surface of Mercury. In this case, the properties of the plasma streams and ejections investigated by Metis can provide essential input to the analysis of the physical properties of the phenomena induced by the solar activity on the magnetised environment of Mercury, detected by the instruments of BepiColombo.

Proba-3 is the first precision formation-flying mission, scheduled to be launched by ESA in 2021 \citep{lamy2010,renotte2014}. The mission will demonstrate formation flying in the context of a large-scale science experiment. The pair of Proba-3 satellites will form in space a 150-m long solar coronagraph - ASPIICS - to study the faint corona of the Sun closer to the solar limb than has ever before been achieved. During its 19-hour highly elliptical Earth orbit, Proba-3 will create a 6-hour-long artificial eclipse, allowing ASPIICS to perform high-spatial-resolution (i.e., 2.5 arcsec) observations of the polarised K-corona and Fe XIV and He D3 E-corona emissions in a field of view ranging from 1.08~$R_\odot$ to 3~$R_\odot$. The imaging of the inner corona will  either overlap or complement the observations of the outer corona obtained with Metis, when it moves from perihelion to  0.5 AU. In this way, the comparison of the solar wind maps derived by Metis with the ASPIICS high-resolution images, outlining the topology of the highly structured inner-coronal magnetic field, will address the question of tracing the sources of the slow solar wind. Similarly, the extended coverage of the corona, made possible by joint data of Metis and ASPIICS, will facilitate the observation of the onset of CMEs in the inner corona and their evolution and propagation in the outer corona. Also, the different view-points of the corona offered by these two platforms will allow, for instance, the study of the 3D
distribution of coronal electrons.

The Advanced Space-based Solar Observatory (ASO-S) is the first Chinese space solar mission, scheduled to be launched in 2022 \citep[][]{gan2015}. The spacecraft, on a polar Earth orbit, will carry a payload consisting of a Lyman-$\alpha$ disk and coronal imager, a full-disk vector magnetograph and an X-ray disk imager, whose main objective is to study the relationship between solar flares, CMEs, and the magnetic field. ASO-S capability of imaging  the solar corona and the disk in Lyman-$\alpha$ will provide the disk brightness for Metis Doppler dimming diagnostics as well as a different coronal imaging point of view that will make it possible to define 3D corona structures.

Aditya-L1 is the first Indian solar space mission, to be launched in 2021 and bound to reach an L1 orbit \citep[][]{seetha2017}. Aditya will be carrying a suite of remote sensing and in-situ instruments, among which an internally occulted coronagraph that will provide images of the lower corona, linking the Metis FoV to the upper solar chromosphere with broadband and narrow-line imaging, spectroscopy, and polarimetry in the visible and near infrared (NIR) wavelength band.

Proba-3, ASO-S, and Aditya coronagraphs will contribute to building a unprecedented coronagraphic network that will efficiently monitor the evolution of the dynamical solar corona.

\begin{figure}
\includegraphics[width=\columnwidth]{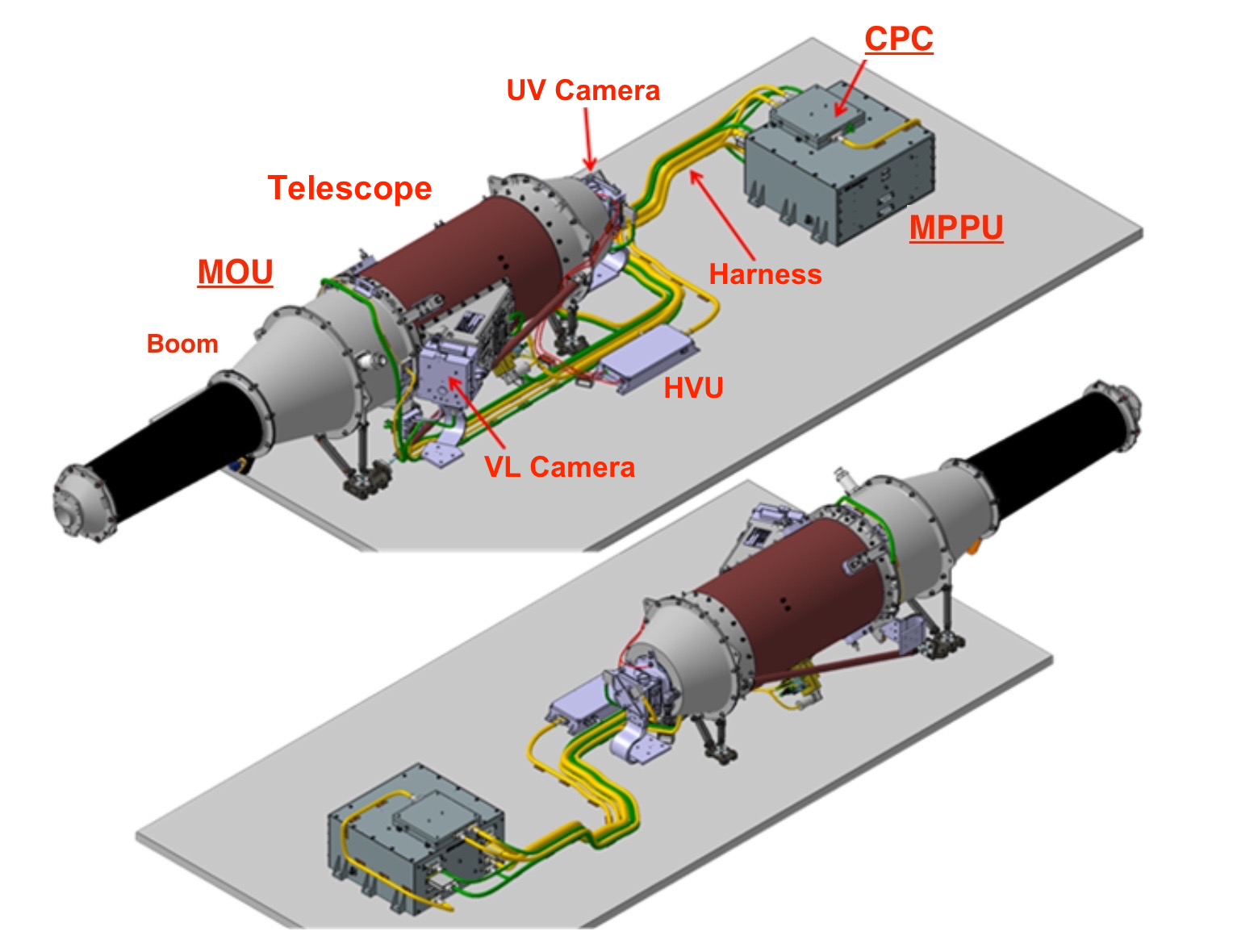}
\caption{Illustration of Metis in its flight configuration, consisting of the Metis optical unit (MOU), the camera power converter (CPC), and  the Metis processing and power unit (MPPU). The high voltage unit (HVU) provides the high voltage to the UV detector.}
\label{fig:metis_view}
\end{figure}

\section{Instrument overview}
\label{sec:instr_overview}

The design of the Metis instrument was driven on the one hand by the scientific objectives described in Sect.~\ref{sec:sci_objectives}, and on the other hand by much more limited resources than those usually available to near-Earth missions and by the very demanding thermal environment encountered by Solar Orbiter. In other words, the Metis design is aimed at obtaining the highest scientific return, while taking into account the effect of the large thermal variations along the Solar Orbiter orbit and minimising the overall resource allocation.  The resources available to Metis are summarised in Table~\ref{tab:metis_resources}. 

\begin{table}[tb]
\caption{Metis resources.}
\label{tab:metis_resources}
\centering
\begin{tabular*}{\columnwidth}{l|l}
\hline
\hline
Total mass & 24.55 kg \\
Telescope envelope & $1430 \times 408 \times 329$ mm$^3$ \\
Electronic box envelope & $282 \times 252 \times 155$ mm$^3$ \\
Average power & 28 W \\
Average telemetry rate & 10.5 kbps\\
Total data volume & 27.2 Gb per orbit\\
\hline 
\hline
\end{tabular*}
\end{table}

During the nominal and the extended mission phases, Solar Orbiter will orbit around the Sun reaching a minimum perihelion distance of 0.28~AU and an aphelion of $\leq 1$~AU. The dramatic variation of heliodistance along the orbit produces strong variations in temperature conditions, thus posing important challenges to the thermal stability of the instrument. Therefore, a key requirement in the design has been defined to minimise the impact of thermo-elastic deformations on the optical performance of the instrument. Moreover, the highly variable distance between the spacecraft and the Earth poses stringent requirements on the telemetry available to the scientific payload. Finally, as Solar Orbiter is an interplanetary mission, there are tight constraints to the mass, volume, power and telemetry data rate available to the payload. For instance, Metis design represented a challenge to limit the instrument total mass, that was 24.55~kg all inclusive, at completion.

In more detail, the coronagraph design was driven by the need to:
\begin{itemize}
\item achieve both VL and UV coronagraphic measurements with a compact instrument suitable to fly aboard a spacecraft reaching the inner heliosphere (minimum perihelion at 0.28~AU), thus respecting stringent mass and volume requirements;
\item cope with the extremely high temperatures ($\sim 400\degr$C) at the entrance aperture,  this being the only instrument of the Solar Orbiter protruding across the heat shield, which in turn reaches temperatures as high as about $500\degr$C;
\item achieve high sensitivity to allow measurements of the weak corona at very different distances from the Sun from 1.7 to 9~$R_\odot$, this range corresponding to the nominal remote-sensing windows;
\item respect the stringent requirement on stray light imposed by broadband VL measurements of the corona, with a stray light to Sun disk brightness ratio lower than $10^{-9}$;
\item achieve the tight tolerances and stabilities of the optical alignment between the telescope and the entrance aperture needed to maintain Sun centre pointing within 1~arcmin for optimal stray-light rejection;
\item attain high precision of the fractional linear polarisation measurement (i.e., minimum detectable linear polarisation) in VL better than or equal to 0.01.
\end{itemize}

\begin{table*}[tb]
\caption{Metis measured key instrumental performances.}
\label{tab:metis_performances}
\centering
\begin{tabular}{l|l}
\hline
\hline
Wavelength spectral ranges & VL: 580--640 nm \\
& UV: $121.6 \pm 10$ nm \\
Field of view & $1.6\degr$--$2.9\degr$/$3.4\degr$ (see Sect.~\ref{sec:fov}) \\
Angular resolution & VL: $\leq 20$ arcsec (except at the edges of the FoV) \\
& UV analogue: $\geq 80$ arcsec \\
Stray light rejection ($B_\text{corona} / B_\text{disk}$) & VL: $ < 10^{-9}$ \\
& UV: $< 10^{-7}$ (by analysis) \\
\hline 
\hline
\end{tabular}
\end{table*}

The instrumental performance of the Metis coronagraph, as measured during the characterisation and calibration campaign on the flight model, is reported in Table~\ref{tab:metis_performances}.

\subsection{Metis hardware units}

 Metis consists of three separate units:
\begin{itemize}
\item Metis optical unit (MOU), the total envelope being $1430\ \text{mm}\times 408\ \text{mm}\times 329\ \text{mm,}$   including the optical head -- consisting of external occultation system, telescope, polarimeter, and two detectors (UV and VL detector) -- and the high voltage unit (HVU), which provides high voltage to the UV detector;
\item Metis processing and power unit (MPPU), being $282\ \text{mm}\times252\ \text{mm}\times 155\ \text{mm}$ in size, and handling all the instrument processes, distributing the power, and providing the data interface with the spacecraft;
\item camera power converter (CPC), a small electronics box to supply low-voltage power to the detectors.
\end{itemize}
The three units are connected by a suitable harness. They are mounted on the -Y panel of the Solar Orbiter spacecraft (S/C) behind the heat shield \citep{scpaper2019}. The telescope receives the coronal light through a dedicated entrance aperture. A schematic representation of Metis in its flight configuration is shown in Fig.~\ref{fig:metis_view}.

\subsubsection{Metis optical unit}
\label{sec:optical_unit}

Metis is an externally occulted solar coronagraph, based on an innovative optical design \citep{fineschi2013}, achieved by combining the characteristics of classical Lyot coronagraphs in visible light with those of UV all-reflective  coronagraphs. The telescope, an on-axis aplanatic Gregorian where both primary and secondary mirrors are aspherical in order to achieve good spatial resolution on both the primary focal plane and the telescope focal plane, is described in detail in Sect.~\ref{sec:opt_unit}. The main idea at the basis of the Metis design is the use of a small circular aperture, instead of a classical large annular one, as instrument entrance aperture, which enables us to minimise the radiation flux inside the instrument, thus minimising thermal problems due to the  proximity to the Sun. 

The other innovative idea in the Metis experiment is the multi-wavelength coating for telescope mirrors. This interference coating (Al/MgF$_2$) has good reflectivity in both the UV (122 nm) and visible light (580-640 nm) spectral ranges. This allows for a compact instrument with two spectral channels in one single optical train. The use of an interference filter (Al/MgF$_2$) acting as a beam splitter allows for simultaneous viewing of both visible - reflected by the filter - and UV - transmitted by the filter - spectral ranges. This scheme was successfully tested with the Sounding Rocket Coronagraphic Experiment (SCORE)  \citep{fineschi2003,romoli2007} onboard the HERSCHEL suborbital mission.

In order to limit the stray light due to the solar disk light entering the instrument aperture and the related thermal issues, Metis is equipped with an on-axis spherical mirror that back-rejects this light through the entrance aperture itself. The internal occulter (IO) is a very critical item for instrument alignment and solar disk stray light suppression. To compensate for possible misalignments induced by launch vibrations or unbalanced thermal environment, Metis was equipped with a mechanism that allows a re-centring of the IO. This is the only internal mechanism present in Metis.

The main optical parameters of this coronagraph are summarised in Table~\ref{tab:metis_opto_parameters}.

\begin{table*}[tb]
\caption{Metis optical parameters.}
\label{tab:metis_opto_parameters}
\centering
\begin{tabular}{l|p{10cm}}
\hline
\hline
Telescope type                   & Externally occulted, on-axis aplanatic Gregorian \\
\\
Overall length                   & 1309 mm \\
\\
Effective focal length           & VL: 200 mm \\
                                 & UV: 300 mm \\
\\
Inverted External Occulter (IEO) & Circular aperture: \diameter\ 40.0 mm \\
                                 & Distance IEO-M0 (boom): 800.0 mm \\
\\
Sun-light rejection mirror (M0)  & Spherical: \diameter\ 71.0 mm \\
                                 & Curvature radius: 1600.0 mm \\
\\
Primary mirror (M1)              & On axis ellipsoidal: outer \diameter\ 160.0 mm, inner \diameter\ 88.0 mm \\
                                 & Curvature radius: 272.00 mm \\
                                 & Conic: $-0.662$ \\
\\                                 
Secondary mirror (M2)            & On axis ellipsoidal: outer \diameter\ 216.0 mm, inner \diameter\ 125.0 mm \\
                                 & Curvature radius: 312.4 mm \\
                                 & Conic: $-0.216$ \\
\\
Internal occulter (IO)           & Distance M1-IO: 154.8 mm \\
                                 & Circular: \diameter\ 5.0 mm \\
\\                                 
Wavelength band-passes           & VL: 580-640 nm \\
                                 & UV: \ion{H}{i} Lyman-$\alpha$ $(121.6 \pm 10)$~nm \\
\\
Detectors                        &VL: APS \\
                                 & - Scale factor 10.7 arcsec/pixel \\
                                 & - Image size: 20.7 mm $(2048\times2048)$ with 10 $\mu$m pixel size\\
                                 & UV: Intensified-APS \\
                                 & - Scale factor 20 arcsec/equivalent pixel \\
                                 & - Image size: 30.7  mm $(1024\times1024)$ with 30 $\mu$m equivalent pixel size at the UV focal plane (microchannel plate), downsized by a factor 2 by a tapered fibre optics to match the sensor 15~$\mu$m pixel size \\
\hline 
\hline
\end{tabular}
\end{table*}

The thermo-mechanical design is one of the most critical issues solved in designing Metis (Sect.~\ref{sec:thermo_design}). The instrument entrance aperture is located on the outside panel of the S/C heat shield, where it reaches temperatures as high as about $400\degr$C. At the same time the coronagraph structure has to hold a detector cooled at $-45\degr$C. The alignment of the instrument, subject to such a large temperature difference, is achieved by constructing a structure divided into the following five sections: 
\begin{itemize}
\item structural boom, which holds the Metis entrance aperture passing through the Solar Orbiter heat shield and is subjected to the highest temperatures; 
\item front cone, which provides an increase of the instrument cross section and a first thermal decoupling; 
\item front cylinder, which introduces a second thermal decoupling;
\item central cylinder assembly, made of carbon fibre reinforced polymer (CFRP), which holds the large majority of Metis subsystems, including the cooled VL detector; 
\item rear cone, which closes the rear part of the instrument and holds the UV detector.
\end{itemize}

\begin{figure*}
\centering
\includegraphics[width=0.75\textwidth]{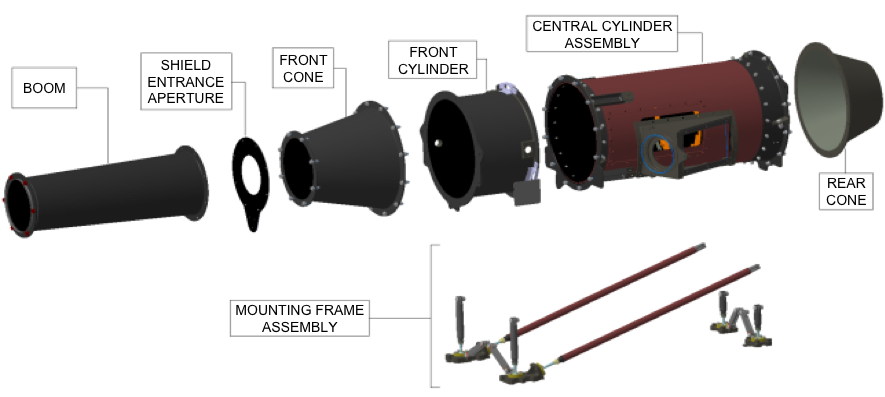}
\caption{Structural parts of the Metis telescope.}
\label{fig:structural_parts}
\end{figure*}

The instrument sections are evident in Fig.~\ref{fig:structural_parts}.

As typical of both coronagraphic and UV space instrumentation, in the development of Metis great care was taken to ensure the cleanliness and contamination control of the optical head: the molecular contamination of the optics was maintained below 100 ng/cm$^2$, and the instrument particulate contamination below 3 ppm. These very stringent values arise from two distinct requirements (Sect.~\ref{sec:cleanliness}). The first derives from the need to limit the possible formation of polymer layers on the optics by UV irradiation of residual molecules, which would reduce the mirror UV reflectivity. The second requirement is to minimise the probability of having particles deposited on critical surfaces, such as the edge of the external occulter, that would be a source of stray light inside the optical path and would greatly reduce the contrast ratio of the coronagraph \citep{fineschi2015,sandri2017a}.

In order to minimise the particle contamination, both a sintered metal filter is used on Metis purging/venting holes, and an ejectable sealing cap is mounted at the Metis entrance aperture during all integration activities and during launch. Thanks to these protecting devices, Metis has also maintained the required clean conditions during the S/C integration activities, which were performed in an ISO 8 environment. The instrument integration was instead performed in ISO 5 environment.

The sealing cap ejection system is a one-shot mechanism planned for use at the very beginning of the mission after the outgassing of the instrument. Afterwards, the Metis aperture remains open and direct Sun disk light protection is provided when needed by 
closing the Solar Orbiter door, which is provided for all the Solar Orbiter remote sensing instruments \citep{scpaper2019}. This is essentially a movable shutter, which does not seal the aperture. The Metis door is fundamental for the safety of the instrument, as it must be closed when the S/C is pointing off Sun centre, since a coronagraph is designed specifically to operate in Sun-centred position. If this is not the case, and sunlight enters the instrument from other angles, there are risks  of both overheating the instrument and damaging the sensors. 

Metis is equipped with one detection system for the visible and one for the UV light channel. The visible light detector assembly (VLDA) consists of a visible light camera equipped with a CMOS 
active pixel sensor (APS), 2k $\times$ 2k, 10 $\mu$m pixel. The detector is installed at the end of the visible light optical path, following the polarimeter assembly, on one side of the central cylinder assembly (Fig.\ref{fig:metis_view}). It is thermally connected to the S/C cold element (CE) which provides a heat sink to allow its operations at $-45^\circ$C. The UV light detector assembly (UVDA) consists of three main elements: a) an intensifier with KBr photocathode in optical contact (via a phosphor screen) with b) a fibre optics 2:1 de-magnifying coupler connected to c) a CMOS 
APS STAR1000 (1k $\times$ 1k format, 15 $\mu$m square pixel). The UVDA can operate either in analogue or photon counting mode by suitably changing the intensifier gain. The HVU mounted on the S/C panel provides the high voltages required by the intensifier. Finally, the CPC mounted on the MPPU provides regulated low voltages to both cameras.
\subsubsection{Metis processing and power unit}\label{sec:processing_unit}

\begin{figure}
\centering
\includegraphics[width=\columnwidth]{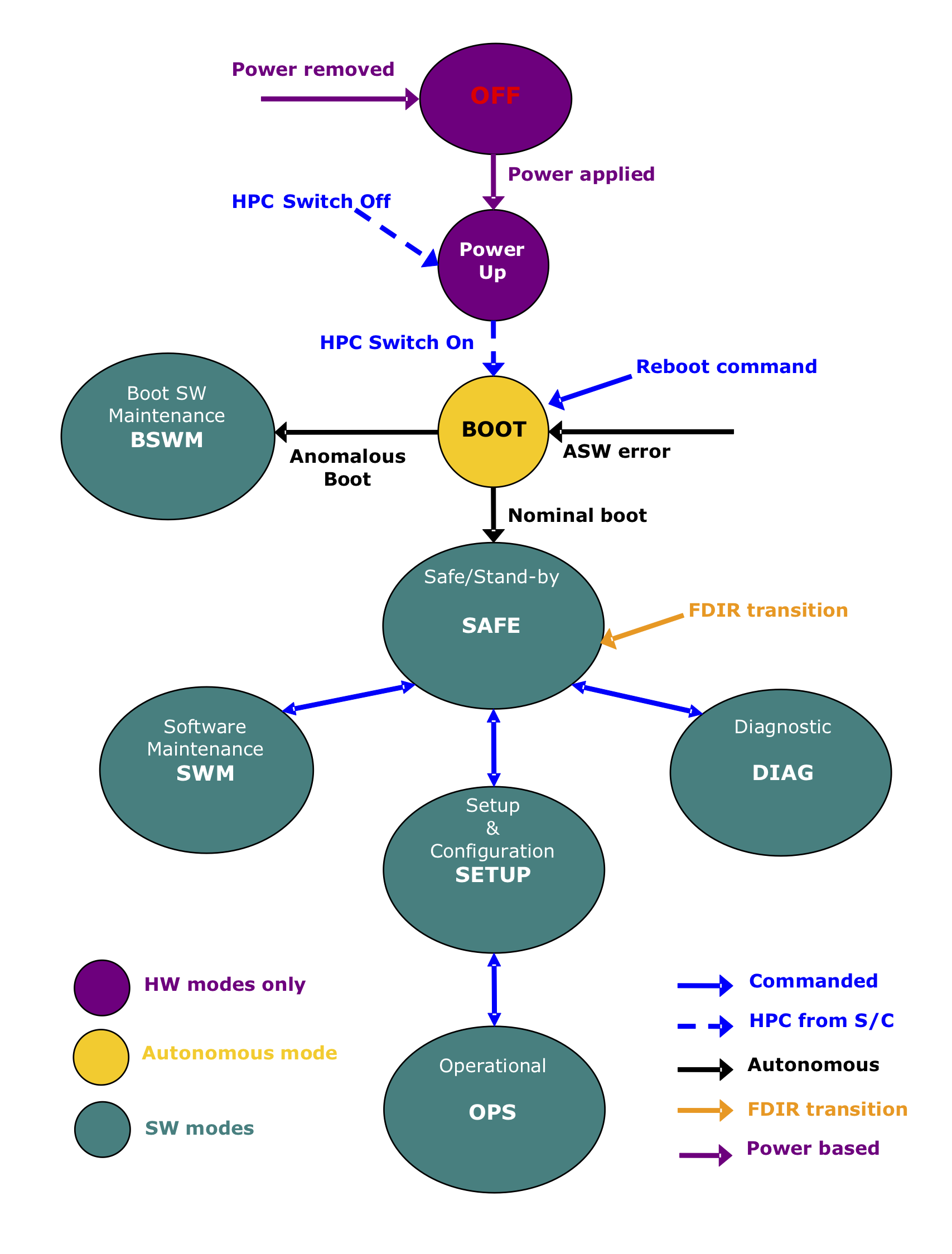}
\caption{Metis operational modes and transitions (in the figure ASW is  Application SoftWare; HPC, High Power Command).}
\label{fig:metis_ops}
\end{figure}

The MPPU is the main electronics unit, described in detail in Sect.~\ref{sec:MPPU}, and is an independent unit also mounted on the $-\text{Y}$ panel of the S/C, in close proximity to the MOU. It consists of the electronic hardware and software that manages the instrument operations, acquires and processes the image data of the VL and UV channels, controls and monitors the telescope mechanism and the thermal hardware installed on the MOU, converts and distributes the power received from the S/C, and provides the data interface with the spacecraft, that is, processing, data storage, and telemetry/telecommand functions. The CPC box is mounted on the MPPU housing top plate (Fig.~\ref{fig:metis_view}).
Metis is designed to operate in different modes (see Fig.~\ref{fig:metis_ops}):
\begin {itemize}
\item Operation mode, used for data acquisitions with both the detectors and for the Metis thermal control system; 
\item Diagnostic mode, applied for possible troubleshooting;
\item Setup \& configuration mode, used for configuring the operative parameters, the diagnostics, and observation setup; 
\item Safe/standby mode, utilised with Metis on, but not operative;
\item Boot and Software maintenance modes.
\end {itemize}

In terms of power budget, Metis uses 24 W during the operation mode, that is, during the large majority of the operative time.

\subsection{Metis and sub-systems model philosophy}

In order to qualify the instrument, a proto-flight qualification approach was adopted at system level. This approach was chosen, instead of a full qualification, because of schedule constraints. Under this philosophy, the following Metis models were developed:
\begin{itemize}
\item Structural thermal model (STM): Metis STM was used for the thermo-mechanical qualification of the telescope.
\item Engineering model (EM): Metis EM was used for verification of the electrical and software interfaces between the instrument and the S/C, verification of the operational modes and procedures, and qualification of the instrument flight software and database.
\item Structural qualification model (SQM): Metis SQM is the refurbished STM used to complete the random and shock qualification tests of the telescope, and to characterise the actual dynamic behaviour of the subsystems to minimise the risk of damage and/or performance degradation during their following qualification.
\item Proto-flight model (PFM): Metis PFM was subjected to environmental tests at acceptance level, to avoid excessive stress on flight hardware.
\end{itemize}

At sub-system level, some components underwent a full qualification process, as was the case for the detectors  for example. Moreover, specific flight spares of sub-systems are available for replacement of possibly failed or damaged equipment at integration and launch site.

\begin{figure}
\includegraphics[width=\columnwidth]{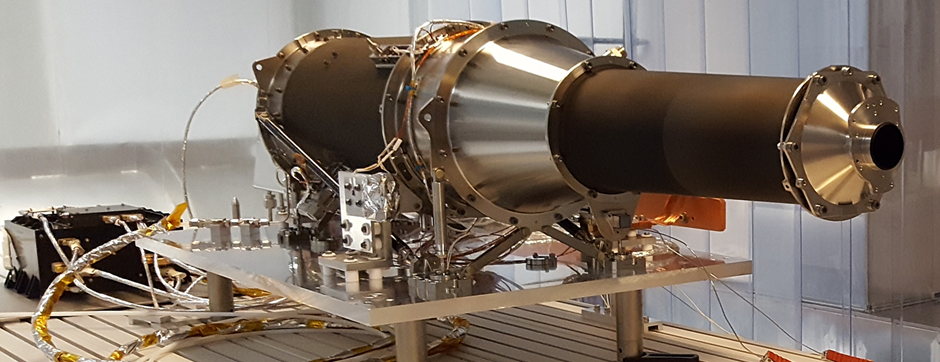}
\caption{Photo of the Metis proto flight model during calibration activities.}
\label{fig:metis_PFM}
\end{figure}

Finally, the Metis PFM was subjected to full functional tests, which demonstrated its flawless functioning, and a full calibration campaign. During calibration activities, described in detail in Sect.~\ref{sec:calibrations}, the optical performance of the instrument was verified, with particular emphasis on stray light characterisation. Figure~\ref{fig:metis_PFM} shows the Metis PFM installed on its holding plate on the clean room optical bench during the calibration activities, when the calibration vacuum chamber was open.

\section{Optical unit}\label{sec:opt_unit}

\subsection{Overall optical design}

The classical design for visible-light solar coronagraphs is based on the traditional optical design that includes a Lyot stop blocking the light diffraction off the entrance pupil. In the UV, the disk-to-corona intensity ratio is orders of magnitude lower than that in the visible-light, and the resulting diffraction off the entrance pupil is negligible. Thus, UV solar coronagraphs would not require Lyot-stops. The design of the Metis coronagraph, conceived to image the visible and UV emission of the solar corona simultaneously, combines characteristics of classical Lyot coronagraphs in VL with those of UV all-reflective coronagraphs. 

Whilst a classical externally occulted coronagraph has an annular aperture, defined by the entrance aperture and the outer edge of a disk that acts as external occulter with the telescope objective optics in the shadow of the external occulter disk, in the case of Metis, the disk-light rejecting mirror is positioned internally and is much smaller than that used in a classical externally occulted coronagraph. The main motivation for adopting this occultation scheme derives from the need to reduce the thermal load inside the instrument during perihelion operations. This innovative design reduces by two orders of magnitude the thermal load on the rejection mirror, and in general inside the Metis structure.
Furthermore, this choice is adopted in order to reduce the large dynamic range of the coronal signal that decreases exponentially from the limb to the outer corona. The highly vignetted aperture of the externally occulted coronagraph at lower heliocentric height compensates in part the high dynamic range, making observations of the corona possible at greater heights. 

Taking into account these considerations, Metis was designed as an externally occulted coronagraph, based on the principle of inverted occultation  \citep{fineschi2013,romoli2017}. The coronal light enters through a circular aperture acting as an inverted external occulter (IEO) located on the outside panel of the S/C heat shield. The IEO consists of a small circular aperture (40~mm diameter), the entrance pupil of the instrument, at the narrow end of a truncated cone. This geometry is chosen in order to minimise the stray light. The IEO is held by a boom at 800~mm in front of the telescope. The disk-light entering through the IEO is back rejected by a spherical heat-rejection mirror (M0,  71~mm diameter) up to $1.1\degr$ (i.e. 1.17~$R_\odot$ at 0.28~AU). 

\begin{figure}
\includegraphics[width=\columnwidth]{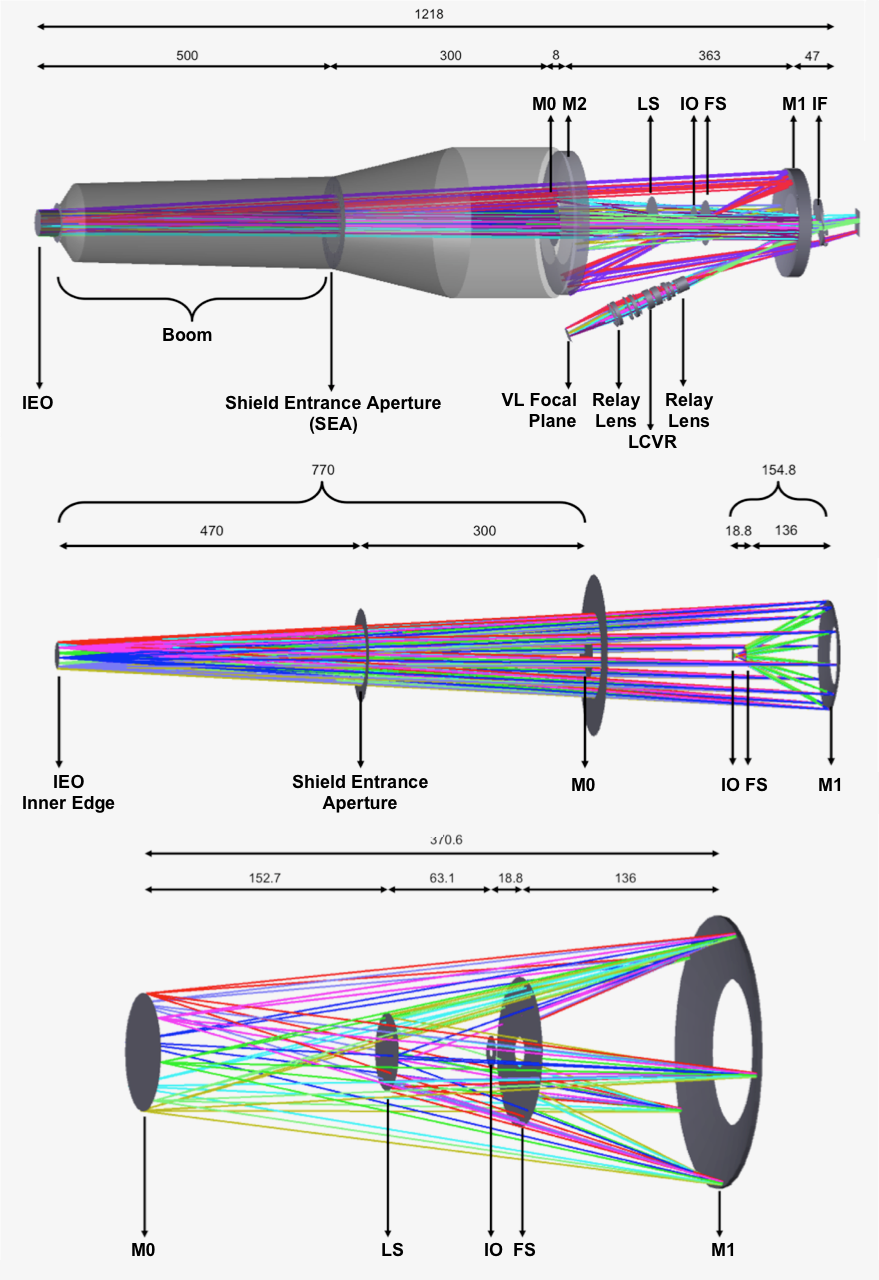}
\caption{Optical layout of the VL and UV light paths in the inverted-occultation coronagraph (measurements in mm). \emph{Top:} Ray-trace of the light path in the Metis coronagraph for both the UV and VL channels. \emph{Middle:}  Primary mirror creates a real image of the diffracting inverted external occulter (IEO) inner edge of the cone at the internal occulter (IO) stop. \emph{Bottom:} Diffraction off the IEO illuminates the edges of the disk-light rejection mirror (M0). This in turn diffracts the disk-light on the primary mirror (M1). Here, M1 creates a real image of the M0 edges that is blocked by the  Lyot stop (LS).}
\label{fig:metis_raytrace}
\end{figure}

The coronal light is collected by an on-axis aplanatic Gregorian telescope (Fig.~\ref{fig:metis_raytrace}, top panel). The suppression of the diffracted light off the edges of the IEO and M0 is achieved with an IO and a Lyot stop (LS), respectively (Fig.~\ref{fig:metis_raytrace}, middle and bottom panels). A field stop (FS) is placed at the primary focal plane, whose function is to limit the outer FoV and provide a further reduction of the stray light. The VL and UV paths are split by a UV interference filter (IF) at $12\degr$ angle of inclination with respect to the optical axis (Fig.~\ref{fig:metis_raytrace}, top panel). The Al/MgF$_2$ coating of the primary (M1) and secondary (M2) telescope mirrors is optimised to enhance the reflectivity at 121.6~nm. The coating also has high reflectivity in the visible light (580--640~nm). The UV narrow bandpass interference filter acts as VL-UV beam splitter by selecting the 121.6~nm UV band in transmission and reflecting the VL to the polarimeter. Inside the polarimeter a broadband filter selects the VL bandpass (580--640~nm). 
On the focal planes of the two optical paths there are the UV and VL detectors. In order to satisfy the scientific requirements, they collect the whole Metis FoV with a 20 arcsec/pixel and 10 arcsec/pixel scale factor, respectively. The two detection systems are described in detail in Sect.~\ref{sec:detectors}.

\begin{figure*}
\centering
\includegraphics[width=0.75\textwidth]{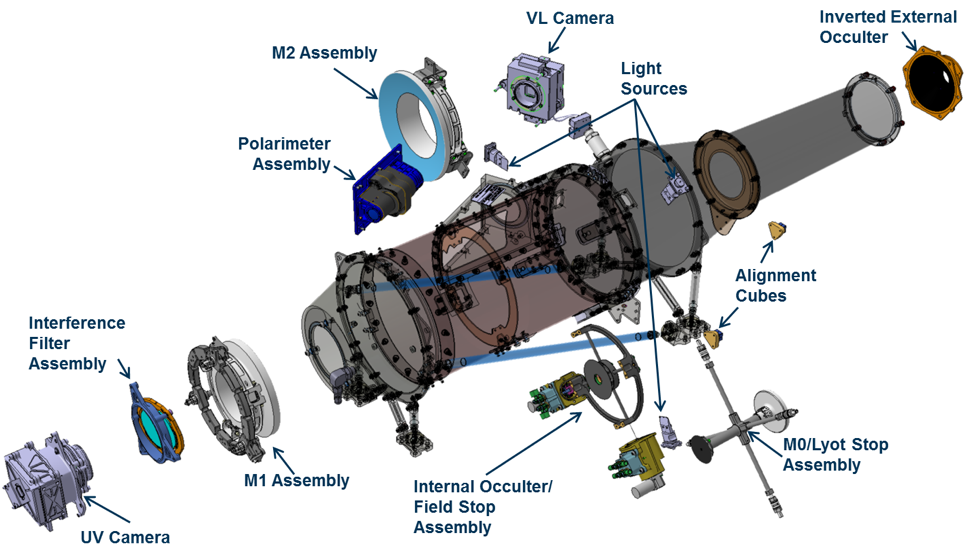}
\caption{Opto-mechanical subsystems of the Metis telescope.}
\label{fig:opto_mechanics}
\end{figure*}

The opto-mechanical subsystems of the Metis telescope are shown in Fig.~\ref{fig:opto_mechanics}.

\subsection{Field of view}\label{sec:fov}
\begin{table}
\centering
\caption{Metis field of view limits expressed in solar radii $R_\odot$ at different FoV angles and S/C heliodistances}\label{tab:fov}
\begin{tabular}{c|c|c|c|c}
\hline
\hline
\multirow{2}{*}{Angular FoV}  & \multicolumn{4}{c}{S/C heliodistance (AU)} \\
    & 0.28    & 0.4    & 0.5    & 0.7 \\
\hline
$1.6 \degr$ & 1.7    & 2.4   & 3.0   & 4.2 \\
$2.9 \degr$ & 3.1    & 4.4   & 5.5   & 7.6 \\
$3.4 \degr$ & 3.6    & 5.1   & 6.4   & 9.0 \\
\hline
\hline
\end{tabular}
\end{table}

\begin{figure}
\centering
\includegraphics[width=\columnwidth]{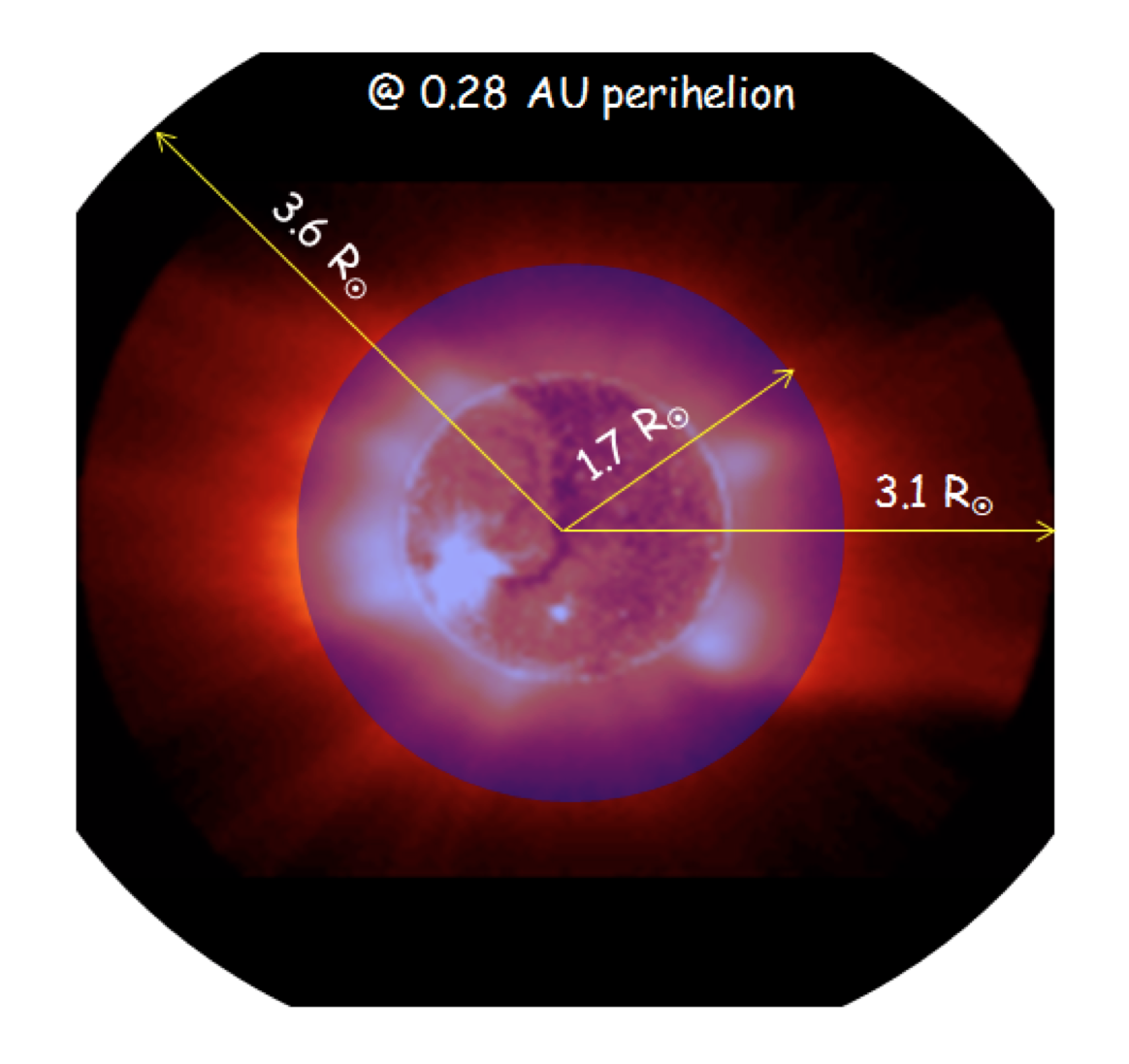}
\caption{Metis field of view for the spacecraft at 0.28~AU from the centre of the Sun.}
\label{fig:FoV}
\end{figure}

The Metis FoV is determined by the following boundaries:
\begin {itemize}
\item Inner FoV, determined by the combined action of the IO and the LS vignetting. It is a circular, smooth boundary at about 1.6\degr.
\item Outer FoV, determined by the detectors size and the FS. The detectors delimit a square of  $\pm 2.9$\degr\ in width. The corners are vignetted by the round field stop at 3.4\degr.
Figure~\ref{fig:FoV} depicts the field of view of Metis in solar radii ($R_\odot$) when Metis is at 0.28 AU perihelion.
Due to the ellipticity of the orbit, the FoV projected on the plane of the sky varies due to a zooming effect. Table~\ref{tab:fov} gives the variation of the FoV in solar radii at selected heliocentric distances of the spacecraft from the Sun.
\end {itemize}

\subsection{Tolerances}

The sensitivity of the optical design to the manufacturing, assembling, and integration, and the short- and long-term stability tolerances was thoroughly investigated. The surface figure accuracy of the mirrors and interference filter takes into account all the contributors: manufacturing, coating deposition, glue shrinkage, transition to zero-g, mismatching of coefficient of thermal expansion between coating and substrate, and thermo-elastic deformation.
All the tolerances, including radius of curvature, conic constant, positions, decentring and tilts, and surface form errors, were statistically combined with Monte Carlo trials, thus obtaining models that allow us to predict the instrument performance at the end of integration and in orbit \citep{dadeppo2013}.

\subsection{Inverted external occulter}
\label{sec:ieo}

\begin{table*}
\caption{Main geometric properties of the Metis mirrors. R is the curvature radius and k is the conic constant.}
\label{tab:mirror1}
\centering
\begin{tabular}{c|c|c|c|c|c|c}
\hline
\hline
\multirow{2}{*}{Mirror} & \multirow{2}{*}{Shape of the Surface} & \multicolumn{2}{c|}{Nominal} & \multicolumn{2}{c|}{Measured} & \multirow{2}{*}{Clear Aperture (Diameter, mm)}\\
& & R (mm) & k & R (mm) & k & \\
\hline
M1 & concave ellipsoid with a central hole & 272.000 & $-0.662$ & 272.003 & $-0.662$ & Annular from 88 to 160 \\
M2 & concave ellipsoid with a central hole & 312.385 & $-0.216$ & 312.388 & $-0.216$ & Annular from 131 to 216 \\
M0 & concave spherical & 1600 & 0 & 1596.34 & 0 & 71\\
\hline
\hline
\end{tabular}
\end{table*}

\begin{table*}
\caption{Measured surface form errors of the Metis mirrors.}
\label{tab:mirror2}
\centering
\begin{tabular}{c|c|c|c|c}
\hline
\hline
\multirow{3}{*}{Mirror} & \multicolumn{2}{c|}{At component Level} & \multicolumn{2}{c}{At assembly Level} \\
& Measured PV & Measured RMS & Measured PV & Measured RMS \\
& (nm) & (nm) & (nm) & (nm)  \\
\hline
M1 & 367 & 44 & 423 & 69  \\
M2 & 740 & 145 & 1360 & 272  \\
M0 & 449 & 82 & 449 & 88  \\
\hline
\hline
\end{tabular}
\end{table*}

\begin{figure}
\centering
\includegraphics[width=0.75\columnwidth]{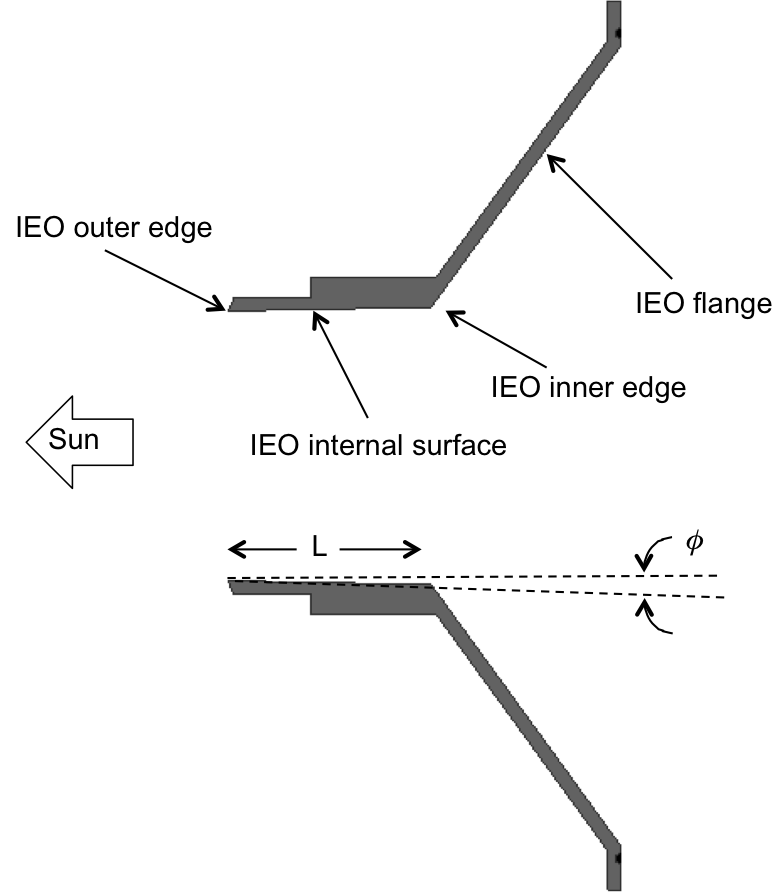}
\caption{Cross-section of the inverted external occulter (IEO) with emphasis on its key features. The shape of the IEO internal surface is conical with length L, and semi-angle $\phi$=1.07\degr}.
\label{fig:IEO_parameters}
\end{figure} 

In order to improve the stray light suppression performance of the entrance aperture, a suitable shape for the IEO was investigated by testing different prototypes. The geometrical parameters that define the chosen shape are the length, $L$=30.0 mm, and the semi-angle, $\phi$=1.07\degr.
The experimental setup and the results are detailed in \cite{Landini-etal:12} and \cite{Landini-etal:13}, showing that the best shape is conical.
Figure~\ref{fig:IEO_parameters} shows the cross-section of the IEO, 
where the design parameters are identified. 

Due to the orbit of the spacecraft, the IEO will undergo severe temperature variations: from $\sim 400\degr$C in the hottest operational case, to $\sim50\degr$C in the coldest operational case, and to $\sim-50\degr$C in the coldest non-operational case, when the external door is  closed. This is the reason for choosing titanium (Ti-6Al-4V) as the material for the IEO manufacturing, due to its low thermal expansion coefficient. The mechanical specification for manufacturing the edge of the IEO was an arc with a maximum radius of 50~$\mu$m. The IEO internal flange is the mechanical connection between the IEO and the boom. The solar disk light scattered by M0 might be reflected towards M1 by the flange surface and by any mechanical edges between the IEO and the boom surface. Therefore, the number of edges was minimised and the tilt of the boom surfaces was accurately designed.

The IEO coating was selected after a dedicated experimental investigation that compared the properties of the bidirectional reflectance distribution function of several coating types \citep{Landini-etal:14}. The selected coating is ACKTAR Magic Black$^\text{\textregistered}$ which guarantees a total integrated scatter below 1\% at normal incidence, and about 1.4\% at an angle of incidence of $50\degr$. ACKTAR Magic Black$^\text{\textregistered}$ was qualified for the thermal environment that Metis is going to face.

The IEO was manufactured by OHB Italia S.p.A..

\subsection{Mirrors}

Metis includes a rejection mirror, M0, and the telescope primary and secondary mirrors, M1 and M2, respectively. The mirrors, manufactured by TOPTEC (Czech Republic) in Schott Zerodur$^\text{\textregistered}$, have the characteristics reported in Table~\ref{tab:mirror1}.

The surface form error of the mirrors was measured at the delivery and during the integration phases in the ISO5 clean room of OHB-Italia by means of a digital interferometer model $\mu$-phase 1000 Trioptics, with an accuracy of $\lambda/20$~PV (peak to valley), a repeatability of $\lambda/500$~PV, and a camera resolution of $1024\times 1024$. 
Results of the form error measurements are listed in Table~\ref{tab:mirror2} and show that the integration of the mirrors inside the telescope structure produces a slight distortion that increases the surface error. Nevertheless, the amount of this distortion does not impact on the instrument performance \citep{sandri2017b}.

The mirror assemblies were manufactured by  OHB Italia. The mirror coating was deposited by Thin Film Physics AG, Switzerland.

\subsection{Internal occulter and mechanism}

The IO is a stop placed at the conjugate plane of the inner aperture of the IEO cone with respect to the primary mirror M1. Its function is to block the primary source of diffracted light from the entrance aperture. The IO is a circular stop with 5 mm diameter that over-occults the image of the IEO inner edge in order to account for the possibility of insufficiently accurate optical images produced by M1 and the budgeted optical misalignment. Given the primary role of the IO in the coronagraph stray light rejection, the stop can be translated perpendicularly to the optical axis by means of the internal occulter mechanism (IOM).

\begin{figure}
\centering
\includegraphics[width=\columnwidth]{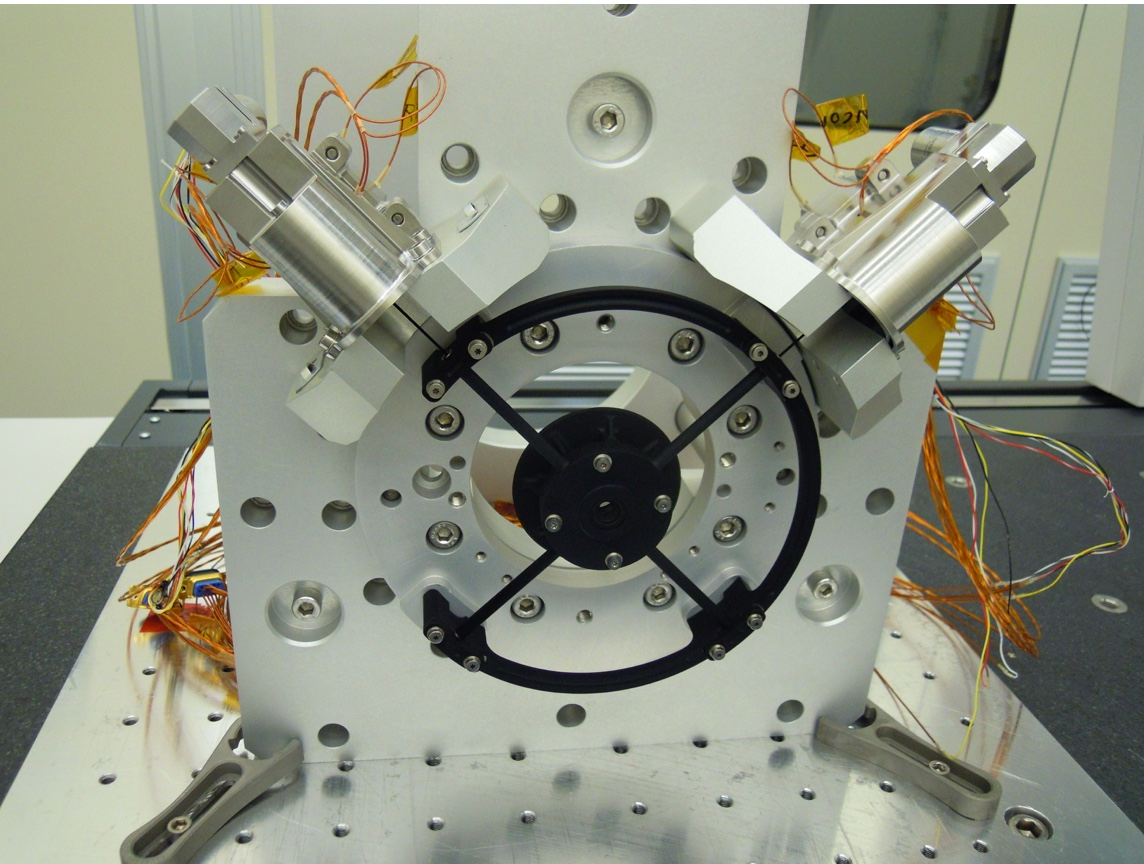}
\caption{The flight model of the internal occulter mechanism (IOM) subsystem after integration in the ISO5 clean room of the Optical Payload Systems facility at ALTEC in Turin. The IOM is the central holed disk, held in position by four spiders; the two translation stages are the top cylinders along the system diagonals.}
\label{fig:iom}
\end{figure}

The IOM subsystem (Fig.~\ref{fig:iom}) is accommodated inside the Metis optical unit, along the optical axis, and is placed at a location corresponding to the intermediate focal plane of the telescope. The IOM function adjusts the IO position to compensate for any external occulter misalignment. After the Metis instrument alignment and calibration, no other IOM activation is foreseen, although the position of the IO can be adjusted, if required, throughout the whole mission.

The IOM is designed to move the IO along two axes within a range of $\pm 1.0$~mm with a step size equal to 6.12~$\mu$m. The material chosen for all parts is titanium Ti-6Al-4V. All external parts are black-coated to reduce stray light. In order to avoid any possible contamination inside the Metis optical unit, the actuation system is external to the telescope structure, and the IO is suspended by flexures. The IOM is composed of fixed and movable parts comprising two main groups: (i) the inner core subsystem and (ii) the actuation system sub-assembly, made of two identical actuation modules. The inner core subsystem parts are all coated by ACKTAR Magic Black$^\text{\textregistered}$ in order to minimise stray light.

The IO and its mechanism were manufactured by OHB Italia.

\subsection{Interference filter assembly}

\begin{figure*}
\centering
\includegraphics[width=0.45\textwidth]{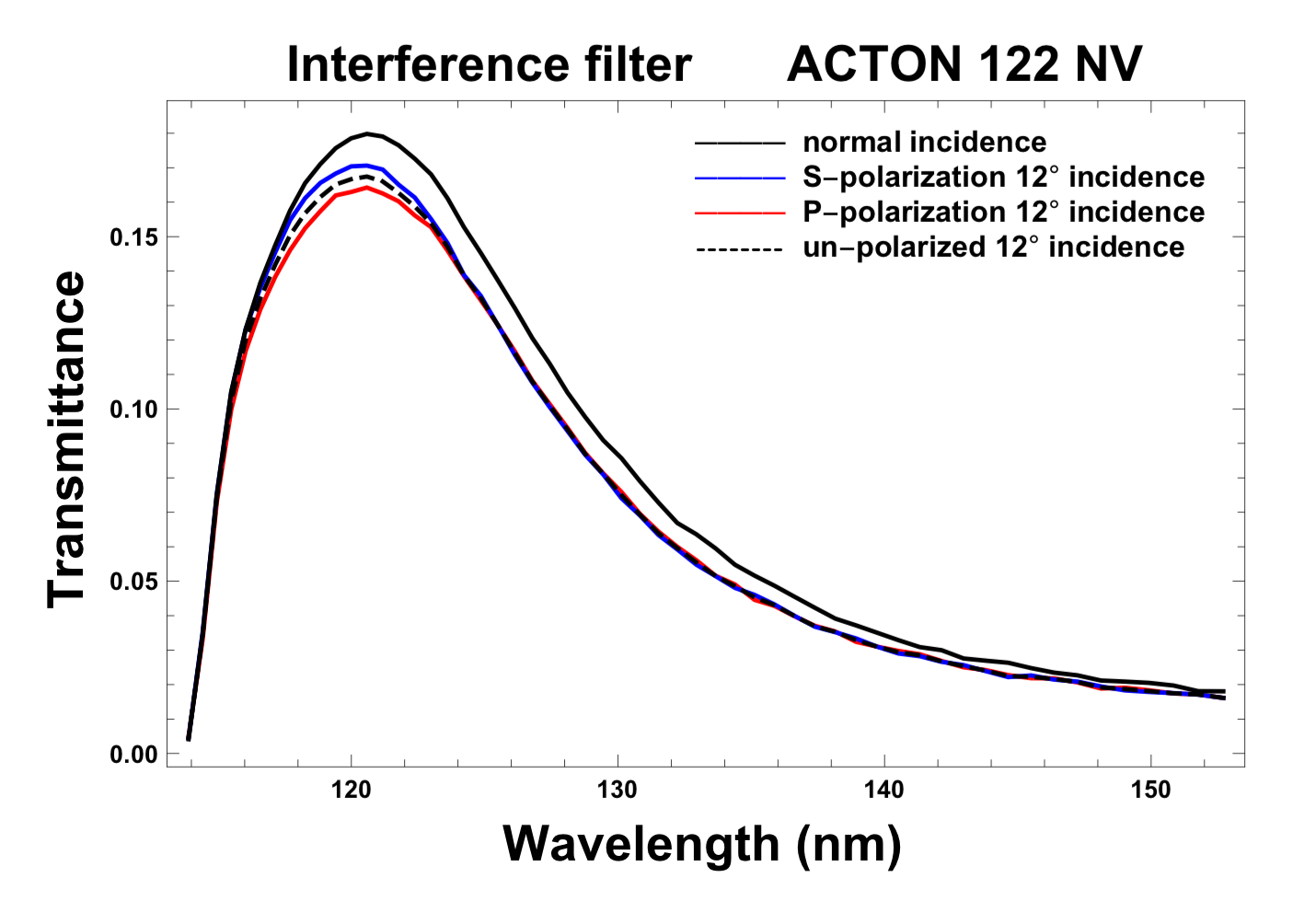}
\includegraphics[width=0.45\textwidth]{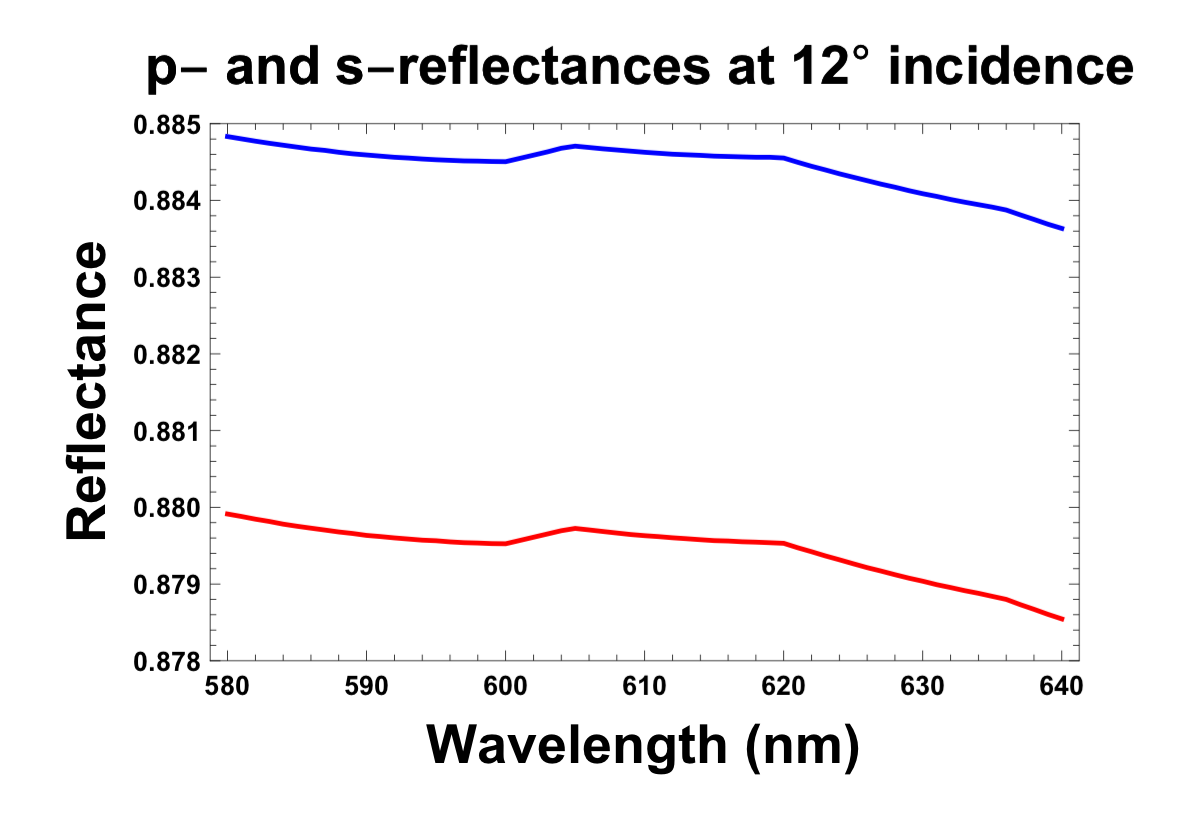}
\caption{UV transmittance (left) and VL reflectance (right) of the interference filter. The interference filter transmits the \ion{H}{i}~Lyman-$\alpha$ radiation at 121.6~nm and  reflects the visible light in the band 580-640 nm. }
\label{fig:ifa}
\end{figure*}

The interference filter assembly (IFA) is located along the optical axis of Metis, downstream from the Lyot occulter. It has the double purpose of selectively transmitting the \ion{H}{i}~Lyman-$\alpha$ radiation (121.6~nm) to be detected by the UV detector assembly, and of reflecting the visible band (580-640 nm) into the Metis polarimeter. The IFA consists of an interferential multilayer filter film (Al/MgF$_2$) deposited on a MgF$_2$ crystal window with the optical axis perpendicular to the coated surface. The IFA is 6~mm thick and 70~mm in diameter, with six flat side pads for the support arms. In order to reflect the VL light to the polarimeter, the IFA is mounted with a 12$\degr$ inclination relative to the telescope optical axis.  This off-axis geometry modifies the transmission properties of the filter with a lower peak transmittance value and a shift towards short wavelengths of the peak value. These properties were carefully checked on test samples. A standard interference filter for 122~nm was tested at the BEAR beamline of Elettra Synchrotron in Trieste. The experimental results of Fig.~\ref{fig:ifa} illustrate these effects. More quantitatively, the results indicate a polarisation coefficient $\mu = 0.017$ at 122~nm. The measured transmittance of the flight model at 122 nm is $0.24\pm 0.01$.
Polarisation is a far more critical issue for the visible light reflecting channel. Here however, the contribution to polarisation is very low, thanks to the limited
angle of incidence of the radiation onto the multilayer film \citep{crescenzio2012}. Measurements performed with a Cary spectrophotometer indicate a polarisation contribution to the reflected beam of $\mu = 0.002\pm 0.003$ with a reflectivity exceeding 0.881. These figures satisfy both visible and UV Metis specifications.

The IFA was manufactured by OHB Italia. The interferential filter (IF) was manufactured by Acton Optics \& Coatings, USA.

\subsection{Polarimeter}

\begin{figure*}
\centering
\includegraphics[width=0.35\textwidth]{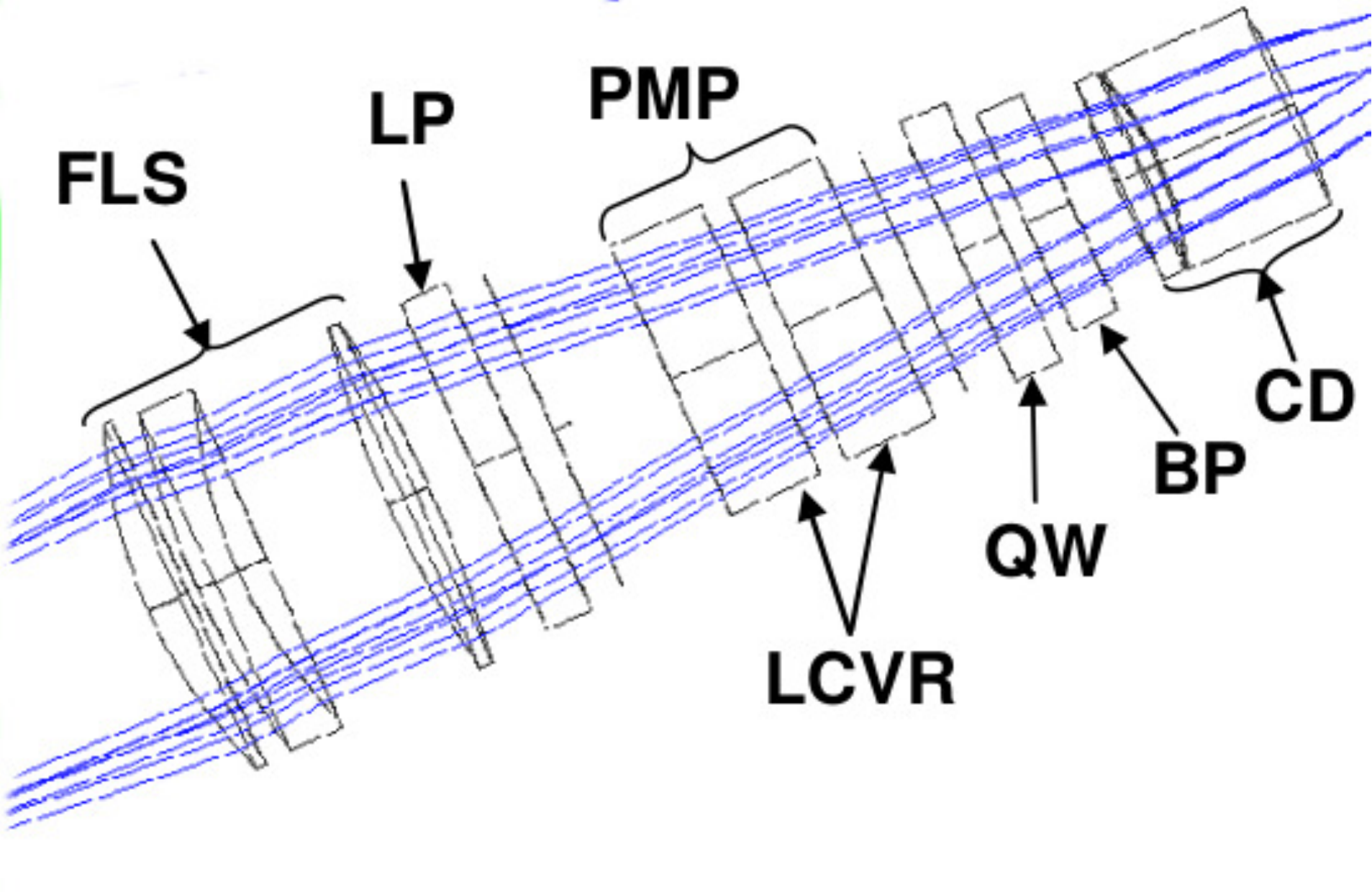}\hspace{2cm}
\includegraphics[width=0.35\textwidth]{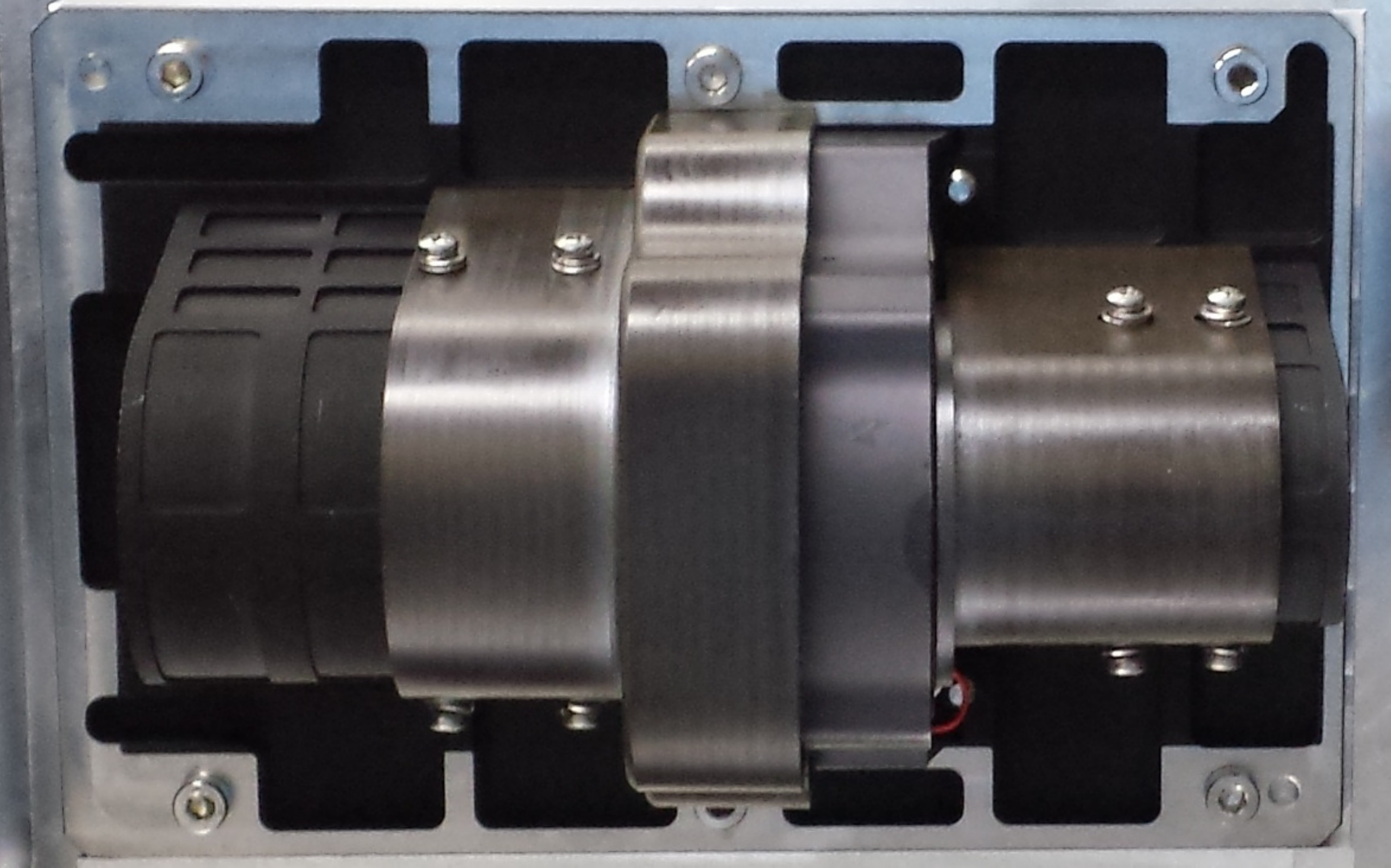}
\caption{\emph{Left:} Schematic view of the Metis polarimeter consisting of the following elements: a focus lens system (FLS),  a linear polariser (LP), a polarisation modulation package (PMP) composed of two liquid crystals-based variable retarders (LCVRs), a quarter-wave (QW) plate retarder, a  bandpass filter (BP), and a collimating doublet (CD). \emph{Right:} Picture of the Metis polarimeter.\label{fig:polarimeter}}
\end{figure*}

The Metis polarimeter, based on a design of a similar K-polarimeter on the SCORE coronagraph \citep{fineschi2005}, measures the linearly polarised solar K-corona in the visible light band from 580 to 640~nm.
The polarimeter consists of the polarisation modulation package and the relay-optics system. The polarisation package electro-optically modulates the intensity of the linearly polarised K-corona. The relay optics collimates the light from the image formed by the telescope on the intermediate VL focal plane and re-focuses the image on the VL detector to match the telescope plate scale with the APS pixel size. 
The Metis polarimeter consists of the following elements (Fig.~\ref{fig:polarimeter}): collimating doublet (CD); bandpass filter (BP), reducing the instrumental bandpass in the range from 580 to 640~nm; quarter-wave plate retarder (QW); polarisation modulation package (PMP), composed of two liquid crystals-based variable retarders (LCVRs) \citep{alvarez2015}; linear polariser (LP) working as analyser; and a focus lens system (FLS). 
The Metis polarimetric measurements provide the total brightness ($tB$) and the polarised brightness ($pB$) of the coronal visible light. 

\begin{table}
\caption{Optical parameters of the Metis polarimeter.}\label{tab:optical_perf}
\centering
\begin{tabular}{l|c}
\hline
\hline
Operating bandpass & 580--640 nm \\
Optical magnification & 0.67 \\
F-number & 4.2 \\
Field of view  & 0.5$\degr$ \\
\hline
\hline
\end{tabular}
\end{table}

\begin{figure*}
\centering
\includegraphics[width=0.85\textwidth]{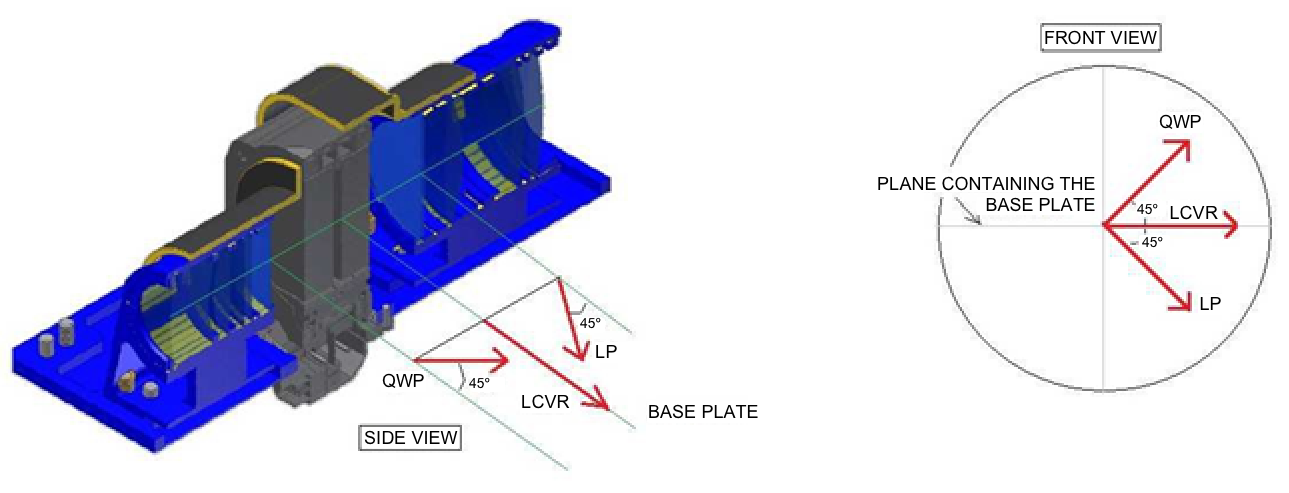}
\caption{Schematic representation of the fast axes alignment of the polarimetric elements of the polarisation modulation package.}\label{fig:axis}
\end{figure*}

The optical parameters of the Metis polarimeter are summarised in Table~\ref{tab:optical_perf}.
The wave-front error, measured close to the external edges of the FoV, has a rms value of 0.38$\lambda$ and a peak-to-valley value of 1.94$\lambda$. The measured intensity ratio of the ghosts, generated by retro-reflection of the optical surfaces of the polarimeter, is 0.7\%, and the measured geometric distortion is lower than $0.1$\%.
The modulation scheme adopted for Metis acquisitions is a modulation package composed of two anti-parallel nematic  LCVRs with the fast axes aligned. 
The modulation state is changed by applying different voltages to the two LCVRs. The polarimeter is in a Senarmont configuration with the orientation of the fast axes of the QW, PMP, and LP as in Fig.~\ref{fig:axis}.

The PMP was developed by the Instituto Nacional de T\'ecnica Aeroespacial, Spain, and the polarimeter was manufactured by OPTEC, in Italy.

\subsection{Detectors}
\label{sec:detectors}

\begin{figure*}
\centering
\includegraphics[width=0.75\textwidth]{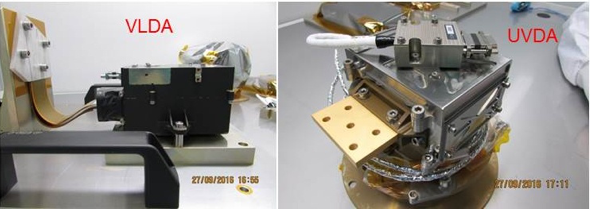}
\caption{Detector units at delivery at the Optical Payload Systems facility at ALTEC in Turin. The two cameras, VLDA, and UVDA  are mounted face down on the transport plates.}\label{fig:detectors}
\end{figure*}

The Metis detection subsystem consists of the  VLDA and of the UVDA (Fig.~\ref{fig:detectors}), each located at the end of its respective optical path. The detection subsystem also includes two service units: a CPC, providing regulated power to both detectors, and a HVU, providing the high voltages needed by the intensifier of the UVDA.

\subsubsection{Visible light detector assembly}

\begin{figure*}
\centering
\includegraphics[width=0.75\textwidth]{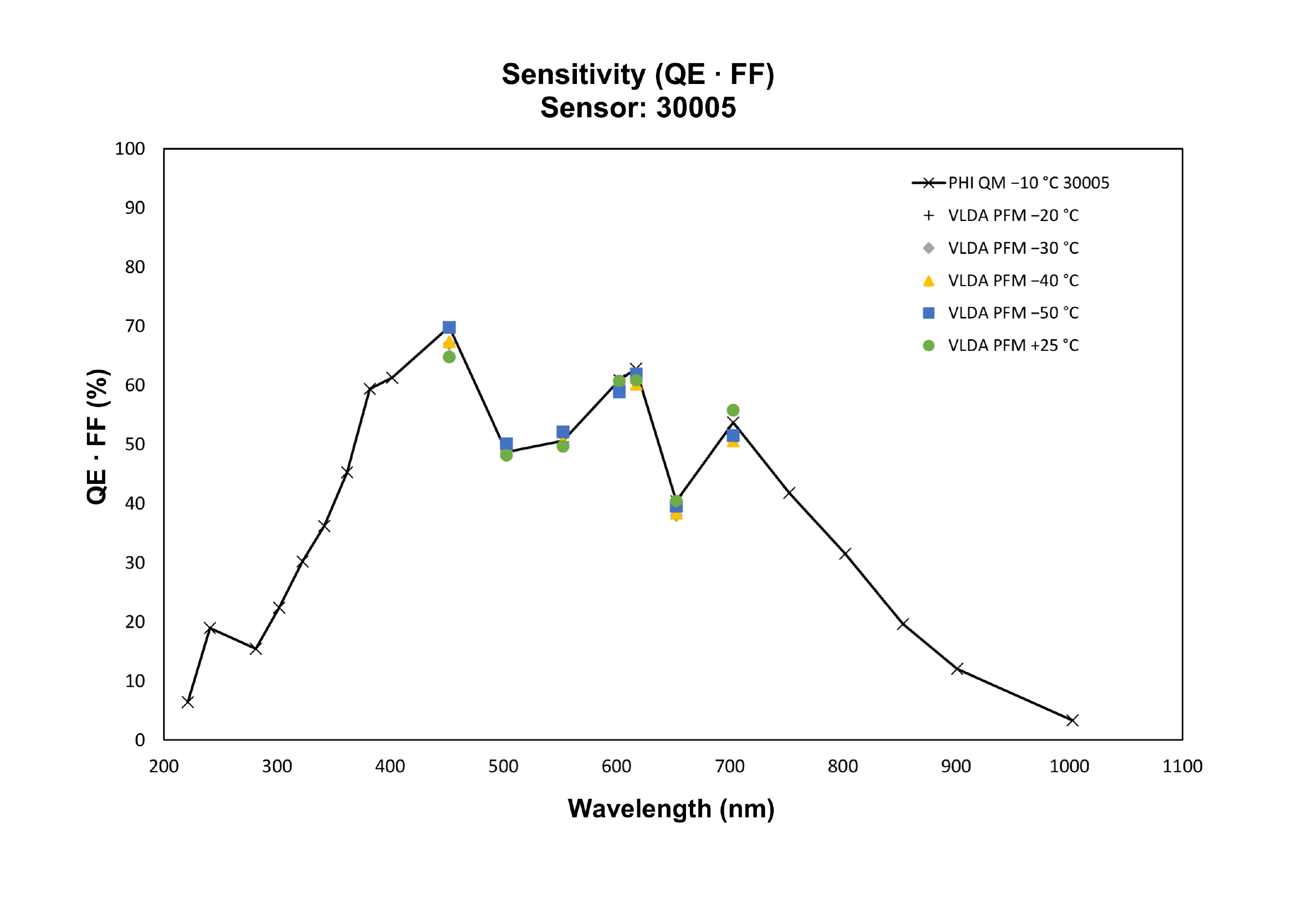}
\caption{Quantum efficiency (QE) of the VLDA sensor, including the pixel fill factor  (FF), as a function of wavelength. The sensitivity values refer to the Metis VLDA proto-flight model  and are compared with the sensitivity of the qualification model of the detector of the Solar Orbiter PHI.}\label{fig:vlda_sensitivity}
\end{figure*}

The Metis VLDA camera is almost identical to the camera employed by the Polarimetric and Helioseismic Imager (SO/PHI) on Solar Orbiter \citep{solanki2019}.

The visible light detector uses a custom-made CMOS-APS sensor with 2048 by 2048 pixels of 10-$\mu$m size, developed by CMOSIS Imaging Sensors (now AMS) in Belgium.
The camera is built around a large beryllium alloy plate that carries the sensor, is attached with thermal glue, and protrudes from the housing to form the camera cold finger. The sensor is connected, via pins passing through the cold finger, to a custom-made sensor socket, soldered to the front end electronics printed circuit board (PCB). Two fast 14-bit analogue-to-digital converters (ADC) are directly coupled to the sensor outputs.
A second PCB contains the components for the command interface to the Metis power and processing unit, the control signals to operate the detector and its support electronics, and to acquire and transfer the image data to the MPPU.

The camera housing is made of aluminium parts with conductive conversion coating; all external surfaces are black-coated. Back-shells and baffles are used to ensure the light tightness of the subassembly. A labyrinth-type structure is built around the sensor to minimise potential contamination from the electronics while avoiding contact with the mounting interface, thus ensuring thermal isolation. The upper part of the housing provides a point-line-plane fixation that guarantees a stress-free mounting over a wide temperature range and at the same time ensures low repeatability error between consecutive mountings. 

The VLDA has an average quantum efficiency (QE; including the pixel fill-factor, FF) of about 55\% in the 580--640~nm range (Fig.~\ref{fig:vlda_sensitivity}) and a dark signal of about 25~e$^-$/s below $-10\degr$C, fully in line with the demanding Metis requirements.
The characterisation campaign was performed at the Max Planck Institute for Solar System Research in G\"ottingen, in Germany.

\begin{table*}
\caption{VLDA sensor characteristics.}\label{tab:vlda_sensor}
\centering
\begin{tabular}{l|l|l|l}
\hline
\hline
Parameter & Unit & Requirement & Value \\
\hline
Format & pixels & 2k x 2k & By design \\
Pixel size & $\mu$m & 10 & By design \\
Shutter & & Rolling & By design \\
Full well capacity & ke- & $>70.0$ &Low gain: $\sim 90.0$ \\
& & & High gain: 38.0 \\
Non linearity & \% & $<2$ & $<2$ \\
Sensitivity (QExFF) within 580--640 nm & \% & $> 40$ & $\sim 55$ \\
Defective complete rows/columns & & None & None \\
Pixel response non uniformity (local) & \% & $<1.5$ & Typical $\sim 1$ \\
Dark current at beginning of life (BoL) at $-40\degr$C & e$^-$/s/pixel & $\leq 25$ & $\sim25$ below $-10\degr$C \\
Readout noise & dB & N/A & $>60$ \\
Pixel gain & DN/e$^-$ & N/A & Low gain: $\sim 0.12$ \\
& & & High gain: $\sim 0.36$ \\
\hline
\hline
\end{tabular}
\end{table*}

The VLDA can be operated at 7.5 and 15~MHz. The 15~MHz mode is the Metis default mode. During characterisation at subsystem level, the parameter settings providing the best performance for the Metis scientific requirements were found. Two settings, one at low gain and one at higher gain, were determined. Table~\ref{tab:vlda_sensor} shows the sensor characteristics and its performance with the recommended parameter settings.

\subsubsection{Ultraviolet light detector assembly}

\begin{figure*}
\centering
\includegraphics[trim=2cm 2cm 10cm 2cm,width=0.45\textwidth]{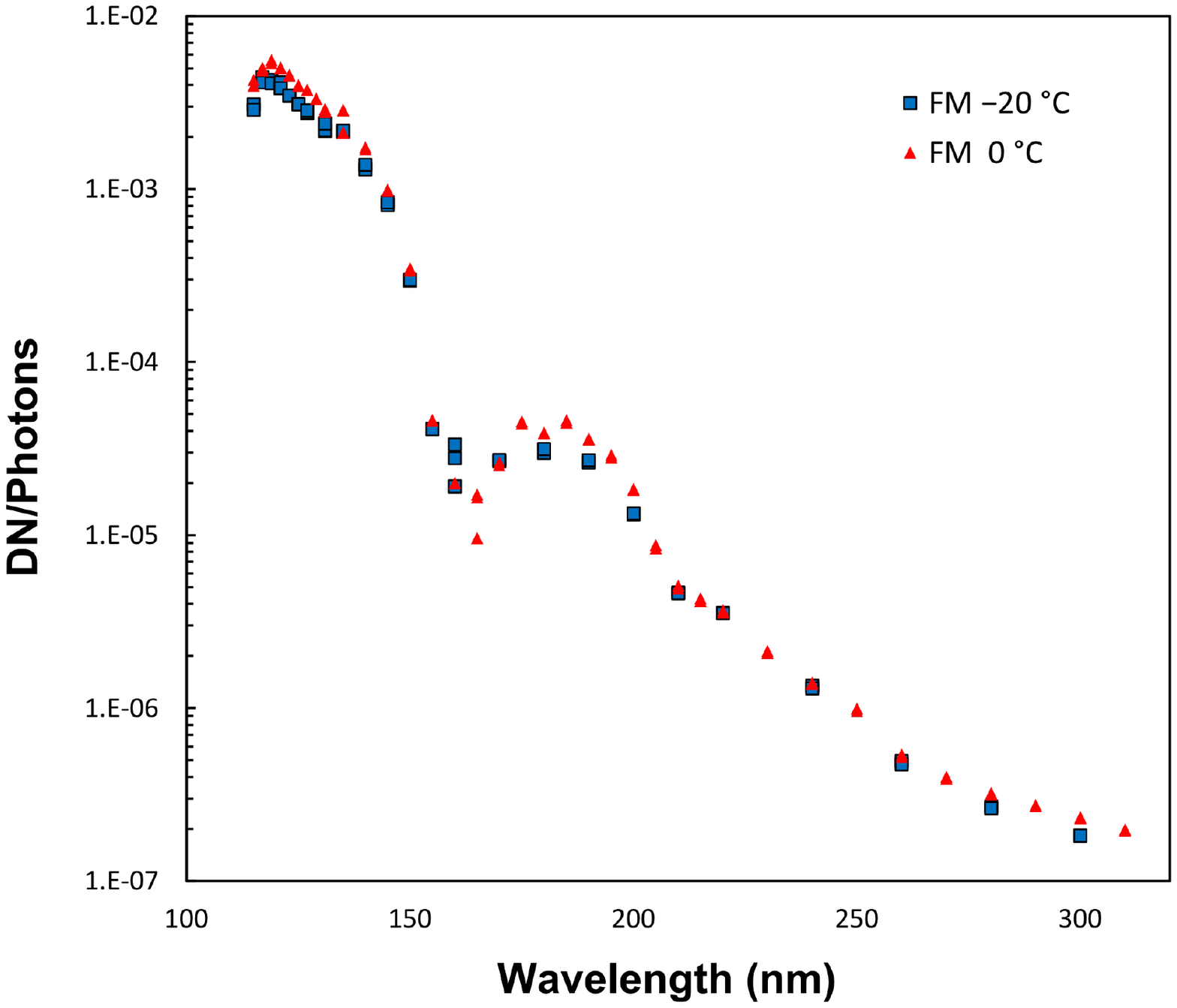}
\includegraphics[trim=2cm 2cm 10cm 2cm,width=0.45\textwidth]{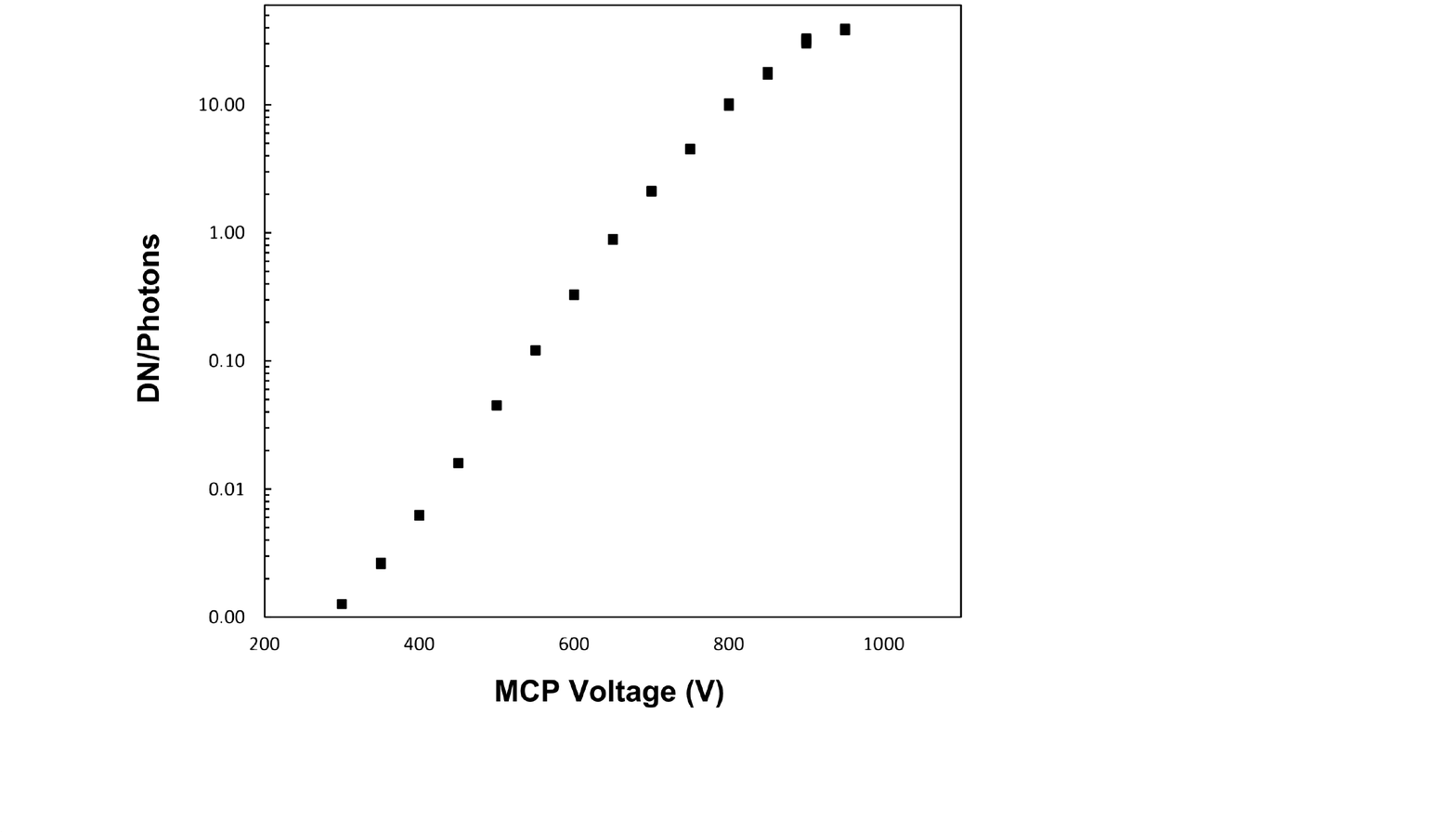}
\caption{\emph{Left:} Spectral response of the UVDA flight model at $0\degr$C and $-20\degr$C. The voltages applied to the microchannel plate (MCP) and to the screen were 400~V and 6~kV, respectively. Screen voltages of 5~kV and 4~kV would reduce the response by a factor of 0.68 and 0.36, respectively. \emph{Right:} UVDA flight model conversion efficiency at 121.6 nm as a function of the voltage applied to the MCP. The screen voltage was set to 6~kV.}
\label{fig:uvda_response}
\end{figure*}

The detailed design of the UVDA is described in \citet{uslenghi2012} and \citet{schuehle2018}. It consists of a microchannel plate (MCP) intensifier and the camera electronic subassembly. The intensifier is a single MCP enclosed under vacuum in a body sealed by a magnesium fluoride entrance window and a fibre-optic output coupler. The  microchannel plate intensifier and the fibre optic coupling with the sensor were made by ProxiVision GmbH in Germany. 

The MCP has a photocathode coating of potassium bromide (KBr), selected for its high efficiency at 121.6~nm. It detects UV radiation passing through the tilted interference filter and transforms it into electrons that are accelerated against a phosphorus screen. The visible radiation emitted by the screen is led onto the APS via a fibre-optic taper. The taper reduces the size of the focal plane image by a factor of two to match the size of the sensor. 

The camera system captures images using a STAR1000 image sensor ($1024\times 1024$, 15-$\mu$m pixels) controlled by an ACTEL field-programmable gate array (FPGA). Image data are read out by a 14-bit analogue-to-digital converter and transferred via a channel-link differential interface. The camera control is provided via a low-voltage differential signalling (LVDS) serial peripheral interface (SPI). The intensifier is supplied with high voltage by the HVU, which is mounted on the spacecraft panel close to the Metis telescope.
The MgF$_2$ entrance window has a thickness of 4~mm and has the effect of moving the focus backward a distance of about 1.5~mm.

The calibration of the spectral radiometric response of the UVDA was carried out at the metrology light source (MLS) of the electron storage ring of the Physikalisch-Technische Bundesanstalt (PTB) in Berlin. The UVDA has a conversion efficiency (digital number, DN, per photons incident on the unit, including the quantum efficiency of the photocathode and the transmission of the input window) of about 1 DN/photon at 121.6~nm and a microchannel plate voltage of 650~V (screen voltage of 6~kV). 

The intensifier assembly has the advantage of having very low sensitivity to radiation above 150 nm (because of the photocathode response) and below 115~nm (because of the MgF$_2$ entrance window), thus greatly helping in isolating the wavelength range of interest. Figure~\ref{fig:uvda_response} (left panel) shows the spectral response of the UVDA alone. On Metis, one Lyman-$\alpha$ interference filter in front of the camera is sufficient to ensure the required spectral purity. Figure~\ref{fig:uvda_response} (right panel) shows the response of the UVDA camera as a function of the voltage applied to the microchannel plate (screen voltage set to 6~kV). Table~\ref{tab:uvda_sensor} shows the sensor characteristics and its performance with the recommended parameter settings.

The VLDA, UVDA, CPC, and HVU were manufactured by the Max Planck Institute for Solar System Research in G\"ottingen, in Germany.

\begin{table*}
\caption{Characteristics of the UVDA.}\label{tab:uvda_sensor}
\centering
\begin{tabular}{l|l|l|l}
\hline
\hline
Parameter & Unit & Requirement & Value \\
\hline
\multicolumn{4}{c}{Sensor} \\
\hline
Format & pixels & 1k x 1k & By design \\
Pixel size & $\mu$m & 15 & By design \\
Shutter & & Rolling & By design \\
Capacity with linearity within 5\% & ke$^-$ & $\ge 99$ & $\approx 102$\\
Dark current & e$^-$/s & $\leq 3135$ at $20\degr$C & $\approx 370$ at $-20\degr$C \\
 & & & $\approx1100$ at $+20\degr$C \\
Sensor gain & e$^-$/DN & & $\approx 9.3$\tablefootmark{a} \\
\hline
\multicolumn{4}{c}{Fibre optic coupler} \\
\hline
Focal plane to sensor demagnification & & $(2\pm 0.04):1$ & $(1.98\pm 0.01):1$\tablefootmark{b} \\
\hline
\multicolumn{4}{c}{Intensifier} \\
\hline
Entrance window & & MgF$_2$ & MgF$_2$ \\
Photocathode & & KBr & KBr \\
Applicable voltages & V & & $0\le V_\text{MCP}\le 1000$ \\
& & & $0\le V_\text{SCREEN}\le 6000$ ($V_\text{MCP} \le V_\text{SCREEN}$) \\
Conversion efficiency & DN phot.$^{-1}$ & & $\approx 1$ ($V_\text{MCP}=660\ \text{V,}\ V_\text{SCREEN}=6\ \text{kV}$) \\
& & & $\approx 30$ ($V_\text{MCP}=900\ \text{V,}\ V_\text{SCREEN}=6\ \text{kV}$) \\
Dark noise at $20\degr$C & Count pixel$^{-1}$ cm$^{-2}$ & 10 & $<0.1$\tablefootmark{b,c} \\
\hline
\hline
\end{tabular}
\tablefoot{
\tablefoottext{a}{Measured on Laboratory Model at optimal sensor settings.}
\tablefoottext{b}{Measured on Qualification Model.}
\tablefoottext{c}{Measured on Flight Spare.}}
\end{table*}

\section{Processing and power unit}
\label{sec:MPPU}

\subsection{Electronics layout}
As mentioned in Sect.~\ref{sec:processing_unit},
the MPPU manages the instrument operations, acquires and processes the detectors data, controls the mechanisms and the thermal hardware installed on the Metis optical unit, and interfaces to the Solar Orbiter spacecraft for instrument power and data transmission (telecommand/telemetry, TC/TM, and science data). The Metis flight software is loaded in the MPPU and implements the instrument operation, control, and fault detection, isolation and recovery (FDIR) functions.
The MPPU implements the following functions:
\begin{itemize}
\item Interface to the S/C power distribution unit ($+28$~V), and regulate and distribute the instrument power to all its subsystems (MPPU itself, detector units, high voltage unit, PMP heaters, IOM motor).
\item Exchange telecommands, telemetries, and science data with the S/C by means of redundant,  internally cross-strapped SpaceWire channels.
\item Deliver commands and receive telemetries and images from the two detector units (VLDA and UVDA) simultaneously and independently by means of SPIs.\item Control the IO position by means of step motors.
\item Control the polarimeter phase and temperature stability.
\item Acquire and monitor the MOU temperatures.
\item Execute all the instrument on-board processing (frame averaging, variance, radiation artefacts removal, data compression and packet formatting).
\item Manage all the instrument operating modes.
\item Implement an instrument autonomous FDIR function. 
\end{itemize}

To perform all these functions, the MPPU consists of the following functional blocks and boards:
\begin{itemize}
\item Single board computer (SBC), based on the following elements: 
\begin{itemize}
\item LEON2 AT697F microprocessor, running at 50~MHz (40~MIPS), 
\item RTAX2000 FPGA, used as coprocessor at 200~MIPS, running at 66~MHz with a 64~KB boot PROM, 
\item non-volatile 4~MB MRAM for ASW code, 
\item 4~MB SRAM, data and program memory, and 
\item 320~MB SDRAM, memory buffer for frame data processing.
\end{itemize}
\item Power supply board (PSB), based on a forward DC/DC converter switched with a stabilised  125~KHz oscillator; this has an active clamp reset circuit and overvoltage protections on all the secondary outputs. The DC/DC is activated ON/OFF by dedicated S/C high voltage pulsed current (HVPC) commands and provides the status of the switch to the S/C.
\item VLDA board, based on FPGA, managing the VL channel and its related subsystem (PMP).
\item UVDA board, based on FPGA, managing the UV channel and its related subsystem (HVU).
\item Backplane board, connecting all the MPPU boards.
\end{itemize}

\begin{figure*}
\centering
\includegraphics[width=0.65\textwidth]{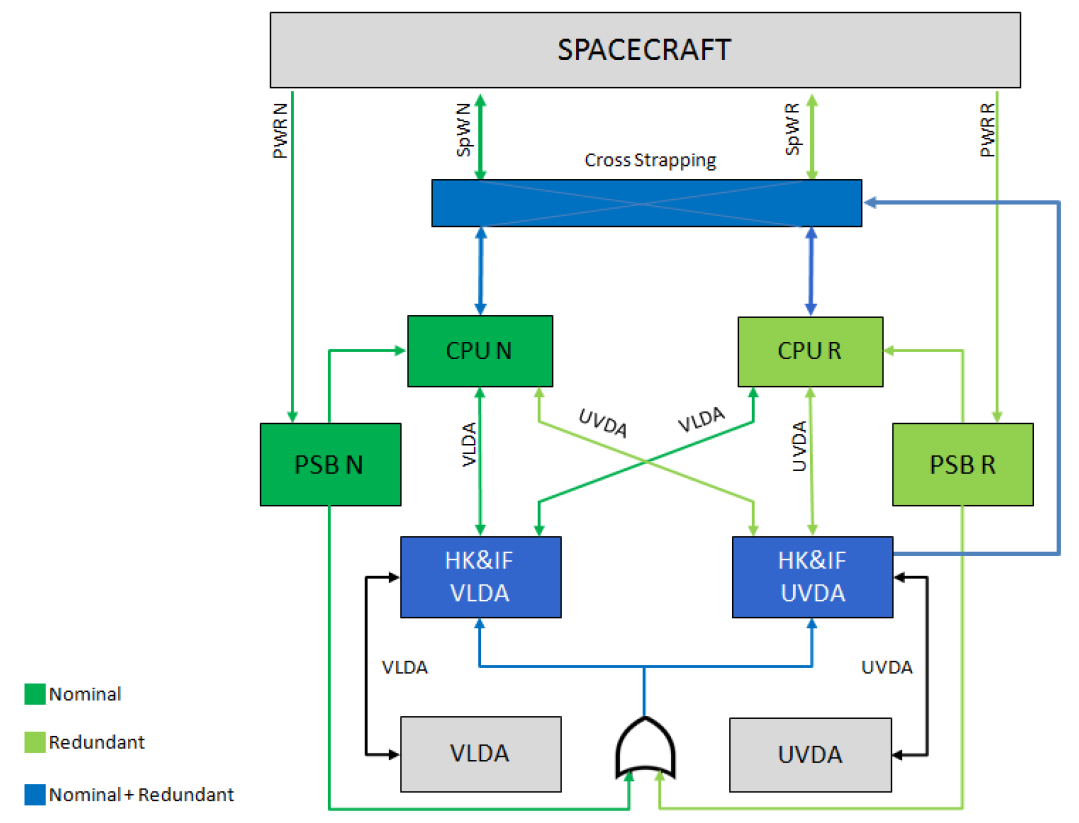}
\caption{Metis processing and power unit block scheme.\label{fig:mppu_scheme_1}}
\end{figure*}

\begin{figure*}
\centering
\includegraphics[width=0.65\textwidth]{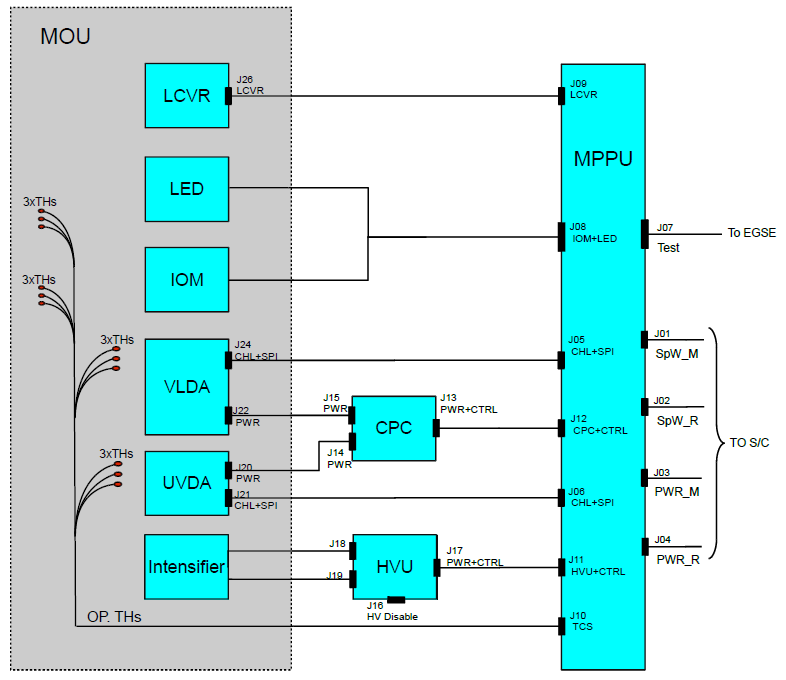}
\caption{Diagram of the Metis interconnections.\label{fig:mppu_scheme_2}}
\end{figure*}

The MPPU architecture foresees a high level of redundancy: the single board computer and power supply board are cold redundant; the SpaceWires are cross-strapped so that each redundant MPPU section can communicate to the S/C with two SpaceWire channels.
The MPPU block scheme and the Metis interconnetion diagram are shown in Figs.~\ref{fig:mppu_scheme_1} and~\ref{fig:mppu_scheme_2}, respectively.

One FPGA interface board is dedicated to each of the two channels to acquire and perform a preliminary manipulation of the detector frames and the housekeeping data. Both FPGA interface boards then send the acquired data to the LEON CPU board, where a further FPGA is resident and implements processing functions in hardware.

The maximum data rate the MPPU is able to receive from the detectors is:
\begin{itemize}
\item one VL detector frame per second when operating in fixed polarisation for a maximum duration of one minute, or one VL detector frame every 15~s when operating in multiple polarisations for an indefinite duration;
\item one UV detector frame every 97~ms when operating in photon counting mode, or one UV detector frame per second when operating in analogue mode, both for an indefinite duration.
\end{itemize}

The MPPU has a mass of 5.56~kg and a volume of $282\times 252\times 155$~mm$^3$ and
was manufactured by SITAEL, Italy.

\subsection{On-board software}
\label{sec:onboardsw}

Metis scientific and engineering activities are converted into telecommand sequences to be executed by the on-board software (OBSW). The instrument functionalities are implemented within a software architecture that complies with standard quality for real-time HW-embedded systems. The application software comprises two main blocks: the process controller software (PCSW) and the instrument controller software (ICSW). The PCSW manages the Metis communication with the Solar Orbiter platform services, by using the packet utilisation standard library (PUSL). The PCSW also manages the  FDIR events and the thermal control, granting autonomy in case of failure or instrument internal heating due to solar radiation. The ICSW is a collection of routines related to instrument command and control, and is in charge of acquiring, processing, and compressing the scientific data. The acquisition tasks acquire and read data from the UVDA and the VLDA. The processing tasks perform actions on the obtained images and remove signals due to cosmic rays (CRs) and solar energetic particles (SEPs). The compression task applies compression algorithms in order to reduce the amount of data to be sent to ground. Each of these steps runs in a different task that works independently.

In most cases, the VLDA and UVDA operate simultaneously creating two concurrent data streams. The VLDA works in analogue (or integration) mode only, whereas the UVDA can work either in analogue or in photon counting mode. In analogue mode, both detectors collect the charge created by the incoming photons over the integration time set for detection. In photon counting mode, the UVDA intensifier operates at high gain and the APS is read out by a dedicated hardware task, able to discriminate single photons hitting the sensor, and produces either a list of photo-events coordinates ($x$, $y$) and energy, or an accumulation matrix containing the number of events detected per pixel along with the integrated energy amplitude profile. 

Each detector acquisition chain has a dedicated image processing and compression line. When the UVDA is operated in photon counting mode, the on-board data processing is limited to the compression of the accumulation matrix and to the telemetry packetisation. In analogue mode, the data processing is similar for the two detectors. An acquisition consists of a programmable number of cadences. For each cadence, one UV and up to four VL images are produced. The cadence is the time necessary for Metis to acquire and average all frames composing an image. 

Each frame is fetched by the acquisition step and provided to the processing step. 
In the image processing chain, each task is defined as data-producer for the next task and data-consumer of the previous task. The processing step acquires many frames and combines them in order to produce the mean image of the set of frames. This approach provides the possibility to minimise the amount of transmitted data, while assuring the information semantic relevance. In addition, it filters errors and frame imperfections.

\begin{figure*}
\centering
\includegraphics[width=0.75\textwidth]{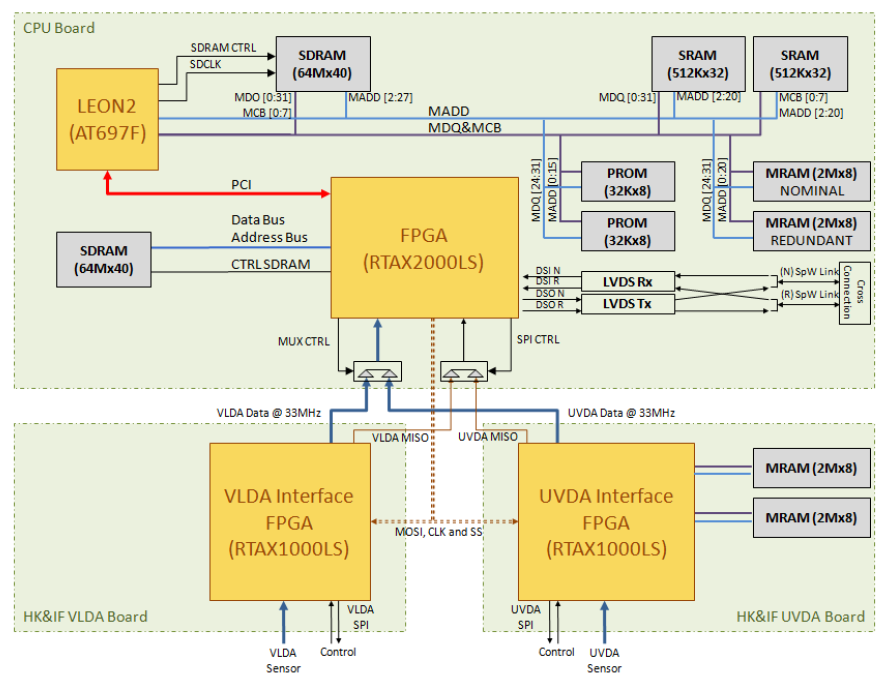}
\caption{Scheme of the Metis on-board data processing electronics.}
\label{fig:onboard_proc}
\end{figure*}

A detector scientific acquisition consists of the following operations: read and, only for the VLDA, reorder the pixels to reconstruct their correct sequence in the acquired frame. The UVDA performs a pre-addition operation among frames. An algorithm filters out CRs and SEPs.  A number of filtered frames is then selected and these frames are averaged together providing the final VL/UV image. Statistical computations are performed to determine the image spatial noise, before passing the final exposure to the compression algorithm. Finally, the compressed data are split into packets and sent to ground. 

A further special configuration can be used to produce high-frame-rate acquisition (up to 1~frame/second), by operating only the VL detector for a limited time interval. In this configuration, the processing can be reduced by applying a masking or a binning step, in addition to the usual data compression. Figure~\ref{fig:onboard_proc} shows a scheme of the Metis on-board data processing electronics.

The on-board software was developed by OHB Italia, as responsible for the  Basic Software (BSW), and by Thales Alenia Space Italia, as responsible for the Application Software (ASW).

\section{Thermo-mechanical design}
\label{sec:thermo_design}

\subsection{Mechanical design}

The structural design of the Metis telescope is driven by the following main requirements:

\begin{itemize}
\item withstand the specified qualification loads while preserving the integrity of the internal components (optics, detectors);
\item ensure that all fundamental resonance frequencies are above 140 Hz, while minimising the contribution to the instrument overall mass;
\item maintain the alignment of the VL and UV channels under the launch loads, the 1-g to 0-g transition, the moisture release and the large range of temperatures experienced along the orbits, so that the image quality degradation (expressed as size of the ensquared energy area) throughout the mission lifetime is $<40$\%;
\item maintain the instrument internal contribution to the absolute pointing error of the line of sight (LoS) towards the centre of the Sun within 1~arcmin (cone semi-aperture angle) with 95\% confidence level during the observations.
\end{itemize}

To combine the strength, alignment, and thermal stability performance with a lightweight mass, the structure of the telescope is built around a central cylinder made of CFRP, characterised by a thermo-elastic coefficient $\text{CTE}\sim 10^{-6}$~K$^{-1}$, where all the optical elements and the VL camera are integrated by means of titanium flanges (Fig.~\ref{fig:structural_parts}). The front part of the telescope consists of a front cylinder, a front cone, and a boom, all made from titanium, since these parts experience higher temperatures. The external occulter is positioned on the sunward side of the boom at the surface of the heat shield. The boom penetrates the heat shield and is surrounded by a feedthrough. The rear part of the telescope consists of a truncated cone of titanium carrying the UV camera. The telescope is interfaced to the spacecraft panel through a mounting frame made of rods in titanium and CFRP, which provides an efficient decoupling from the thermo-elastic deformations of the spacecraft structure. The structure segmentation facilitates its manufacturing and the integration of the optical elements; moreover, it allows us to introduce thermal cuts along the path from the external occulter, which is the outermost element of the coronagraph, to the central cylinder in order to limit the heat flux from the aperture that reaches the interior of the instrument.

The structure of the telescope was manufactured by DTM in Italy.

\subsection{Ejectable cap}
\label{sec:cap}

\begin{figure*}
\centering
\includegraphics[width=0.75\textwidth]{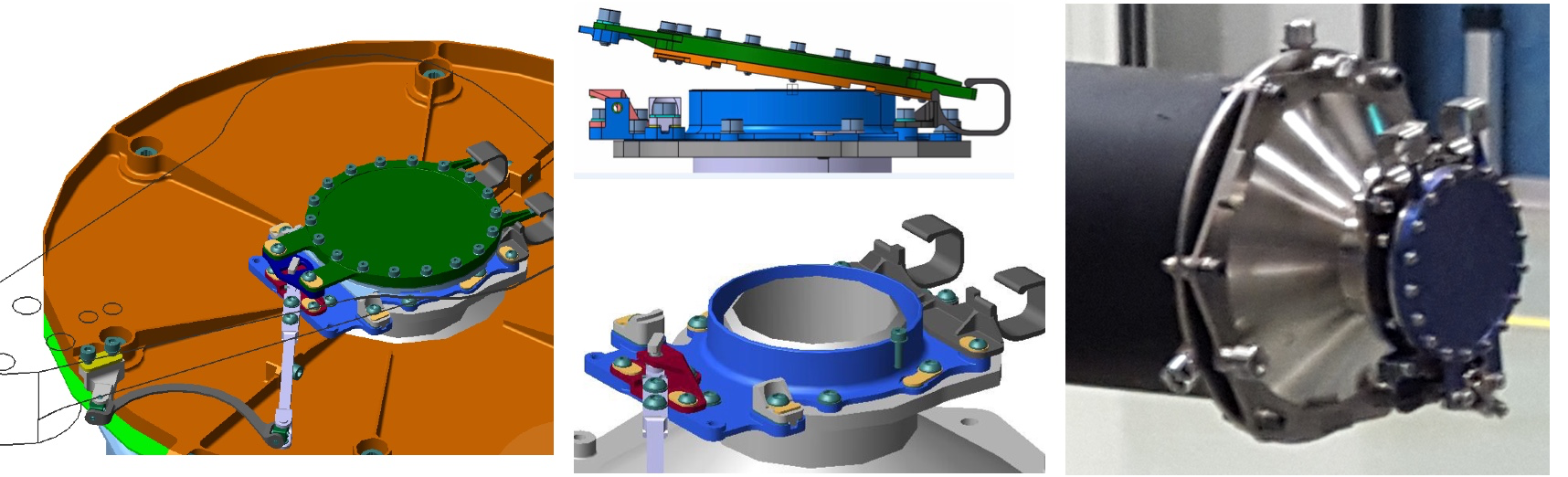}
\caption{Operating principle of the ejectable cap and its qualification model installed on the inverted external occulter.}
\label{fig:cap}
\end{figure*}

Like any coronagraph operating in visible light, Metis is very sensitive to particulate contamination: any particle with size larger than 100~$\mu$m deposited on the edge of the IEO aperture or on the mirror M0 can produce scattering of the intense solar disk light, washing out faint coronal emissions. To prevent the deposition of dust particles on the IEO edge and/or their entrance inside the instrument during on-ground integration and test activity and during launch and early orbit phase, its aperture is protected by a cap (Fig.~\ref{fig:cap}) that will be ejected at least 20~days after launch and early orbit phase (LEOP; see Sect.~\ref{ops:comm}). This cap, produced by SENER in Spain, is ejected by a spring that is released when the external door that covers the Metis aperture is opened for the first time; the rotation of the shaft of the door actuates a leverage system that disengages the spring retention device.

\subsection{Thermal design}

\begin{figure*}
\centering
\includegraphics[width=0.75\textwidth]{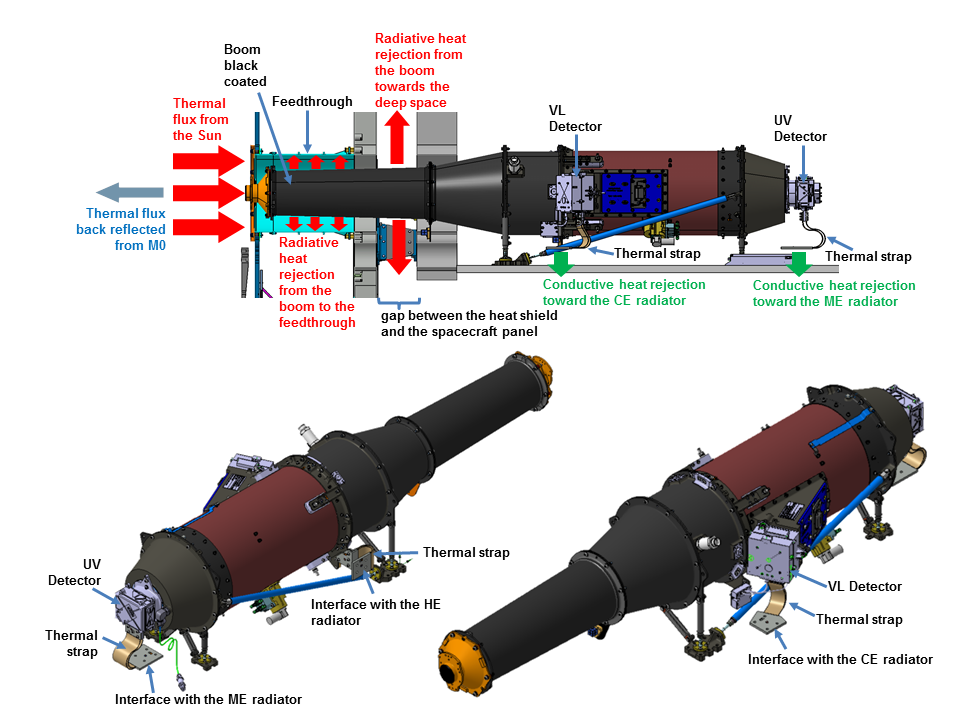}
\caption{Concept of the thermal control of the Metis telescope (above). Interfaces between the Metis thermal straps and the cold element (CE), medium element (ME), and hot element (HE)  radiators of the S/C (below).}
\label{fig:thermal_design}
\end{figure*}

The thermal design of the Metis telescope was driven by the following main requirements.
\begin{itemize}
\item The temperature of the VL and UV imaging sensors must be kept above the minimum design operational limits and below the maximum acceptable limits for the scientific performance during the observation periods. The resulting temperature control ranges are [$-45\degr$C, $-5\degr$C] for the VL sensor and [$-35\degr$C, $-5\degr$C] for the UV sensor.
\item The temperature of the liquid crystal variable retarder of the polarimeter must be kept around any value inside the operational range [$+30\degr$C, $+75\degr$C] within $1\degr$C over time intervals of at least one hour.
\item The thermal flux dissipated by the instrument at the conductive and radiative thermal interfaces inside the spacecraft must be minimised.
\end{itemize}
The main challenge for Metis thermal control is the extremely variable thermal flux incident upon the spacecraft and entering the instrument aperture along the operational orbits, ranging from about 1 solar constant near the aphelion up to about 13 solar constants at perihelion (0.28~AU). 
The thermal control of Metis is achieved by passive means, with the exception of the LCVR, which is controlled by a thermistor-heater loop. Radiation from the solar disk entering the instrument aperture is intercepted by the mirror M0 and reflected back towards space. Most of the heat flux collected by the IEO external surface is dissipated radiatively by the boom (which is black-coated to act as a radiator) toward the internal wall of the feedthrough and toward the gap between the heat shield and the front panel of the spacecraft, before it reaches the inner parts of the instrument (Fig.~\ref{fig:thermal_design}).
This residual heat flux is dissipated radiatively inside the spacecraft cavity by the walls of the telescope (which for this reason are not covered by thermal insulator), in order to minimise the heat flux conveyed towards the conductive interfaces with the spacecraft radiators. The cold-element (CE) and medium-element (ME) conductive interfaces are used to cool the VL and UV imaging sensors via high-conductivity thermal straps connected to their cold fingers (Fig.~\ref{fig:thermal_design}), while the hot-element (HE) conductive interface is exploited to drain some heat from the telescope structure via another thermal strap.

\section{Instrument integration, alignment, and test campaigns}
\subsection{Instrument integration and alignment}

\begin{table*}
\caption{Surface particle contamination (PAC) and surface molecular cleanliness (MOC) requirement budgets for the most critical surfaces of Metis, at delivery and at EoL.}
\centering
\begin{tabular}{l|c|c|c|c}
\hline
\hline
\multirow{2}{*}{Sensitive Area} & \multicolumn{2}{c|}{PAC (p.p.m.)} & \multicolumn{2}{c}{MOC (ng cm$^{-2}$)}\\
& Delivery & EoL & Delivery & EoL \\
\hline
Inverted external occulter & 3.3 & 100 & 200 & 1500 \\
Telescope UV mirrors and stops (M1, M2, IO, Lyot stop) & 3.3 & 31 & 100 & 250 \\
Telescope UV filter & 3.3 & 31 & 100 & 250 \\
UV detector window & 3.3 & 31 & 100 & 250 \\
Visible light polarimeter optics and VL detector & 3.3 & 31 & 200 & 1000 \\
Telescope interior & 3.3 & 54 & 200 & 1000 \\
\hline
\hline
\end{tabular}
\label{tab:pacmoc}
\end{table*}

\begin{figure*}
\centering
\includegraphics[width=0.75\textwidth]{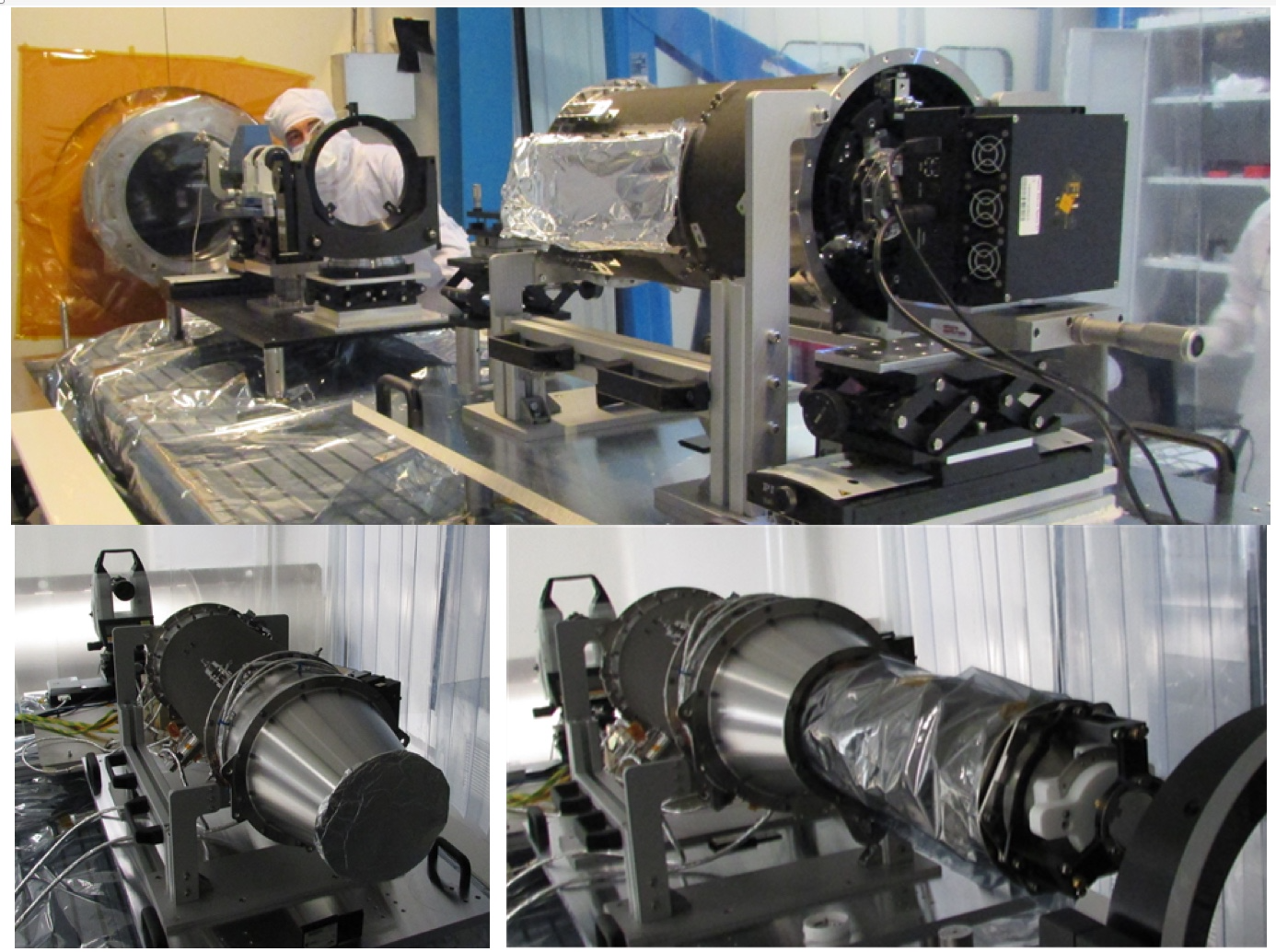}
\caption{Various phases of the Metis telescope integration and alignment in the Optical Payload Systems facility, at ALTEC in Turin.}
\label{fig:ait}
\end{figure*}

The integration, alignment, and calibration of the flight model of Metis (Fig.~\ref{fig:ait}) was performed in a dedicated facility, the Optical Payload Systems, hosted at ALTEC premises in Turin, Italy. Originally implemented by the INAF Astrophysical Observatory of Turin, the facility was then upgraded specifically to perform the  Metis activities (Sect.~\ref{sec:calibrations}). The facility is designed to accommodate the stringent contamination control requirements of space flight instruments. The controlled environment of the $\approx 100$~m$^2$ facility is divided into two parts: a grey (ISO 8) room and a clean (ISO 7) room. Inside the ISO 7 room, the facility hosts the vacuum space optics calibration chamber, providing a cleaner ISO 5 area.

The optical alignment procedure is described by \citet{dadeppo2018}. The alignment of the telescope was performed with the help of reference targets installed in the central holes of the mirror M1, M2, and the IEO, whose centres were co-aligned by means of a sighting telescope and a theodolite. An auxiliary visible light source and an  imaging camera were used to pre-adjust the focus of the UV channel;  the focus of the VL channel was then tuned directly on the flight camera by operating on the position of the interference filter. 

\subsection{Cleanliness}
\label{sec:cleanliness}

Coronagraphic observations define the need for very low levels of stray light and minimal absorption by molecular condensates.
Stray light in Sun-pointed telescopes is dominated by diffraction at the entrance aperture. However, in a coronagraph this source is generally trapped by stops and baffles. Therefore, the residual sources of stray light, which may detrimentally impact the performance of the instruments, are particles within the optical train -- particularly 1) those located near the front aperture, where illumination by full sunlight or diffracted light may cause scattering that cannot be trapped further down the optical path, and 2) those on the optical surfaces, in particular on  the primary mirror M1, where particles generate an increase in scattering of diffracted light equivalent to increased mirror surface roughness. 

The amount of scattering increases with wavelength, hence, the visible light path requirements set the limits on particulate cleanliness. 
Absorption, by molecular condensates on the Metis mirrors, filters and detectors, reduces the throughput and thus degrades the signal-to-noise ratio, especially in the UV. The end-of-life (EoL) maximum total molecular condensate equivalent path length is 16 nm. The above science requirements are expressed as a cleanliness budget in Table~\ref{tab:pacmoc}.

The Metis telescope is completely contamination tight, with a purging inlet and two venting ports, both protected with sintered filters that prevent the entrance of particles and molecular contamination. The external occulter is protected by the ejectable cap until outgassing is completed in flight.

\begin{figure*}
\centering
\includegraphics[width=0.35\textwidth]{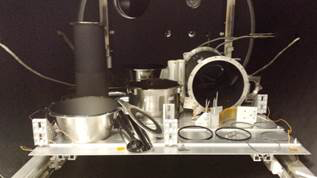}
\includegraphics[width=0.3725\textwidth]{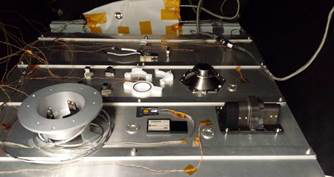}
\caption{Metis structure elements (left) and optical components (right) prepared for the thermal-vacuum bake-out in the “PESCha” facility at Thales Alenia Space in Italy.}
\label{fig:bake_out}
\end{figure*}

\subsection{Subsystem bake-out}

As part of the contamination control plan, all Metis subsystems were subjected to thermal-vacuum bake-out (TVBO) before their integration into the flight model (Fig.~\ref{fig:bake_out}). The TVBO consisted of heating the flight hardware in a clean certified vacuum chamber at the maximum temperature predicted during the mission.
The TVBO was monitored in real time by means of thermoelectric quartz crystal microbalances (TQCMs) until the outgassing rate was reduced to between 0 and 5\% per hour, when measured over a period of not less than 3~hours, and for a total duration not shorter than 72~hours. This procedure allowed a reduction of the evolution of outgas products and  removal of the molecular surface contamination to  below the acceptable limits.

\subsection{Vibration and shock tests}

\begin{figure*}
\centering
\includegraphics[width=0.75\textwidth]{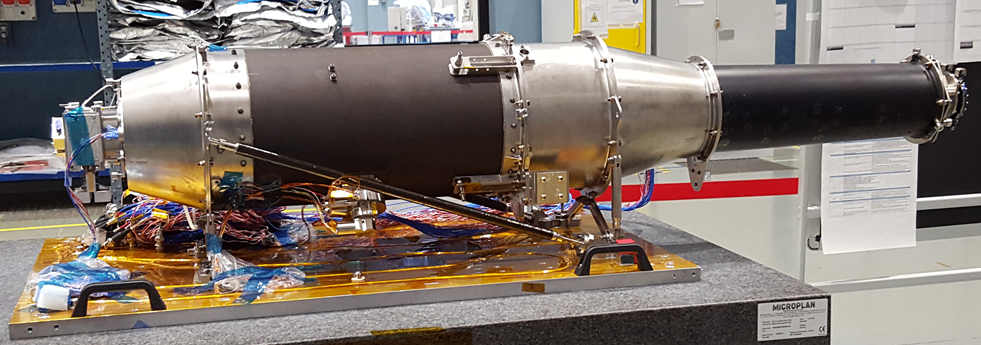}
\caption{Structural qualification model of the telescope prepared for the sine and random vibration tests.}
\label{fig:stqm}
\end{figure*}

\begin{figure*}
\centering
\includegraphics[width=0.75\textwidth]{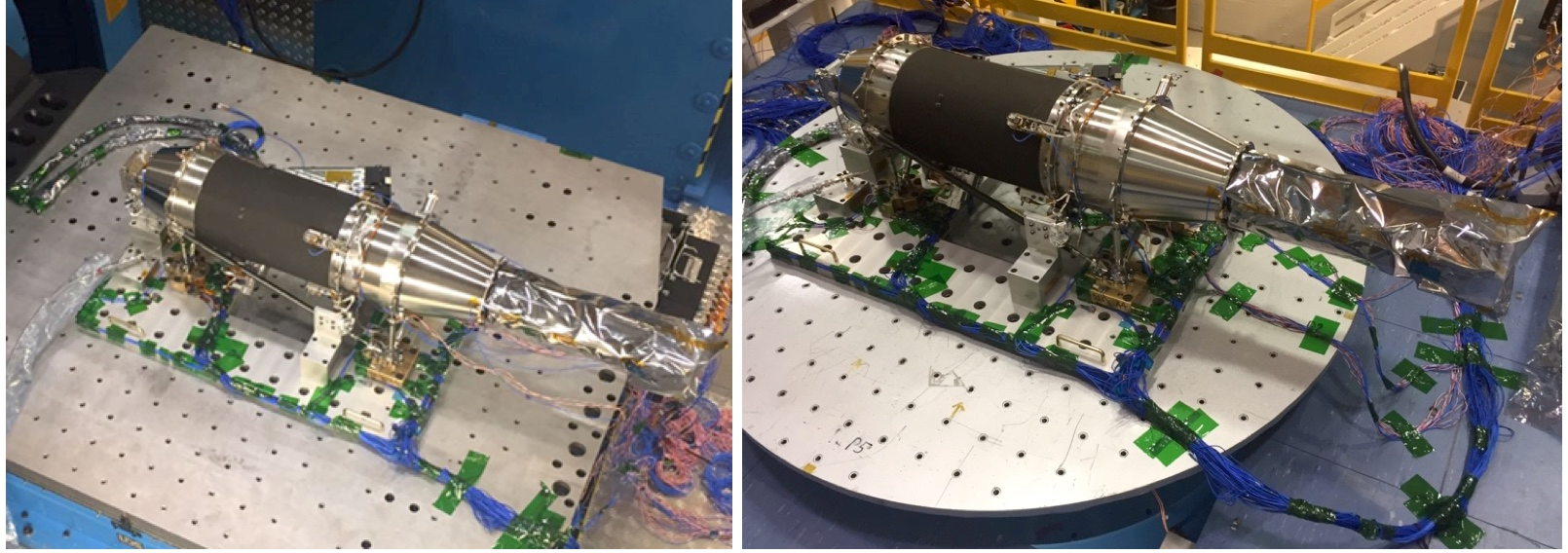}
\caption{Metis flight model on the shaker for the in-plane (left) and out-of-plane (right) acceptance vibration tests.}
\label{fig:shocks}
\end{figure*}

The qualification of the Metis telescope under the launch environment (by applying sine and random loads, as well as shock pulses) was achieved using a dedicated structural qualification model, SQM (Fig.~\ref{fig:stqm}). This model enabled us to qualify as a system the telescope structure, the optics subsystems, and the UV camera (the VL camera instead followed a PFM approach) by applying  the specified
loads to the mounting interfaces with the spacecraft. 
The flight model of Metis was then subjected only to sine and random vibrations at acceptance level in the facility of Thales Alenia Space (Fig.~\ref{fig:shocks}), without repeating the shock test. The instrument optical alignment was verified before and after the vibrations, by measuring external reference points/surfaces with a laser tracker and a theodolite and by taking images with the VL camera, without finding appreciable variations. The shock test was performed at IABG, Germany.

\subsection{Thermal vacuum tests}

\begin{figure*}
\centering
\includegraphics[width=0.75\textwidth]{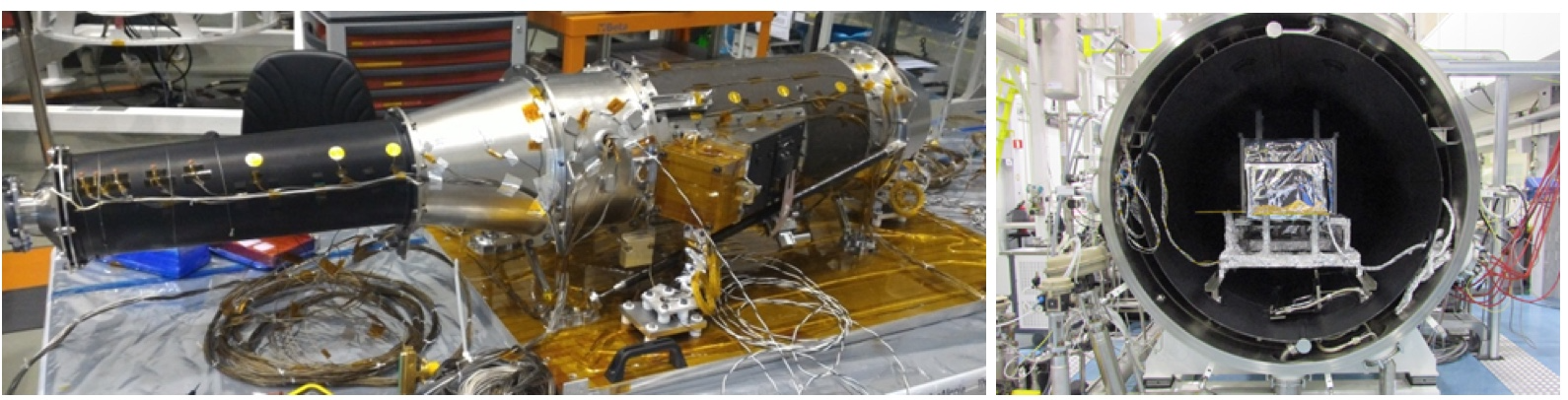}
\caption{Metis structural thermal model prepared for the vacuum thermal-balance test with temperature sensors installed inside and on the external surface (left). Metis structural thermal model inside the thermal-vacuum chamber endowed with  Sun simulator, at ESTEC/ESA (right).}
\label{fig:tvtb}
\end{figure*}

\begin{figure*}
\centering
\includegraphics[width=0.75\textwidth]{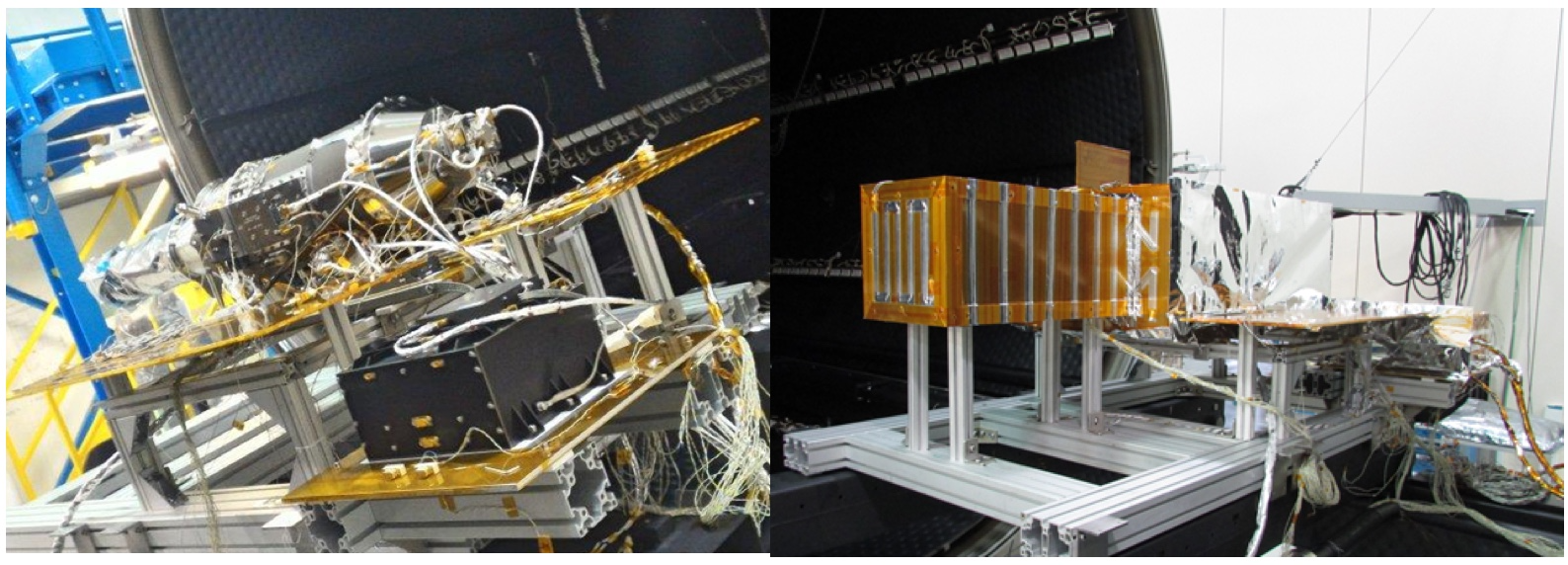}
\caption{Metis flight model  prepared for the vacuum thermal cycling at Thales Alenia Space (left). Metis flight model enclosed in the spacecraft thermal environment simulator before the introduction into the thermal-vacuum chamber (right).}
\label{fig:pfm_tvtb}
\end{figure*}

The thermal design of Metis was qualified during an extensive thermal-balance test campaign carried out in March 2016 on the structural thermal model (STM) in the thermal-vacuum facility of the European Test Services, at ESTEC, equipped with a Sun simulator capable of reproducing the illumination conditions that the instrument will experience at various Sun distances, up to 0.28~AU  (Fig.~\ref{fig:tvtb}). For this test, the STM was put inside an enclosure specifically designed to simulate the thermal environment of the spacecraft. Additionally, the STM test simulated the portion of the heat shield of Solar Orbiter with the feedthrough that gives the instrument access to the Sun view.

The flight model of Metis was then subjected to vacuum thermal cycling (without Sun simulator) to verify the behaviour of the whole integrated system under a representative mission environment, in the facility of Thales Alenia Space (Fig.~\ref{fig:pfm_tvtb}). During the thermal cycling, functional tests were performed to check the operation of the instrument at both the hottest and coldest predicted operational  temperatures. The parameters of the LCVR thermal control loop were also tuned during this test. Images taken with the VL camera and measurements on external reference points/surfaces before and after the test prove that the instrument alignment was not modified by the thermal cycling.

\subsection{Electromagnetic cleanliness}

The electromagnetic compatibility (EMC) was a design driver for the Solar Orbiter mission under control of an EMC working group. The first task was to translate the challenging science requirements imposed by two instruments, the magnetometer \citep[MAG; see][]{horbury2019b} and Radio and Plasma Waves \citep[RPW; see][]{maksimovic2019}, into a measurable, meaningful engineering requirement which could be practically specified and verified. A specific Metis EMC control plan has been  established in order to perform early characterisation of identified specific critical elements with respect to magnetic and electromagnetic cleanliness and to confirm or adapt the proposed instrument design. For DC magnetic cleanliness (MAG requirement), early characterisation of the motor/gear box assembly part of the IO assembly subsystem, and of the M1 and M2 mirror assembly was performed using the ESA Mobile Coil Facility (MCF).

\begin{figure*}
\centering
\includegraphics[width=0.75\textwidth]{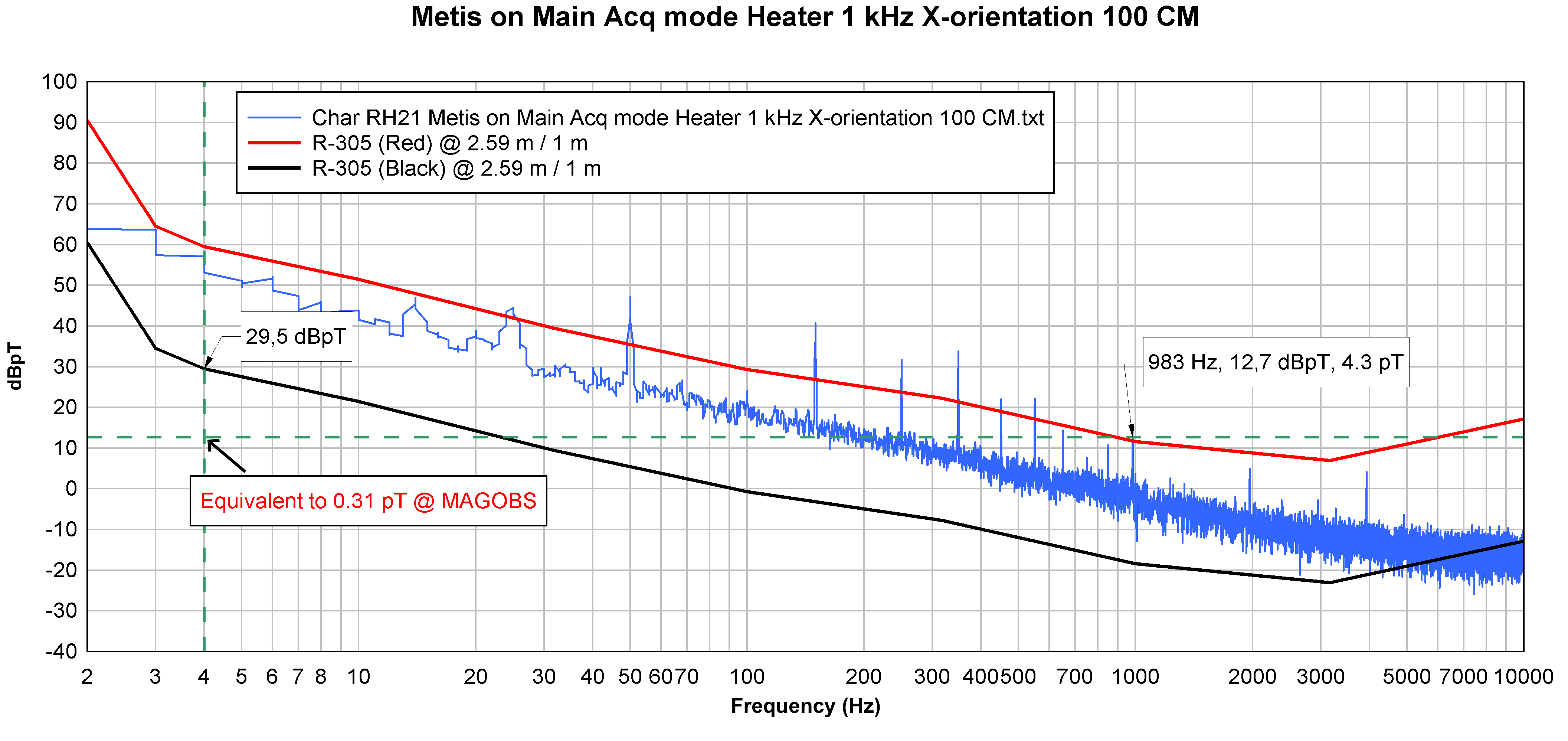}
\caption{PMP heater characterisation rescaled from 1 KHz to 4 Hz.}
\label{fig:laget002}
\end{figure*}

The PMP heater control required specific attention as well. The early characterisation allowed for an optimised control strategy (optimised control frequency) of the heater fully compliant with MAG and RPW magnetic requirements (Fig.~\ref{fig:laget002}). 

\begin{figure}
\includegraphics[width=\columnwidth]{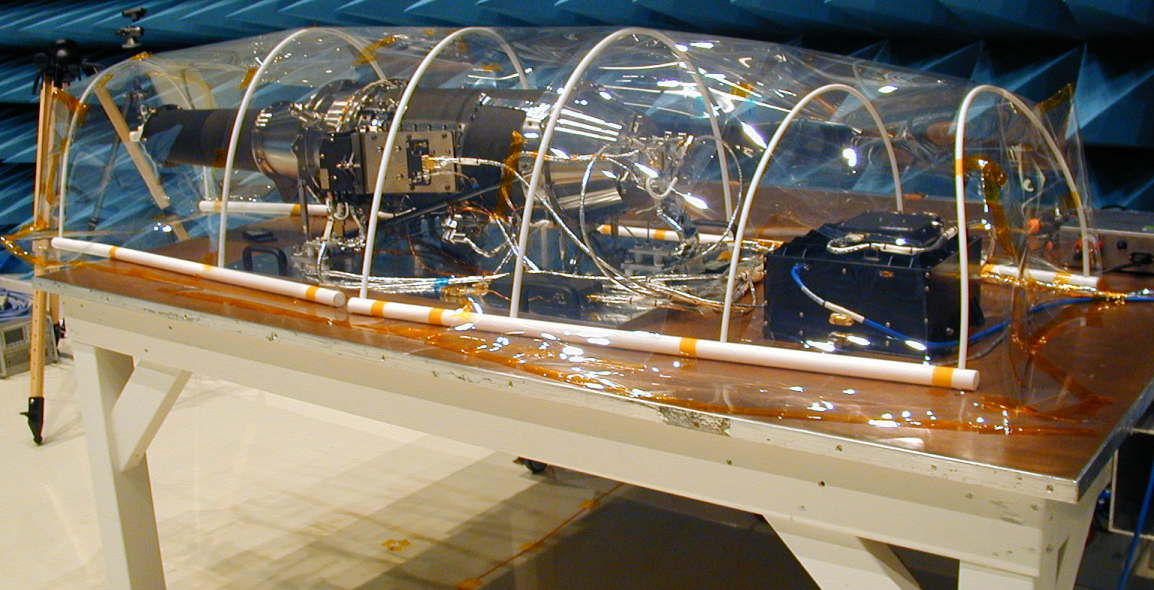}
\caption{Metis flight model during final electromagnetic cleanliness test at the ESA/ESTEC Test Center.}
\label{fig:laget003}
\end{figure}

\begin{figure*}
\centering
\includegraphics[width=0.75\textwidth]{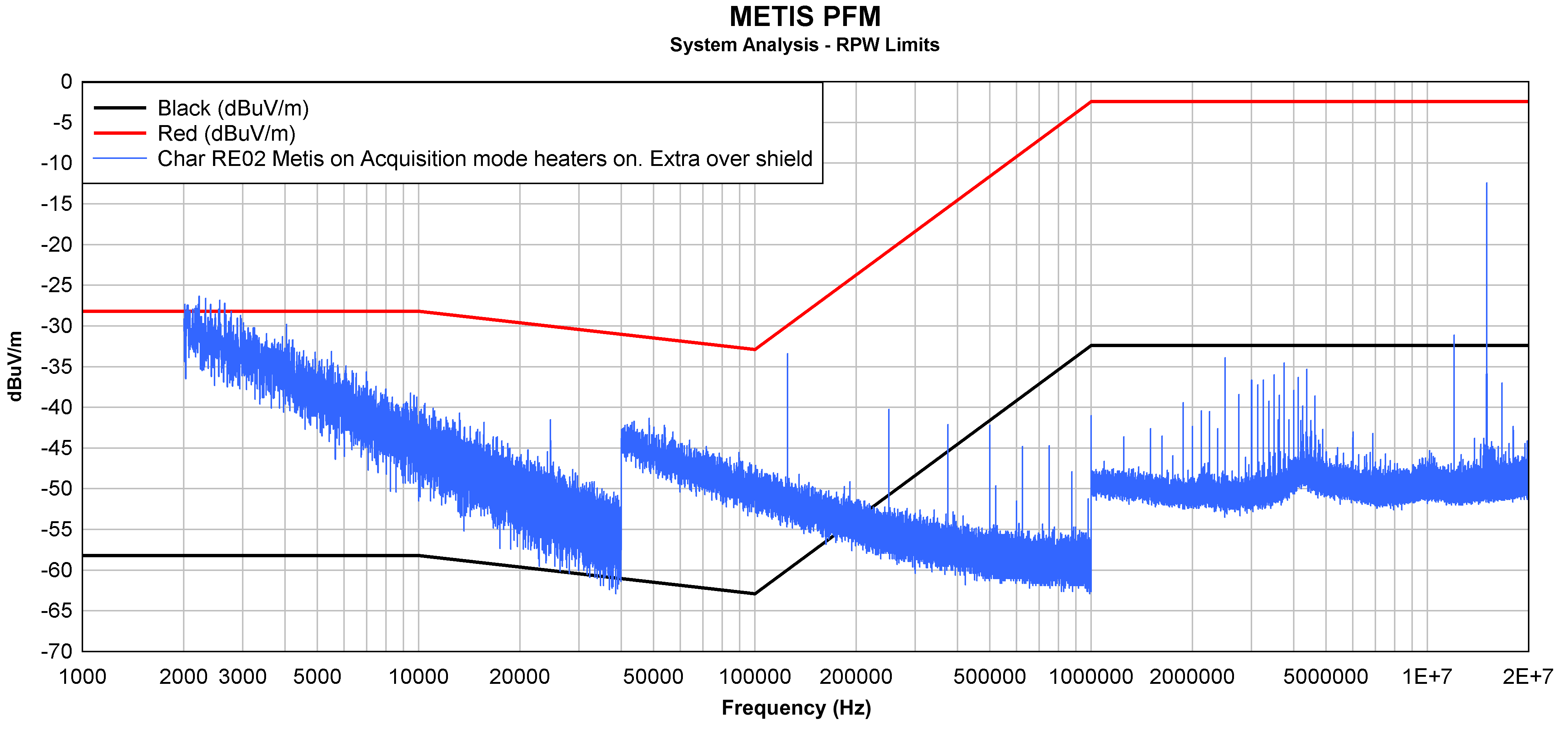}
\caption{Metis electric field emission. Results of the system analysis.}
\label{fig:laget004}
\end{figure*}

During the final instrument EMC test, performed at the ESTEC Test Centre (ETS) facility (Fig.~\ref{fig:laget003}), Metis was operated in the in-flight mode representative of the worst case. The measurements and characterisation results prove that Metis is an EMC quiet instrument, having most of the EM power spectrum below the disturbance levels specified by the MAG and RPW instruments of the Solar Orbiter.
Few spectrum lines exceed such levels, but they are clearly identified and stable in both frequency and amplitude (Fig.~\ref{fig:laget004}). 
At the same time, no degradation of the performance has been observed over the susceptibility test performed on the PFM, in the radiated mode only.

\section{Instrument characterisation and performance}
\label{sec:calibrations}

\subsection{Calibration facility and test configuration setup}
\label {sss:testsetup}

\begin{figure*}
\centering
\includegraphics[width=0.75\textwidth]{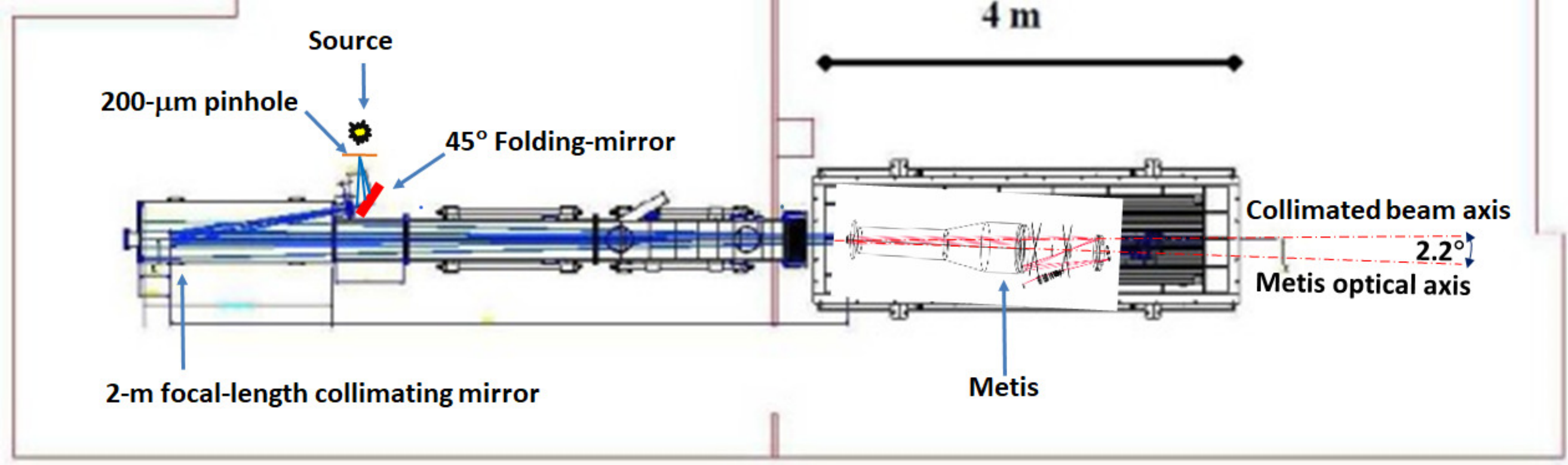}
\caption{Optical Payload Systems facility used for Metis characterisation and calibration tests consisting of: Sun simulator and pipeline located in an area ISO8 (on the left in the figure), and test chamber, where ISO 5 clean conditions can be reached, located in an area ISO7 (on the right in the figure). In this figure, Metis is represented in a $2.2\degr$ tilted setup with respect to the collimated source beamline.}
\label{fig:SPOCC}
\end{figure*}

Metis characterisation and calibration were performed in the INAF Optical Payload Systems laboratory at ALTEC in Turin \citep{fineschi2011}. This facility hosts the vacuum space optics calibration chamber (Fig.~\ref{fig:SPOCC}), consisting of three main sections:
\begin{itemize}
\item the Sun simulator section;
\item the pipeline section: the Sun simulator and pipeline section are in an area ISO 8;
\item the ($1.08 \times 0.88 \times 4$~m$^3$) test section, located in the cleanest area: this consists of a cylindrical chamber that contains the optical bench (ISO 5) located in a clean room (ISO 7).
\end{itemize}

The operative vacuum level of 10$^{-6}$~mbar in nominal conditions was reached within 24 hours of pumping, with a liquid nitrogen LN$_2$ line cooling an internal cold plate that has the effect of a cryogenic pump. 

The motorised vacuum-compatible optical bench has dimensions of $0.9 \times 3.5$~m$^2$; the maximum allowed height of the instrumentation that can be installed on the bench is 0.55~m. Two groups of actuators can tilt in pitch and yaw and translate the optical bench perpendicularly to the chamber longitudinal axis, servo-controlled by a remote computer. Metis was fixed on this optical bench during all the in-air and in-vacuum characterisation and calibration activities.

Several optical measurements were obtained to estimate the Metis optical performance, particularly the instrument imaging capability, the radiometric and polarimetric response, and the stray-light rejection. 
In order to accurately test Metis, considering the very different solar angular sizes during the orbits of the  Solar Orbiter, the Metis calibration camera illumination system is designed to allow an extended circular source as seen at infinity with angular semi-divergence selectable in the range of $0.26\degr$--$0.53\degr$, corresponding to the Sun disk observed from $\sim1$~AU down to $\sim0.5$~AU. In addition, for the stray-light tests described in the following, it is also possible to select an annular source as seen at a finite distance by the IEO, with angular semi-divergence selectable in the range $0.50\degr$--$0.89\degr$. The light emitted by the source is projected towards the  Metis entrance aperture by a parabolic mirror surrounded by a light trap simulating a `dark' corona with a level of brightness less than 10$^{-11}$ times the source brightness. 
A series of baffles is  inserted  in the pipeline to suppress stray light coming from the mirror edge.

The experimental setup for the imaging tests is obtained by illuminating a suitable set of pinholes located on the  collimator focus position. One or several collimated light beams are obtained in order to simulate sources at infinity at different incidence angles within the Metis FoV. 
The radiometric response and the vignetting function in the visible light were obtained using a flat-field source that illuminated the full Metis FoV. The same flat-field source coupled with a rotating linear polariser was used for the polarimetric characterisation.

\subsection{Instrument response, effective areas, expected counts}

\begin{table*}[h]
\caption{Measured throughput values of Metis optical components.}
\label{tab:efficiency}
\centering
\begin{tabular}{l|l|l}
\hline
\hline
\multirow{2}{*}{Component} & \multicolumn{2}{|c}{Throughput Parameter} \\
& VLD & UVD \\
\hline
Telescope mirror reflectivity (M1, M2) & $0.90\pm0.02$ & $0.54 \pm 0.1$ \\
Interference filter reflectivity/transmittivity & $0.886\pm 0.001$ & $0.24 \pm 0.04$ \\
Polarimeter throughput & $0.33 \pm 0.05$ & -- \\
Detector quantum efficiency & $0.50 \pm 0.05$ & 0.18 \\
\hline
Overall efficiency ($\eta$) & $0.118\pm 0.022$ & $0.0126 \pm 0.004$ \\
\hline
\hline
\end{tabular}
\end{table*}

\begin{figure}
\includegraphics[width=\columnwidth]{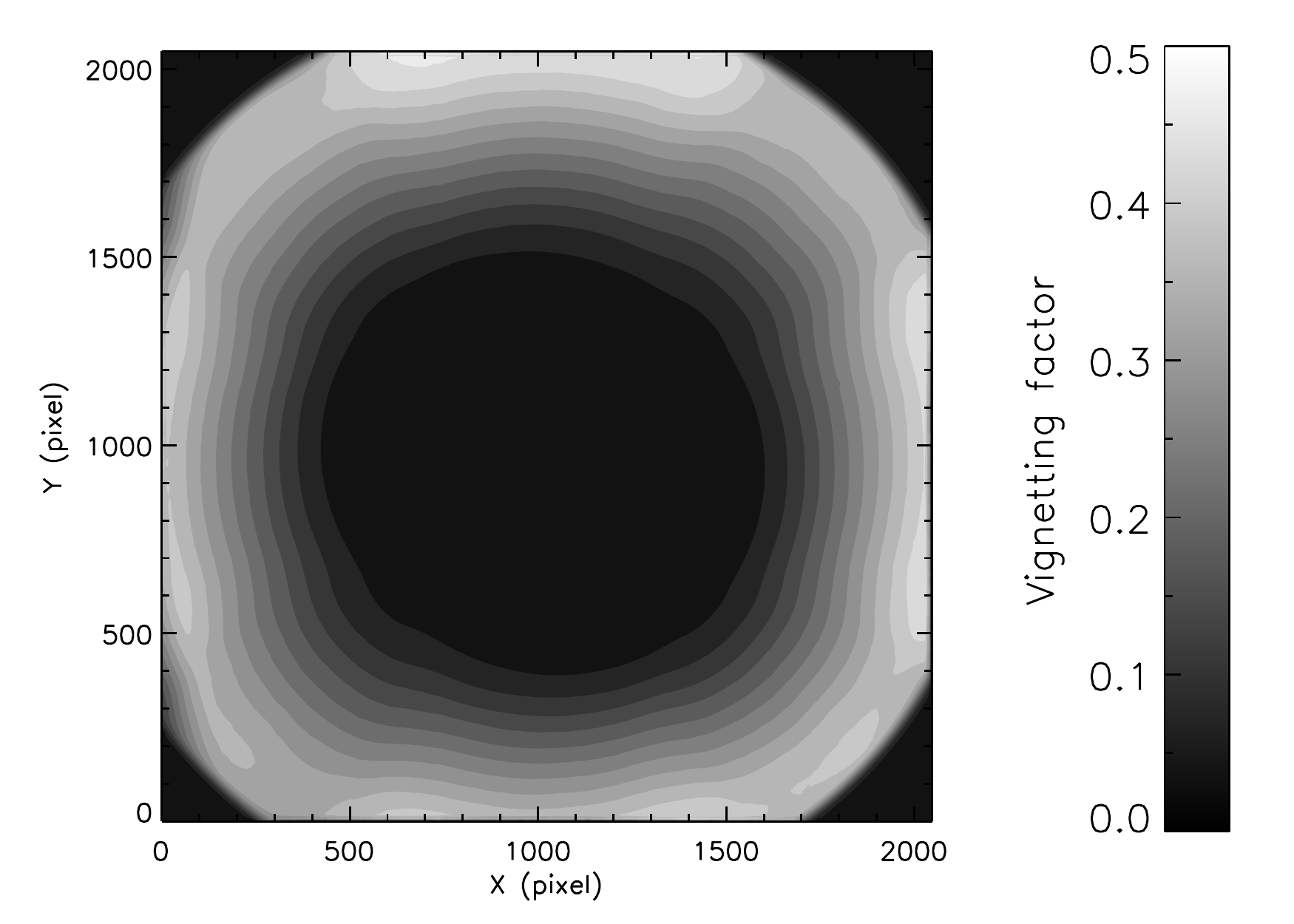}
\caption{The measured two-dimensional vignetting function of Metis.}
\label{fig:vign}
\end{figure}

In order to estimate the Metis radiometric performance, the expected throughput $T$ was determined. This quantity is defined as:
\begin{equation}
T(x,y) = v(x,y) \cdot \eta,
\end{equation}
where $v(x,y)$ is the vignetting function depending on the $(x,y)$ detector spatial coordinates, and $\eta$ is the instrument global efficiency. The latter was derived from the measurements of the properties, reported in Table~\ref{tab:efficiency} at subsystem level, of the Metis components: namely the\ mirror reflectivity, filter reflection/transmission, polarimeter transmission of unpolarised light, and detector quantum efficiency. The two-dimensional vignetting function of Metis, derived from measurements in the visible light, is shown in Fig.~\ref{fig:vign}. The UV vignetting function is the same, since visible and UV light share the same optical path in the telescope, which is the only vignetting optical element. 

The effective area $A_{\it eff}$ characterising the instrument response is given by multiplying the entrance pupil geometrical area $A_T$ ($A_T = 12.57$ cm$^2$) by the measured throughput. 
Table~\ref{tab:areas} reports the resulting $A_{\it eff}$ values for the two imaging channels as a function of the FoV.

\begin{table}[t]
\caption{Effective areas of the VL and UV imaging paths as a function of the distance from the centre of the FoV.}\label{tab:areas}
\centering
\begin{tabular}{c|c|c}
\hline\hline
\multirow{2}{*}{FoV (\degr)} & \multicolumn{2}{|c}{Effective area (cm$^2$)} \\
& VL channel & UV channel \\
\hline
1.6 & $1.14\times 10^{-2}$ & $1.24\times 10^{-3}$\\
1.8 & $8.62\times 10^{-2}$ & $9.34\times 10^{-3}$\\
2.1 & $2.43\times 10^{-1}$ & $2.63\times 10^{-2}$\\
2.4 & $4.06\times 10^{-1}$ & $4.40\times 10^{-2}$\\
2.7 & $5.80\times 10^{-1}$ & $6.28\times 10^{-2}$\\
2.9 & $6.84\times 10^{-1}$ & $7.41\times 10^{-2}$\\
\hline
\hline
\end{tabular}
\end{table}

The overall Metis radiometric performance can be expressed in terms of the count rates $C$ (measured in photons/pixel/s) expected for each channel.
The count rates are obtained by multiplying the coronal radiance $I_C$ (photons/s/cm$^2$/sr) by the effective areas $A_{\it eff}$ (cm$^2$) and the angular pixel equivalent size $\Omega$ (sr/pixel):
\begin{equation}
C = A_{\it eff} \cdot \Omega \cdot I_C,
\end{equation}

where $\Omega_{VL}=2.42 \times 10^{-9}$ sr/pixel for the visible imaging path and $\Omega_{UV}=9.68 \times 10^{-9}$ sr/pixel for the UV imaging path.
For the visible light path, the expected count rates calculated for a Sun-spacecraft distance of 0.28 AU corresponding to the minimum perihelion are of the order of 500--5000 (100--2000)~counts/s/pixel for typical streamer (coronal hole) conditions. For the UV light path, the expected count rates can be lower by more than two orders of magnitude due to the lower efficiency (about one order of magnitude) of the UV optical components and the lower photon flux in Lyman-$\alpha$. However, sufficient statistics can be obtained by integrating the UV count rates for longer time intervals.
We note that in the case of maximum activity, the expected count rates increase by a factor ranging from three to six on average, depending on the distance of the Solar Orbiter from the Sun.

\subsection{Metis optical performance}

Metis VL imaging performance was measured by illuminating the instrument with a multi-beam source. A mask of  200~$\mu$m in diameter with five pinholes positioned at the focal plane of the collimator was illuminated. In this way, five collimated beams, approximately covering a 0.1$^{\circ}$ FoV, illuminate the Metis entrance pupil. The collimator coupled to the Metis VL channel provides an imaging system with a total demagnification of about one tenth. Considering the VL detector pixel size, each pinhole is thus geometrically imaged with the VLDA on a $2 \times 2$ pixel box. Hence, in order to take into account the finite size of the source image used during Metis VL channel characterisation and to provide a more accurate instrumental point spread function (PSF), all the acquired images were deconvolved with the geometrically calculated full width at half maximum (FWHM) broadening.

\begin{figure}
\includegraphics[width=\columnwidth]{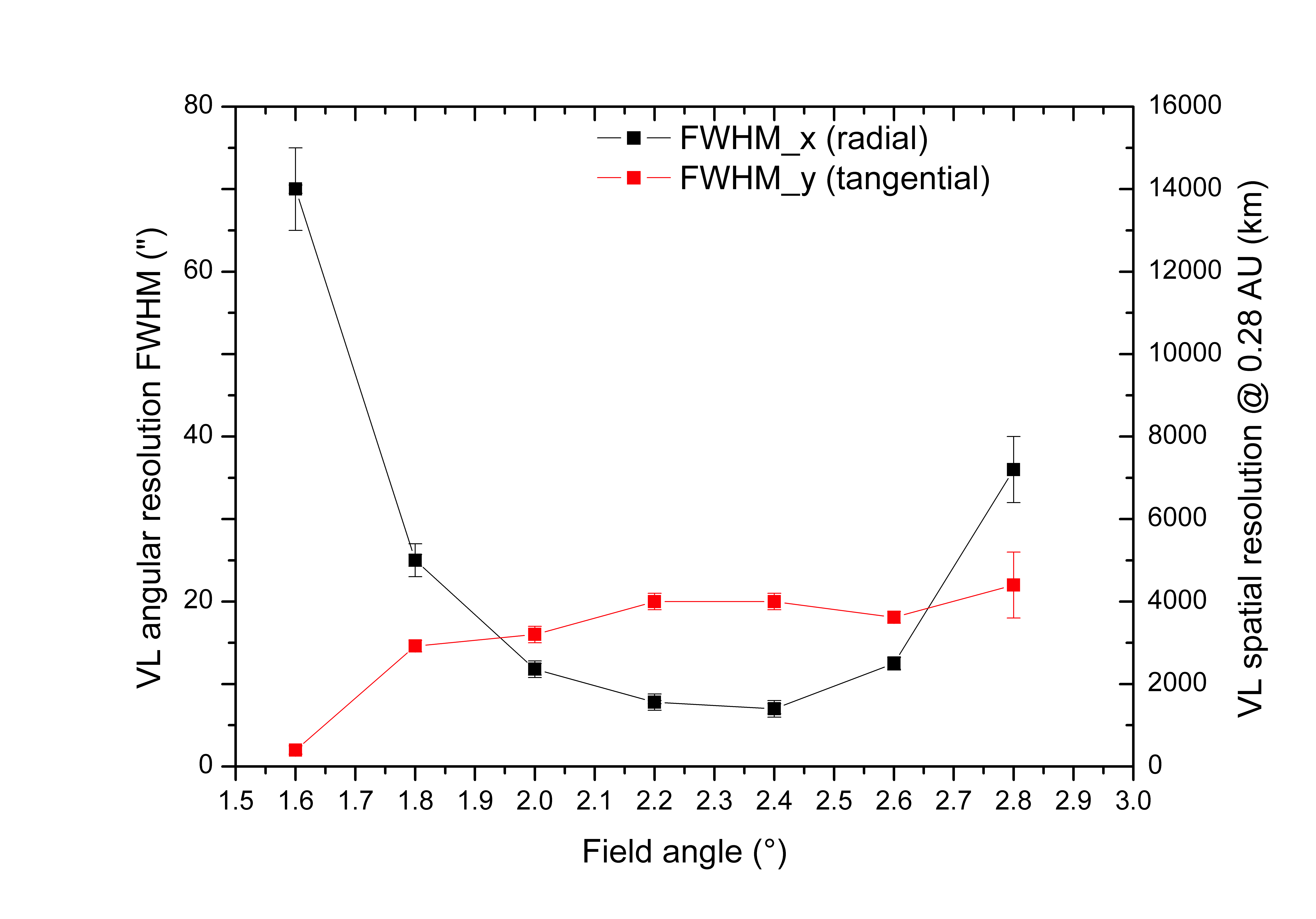}
\caption{Measured Metis VL channel angular and spatial resolution  (measured in arcsec and km, respectively) as a function of the FoV along the radial and tangential directions (i.e. along the direction of a radius from the Sun centre, and the perpendicular one, respectively).}
\label{fig:VL_res}
\end{figure}

As expected, diffraction effects dominate the PSF only in the inner part of the FoV because of the strong vignetting. Since the five pinholes are relatively close to each other, no significant image difference is expected from one to the other; consequently, each pinhole image can be used as a statistical sample, and the Metis VL PSF is derived as the mean value of the five FWHM spots. Six angular positions inside the FoV were measured. 
The measured values of the angular resolution of the VL channel are summarised in Fig.~\ref{fig:VL_res}. A resolution element of the order of $\leq$20~arcsec is obtained over essentially all the investigated FoV, with the exception of the inner portion where diffraction is dominating and the spot size is larger, and the outer portion where astigmatism dominates.

The Metis UV channel was tested with a different setup: instead of the five pinholes mask, a single 25~$\mu$m pinhole was set on the collimator focus and a krypton lamp source was used. In this case, thanks to the much smaller pinhole size, no deconvolution of the measured image was necessary to determine the PSF of the UV channel. The same side of the FoV was sampled, as in the case of the VL channel. The UV detector was cooled at $- 6\degr$C, $-7\degr$C and operated in analogue mode.

\begin{figure}
\includegraphics[width=\columnwidth]{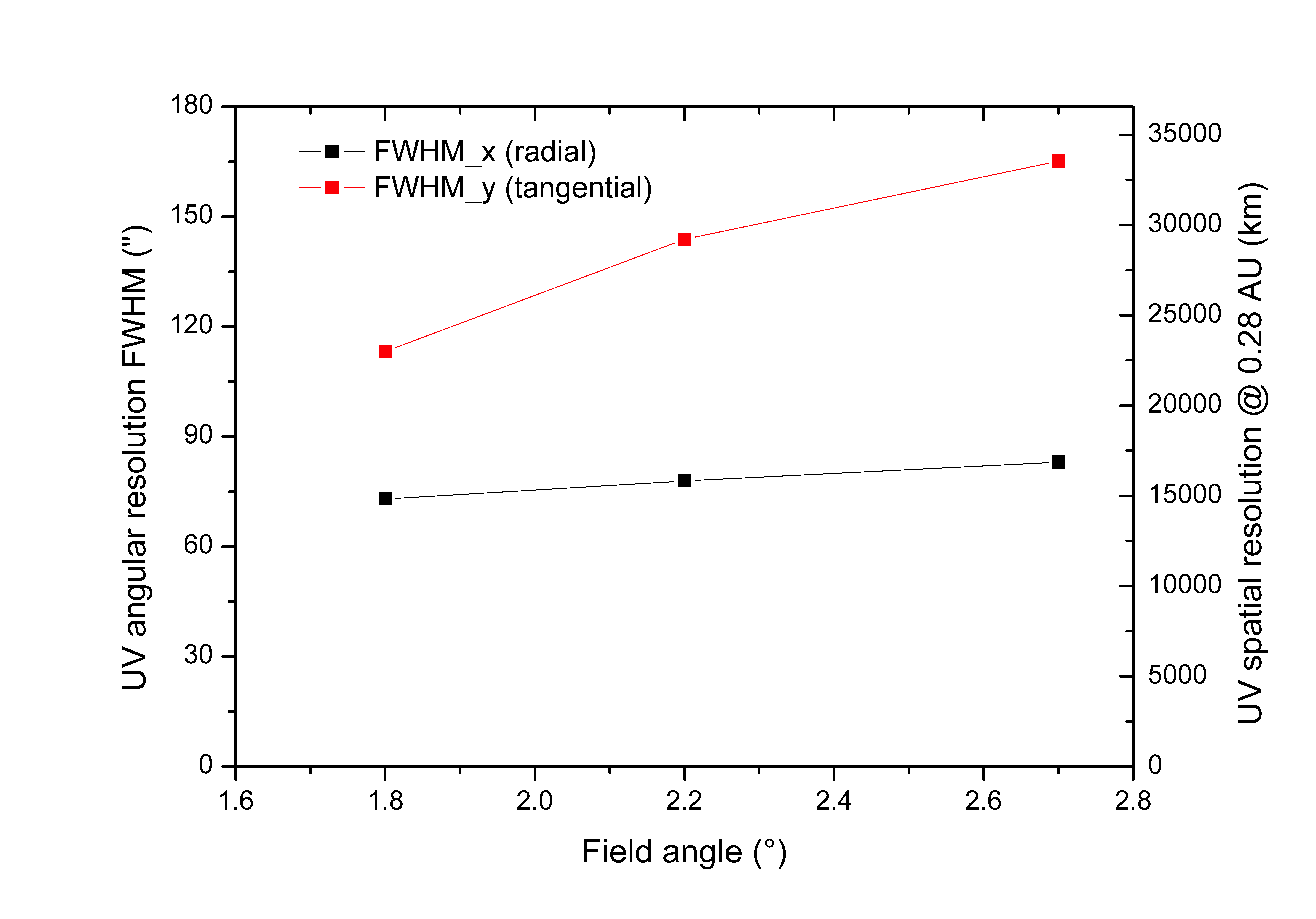}
\caption{Measured Metis UV channel angular and spatial resolution (measured in arcsec and km, respectively) as a function of the FoV along the radial and tangential directions.} 
\label{fig:UV_res}
\end{figure}

The measured values of the angular resolution of Metis UV channel are shown in Fig.~\ref{fig:UV_res}. A resolution element of the order of 80~arcsec is obtained along the radial direction; at minimum perihelion this corresponds to about 15000 km. The resolution is degraded by a factor varying from 1.5 to 1.7 along the tangential direction. By dedicated analyses of the  images acquired at all the steps followed during the procedure for the UV channel alignment, it was possible to infer that the degradation in resolution along the tangential direction is due to two concomitant effects: the UV detector focal plane is probably not exactly in the nominal position, and the UV detector resolving element is slightly larger than expected. The two combined effects generate a degraded spot, resulting in an astigmatic image. We are presently verifying the possibility of partially recovering the UV image quality by a suitable dedicated deconvolution algorithm.
A more detailed presentation of the calibration results on the Metis optical performance can be found in \citet{frassetto2018}.

\subsection{Expected in-flight optical performance}
The expected optical performance of Metis was estimated for the on-orbit case, when subjected to thermal stress. For calculating the Metis angular resolution expected at the end of integration, a statistical analysis by means of Montecarlo simulations was performed, combining the structural manufacturing, the alignment, and the integration tolerances, and then adding the measured wave-front errors of the mirrors assembled in their mechanical structure. All the possible effects that could potentially contribute to a degradation of the image quality were included in the optical models, for both the UV and the VL channels.

To obtain the expected optical performance once in orbit, the effects of the thermo-elastic deformations and of the launch vibrations were also considered. To this end, temperature maps for both hot and cold operative scenarios were processed with structural analyses, producing displacements, tilts, and deformations of the optical surfaces. The displacements and tilts introduced by the launch were instead derived by the vibrational test campaign performed at Metis STM level. This analysis demonstrated that both launch and on-orbit thermal stresses are not expected to significantly change the instrument performance measured at the end of integration \citep{sandri2017b}.

\subsection{Polarimetric performance}

\begin{figure*}
\centering
\includegraphics[width=0.85\textwidth]{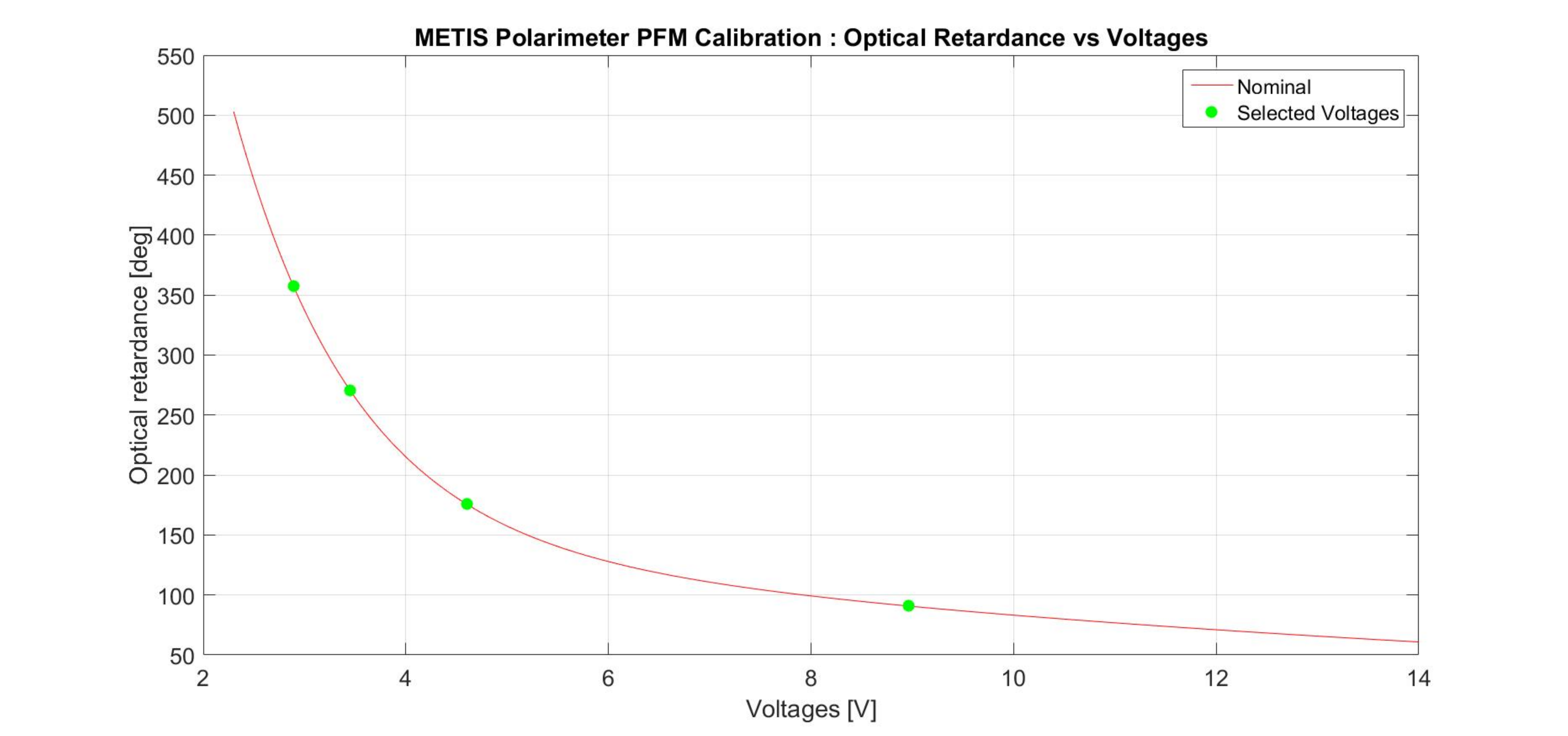}
\caption{Calibrated optical retardance curve of the liquid crystal-based variable retarders of the polarisation modulation package at $30\degr$C.}
\label{fig:modulation}
\end{figure*}

The Metis active polarisation element is the polarisation modulation package, consisting of two anti-parallel nematic LCVRs; the modulation state is changed by applying suitable voltages to both. In Fig.~\ref{fig:modulation} it is possible to see the calibration curve of the optical retardance of the  LCVRs as a function of the applied voltage obtained at $30\degr$C, while the actual LCVR applied voltage for specific multiple quarter-wave retardances is reported in Table~\ref{tab:modulation}.

\begin{table}[t]
\caption{Voltages applied to the  LCVRs of the PMP to obtain multiple quarter wave retardances (green dots in Fig.~\ref{fig:modulation}).\label{tab:modulation}}
\centering
\begin{tabular}{c|c|c}
\hline
\hline
\multirow{2}{*}{Nominal retardance (deg)} & \multicolumn{2}{|c}{Applied voltage (V)} \\
& LCVR1 & LCVR2 \\
\hline
90 & 8.96 & 8.96 \\
180 & 4.60& 4.60\\
270 & 3.45 & 3.45 \\
360 & 2.89 & 2.89 \\
\hline
\hline
\end{tabular}
\end{table}

The linear polarisation Stokes vector $S$ of the coronal light is directly related to the acquired image $M$ and the demodulation matrix $X^+$ by the relation
\begin{equation}
S=X^+ \cdot M.
\end{equation}

\noindent Selecting the voltage values corresponding to the $90\degr$, $180\degr$, $270\degr$, and $360\degr$ retardances provided in Table~\ref{tab:modulation}, the measured demodulation matrix is 
\begin{equation}\label{eqn:measured}
X^{+}_{m}=\left(
\begin{array}{cccc}
0.50 & 0.49 & 0.50 & 0.51\\
1.00 & 0.06 & -1.00 & -0.06\\
0.02 & 0.99 & 0.00 &-1.01
\end{array}\right)
,\end{equation}

\noindent which is in agreement within 6\% with the theoretical demodulation matrix for the same values of retardances.  
This is also confirmed by the measured polarisation modulation vector \citep{deltoro2000}.
A more detailed discussion of the polarimeter calibration results are found in \citet{casti2018} and \citet{capobianco2018}.

\subsection{Stray light}

During Metis calibrations, the stray light value at the VL channel focal plane was measured. When Metis points to the centre of the solar disk simulator (see Sect.~\ref{sss:testsetup}), the VL detector measures only stray light. Its map on the detector focal plane is derived as follows:
\begin{equation}
S_\text{Map}=\frac{S_\text{IM}-Bg}{\hat{I}_{\odot}},
\end{equation}
where $S_\text{IM}$ is the stray-light image, $Bg$ is its corresponding background, and $\hat{I}_{\odot}$ is the mean solar disk brightness on the visible light focal plane. The value of $\hat{I}_{\odot}$ is obtained by off-pointing Metis at $2.2\degr$ and by measuring the average image brightness of the simulated solar disk. The Sun disk image was normalised with the vignetting function to retrieve the solar disk as if it was imaged on-axis without any occulting system. 
Images were acquired at several simulated heliocentric distances, from 0.5~AU to 1.0~AU. 
The measured stray-light level is below $5 \times 10^{-10}$ over most of the Metis FoV, with the exception of  two tiny peripheral zones, where it reaches at most a value of  $10^{-9}$, thus fulfilling the requirements (see Sect.~\ref{sec:instr_overview}). Moreover, the slight enhancement  in these zones has to be ascribed to a residual source identified inside the test facility pipeline, and not to the instrument stray-light reduction capabilities. The activities performed to  minimise the  level of stray-light in the Metis instrument are presented by \citet{landini2018}.

\subsection{Optical element coating performance}
Metis provides images of the corona, both in the visible range and at the \ion{H}{i}~Lyman-$\alpha$ emission line. Therefore, the Metis optical components require optical coatings able to reflect/transmit such spectral components. The selection of the coating material was based on tests performed on different mirror samples, which have been fully characterised. The selected material for the M1 and M2 coating is a standard Al protected by MgF$_2$. In the case of M0, the coating material is Al protected by SiO$_2$. The reflectance was measured at the \ion{H}{i}~Lyman-$\alpha$ line (121.6~nm) and in the whole 130--160~nm spectral range by using a normal-incidence UV reflectometer. More details about this facility can be found in \citet{Zuppella2012}. 

The reflectance measurements in the visible spectral range were performed at a 3$\degr$ angle of incidence.
Such measurements are rather homogeneous and in line with standard reflectivities in the visible spectral range. In the UV, the measurements show relatively large variations in reflectivity depending on the session of coating deposition. 
The flight mirrors were selected by reaching the best compromise between reflectivity in the UV and substrate figuring. Following this criterion, we selected a couple of mirrors with lower reflectivity than nominal (see Table \ref{tab:efficiency}), thus accepting to reduce the foreseen throughput, but allowing us to obtain an overall channel performance in terms of opto-mechanical stability essentially in line with the expectations.

Concerning the interferential filter, whose substrate is MgF$_2$, with the crystal axis perpendicular to the front surface of the filter, it was coated on the side first facing the light (the other being uncoated) with a dichroic layer which both reflects the visible portion of the spectrum and transmits a 10~nm narrowband centred at 121.6~nm. Values of the measured performance of the coating are given in Table~\ref{tab:efficiency}.

\subsection{Metis radiation environment}

An estimate of the dose released by galactic cosmic rays and solar energetic particles in the polarimeter of the Metis coronagraph was carried out with the Fluka Monte Carlo code. The \citet{gleeson1968} model was used to infer the galactic cosmic-ray minimum and maximum projections during the mission on the basis of the interstellar proton spectrum estimated by \citet{shikaze2007} from the BESS experiment data \citep[see][and references therein]{grimani2012}. Minimum and maximum solar modulation levels were considered according to the strongest and weakest solar cycles of the last 100 years. Average and worst case projections for SEP fluxes were also estimated and taken into account for dose calculations. Galactic and solar energetic particles were propagated through the polarimeter geometry environment created in Fluka (protons below 10 MeV and electrons below 1 MeV are shielded by the spacecraft). The simulated geometry includes all main components and materials. Matter distribution around the coronagraph constrains the production of secondary particles that penetrate the instrument. The average matter thickness surrounding the coronagraph was set to 1.2~g/cm$^{2}$ of aluminium-honeycomb and CFRP. The projections of SEP event occurrence were carried out according to the Nymmik model \citep[see][and references therein]{grimani2014}. All details are reported by \citet{telloni2016a}. These projections are at 1~AU. The actual orbit of the Solar Orbiter mission must be considered for dose calculation. A dose ranging between 100~Gy/year and 800~Gy/year is estimated.

Cosmic rays may play a more relevant role than SEPs (average projections), if the average distance of the spacecraft from the Sun is considered over the mission (0.6~AU). It is worth pointing out that the dose reported above may increase by one order of magnitude in the case where one event of intensity 10$^{10}$ particles/cm$^{2}$  occurs when the spacecraft is at distances of less than 0.41~AU from the Sun (i.e. for 14\% of the time).
Moreover, it has to be taken into account that all considered models have uncertainties. The Nymmik model was found in agreement within a factor of two with observations. In addition, SEP actual spatial and energy distributions should be taken into account. All these uncertainties may increase the actual dose released in Metis by another 50\% as a worst case with respect to standard average projections. The aim is to improve this work on the basis of more refined models when the trend of the next solar minimum is known.

\section{Data acquisition and on-board processing}

Metis can acquire data from the two channels (Sect.~\ref{sec:onboardsw}), independently and simultaneously, by implementing the acquisition schemes  listed in Table~\ref{tab:schemes} and described in more detail below.

\begin{table}
\centering
\caption{Metis acquisition schemes}
\begin{tabular}{l|ll}
\hline
\hline

\multirow{4}{*}{\specialcell{VL\\channel}}
& {VL-pB} & Polarised brightness acquisition \\
& {VL-tB} & Total brightness acquisition \\
& {VL-FP} & Fixed polarisation acquisition \\
& {VL-TN} & Temporal noise acquisition \\
\hline

\multirow{4}{*}{\specialcell{UV\\channel}}
& {UV-Analogue} & Analogue mode acquisition \\
& {UV-PC} & Photon counting acquisition\tablefootmark{a} \\
& {UV-PC-Offset} & Photon counting offset mode \\
& {UV-TN} & Temporal noise acquisition \\
\hline
\hline 
\end{tabular}
\tablefoot{
\tablefoottext{a}{Detected events can be delivered in two different formats: 
as a list or accumulated in a matrix.}}
\label{tab:schemes}
\end{table}

\subsection {Visible-light acquisition schemes}
During acquisition, CME monitoring  and Sun disk brightness monitoring can be active to detect  CME events and potentially dangerous off-pointing, respectively 
(Sect.~\ref{ops:sci:cme} and Sect. \ref{ops:offpoint}). Visible-light images can be corrected on-board for cosmic rays and SEP events.

\begin{figure}
\centering
\includegraphics[width=\columnwidth]{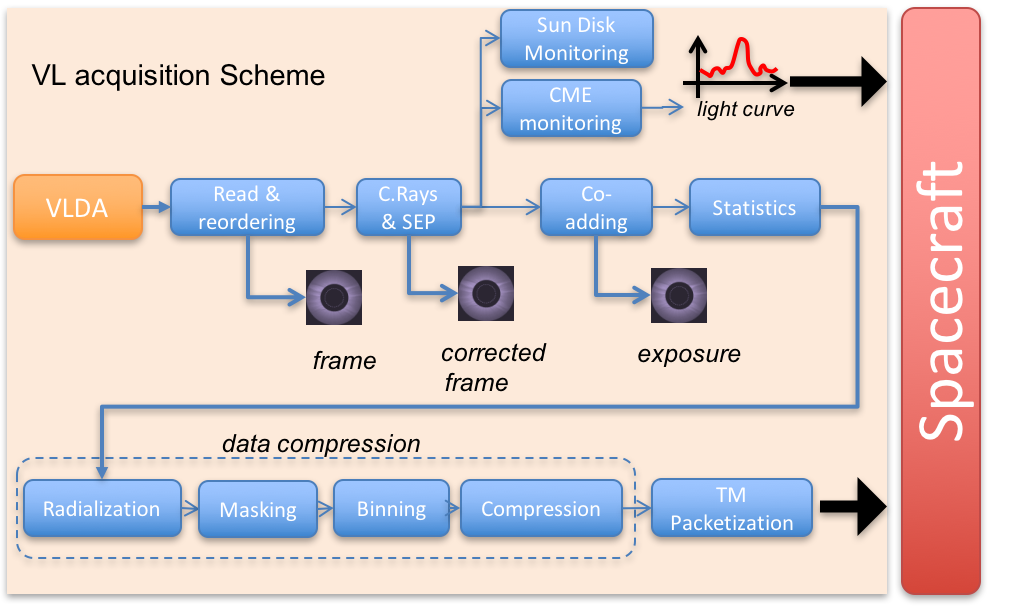}
\caption{Scheme of the VL channel data flow.}
\label{fig:VL_flow}
\end{figure}

The VL channel data flow is depicted in Fig.~\ref{fig:VL_flow} and is common to all acquisition schemes, with a few exceptions, described below. 

\subsubsection {{Polarised brightness of visible light}} In this acquisition scheme, the VL detector acquires and delivers frames at a constant rate and detector integration time. Each frame is acquired at a different known and configurable polarisation angle. The measurement is carried out cycling over a number of specified polarisation angles, thus producing an interleaved data stream (Fig.~\ref{fig:VL_flow}). This number can assume the values  three or four. The on-board data-processing pipeline is capable of  executing the auxiliary computation and monitoring, that is, statistics, CR/SEP removal, and CME monitoring. Eventually, the frames with the same polarisation are averaged and compressed in order to deliver to ground a set of polarisation images at the specified cadence. This acquisition scheme has  a minimum detector integration time of 15~s. 
\subsubsection{Total brightness of visible light} The VL detector acquires and delivers frames at constant rate and detector integration time. The number of polarisation angles can assume only the value of two. All frames are acquired by switching the polarisation angle  in the middle of the detector integration time. The onboard data processing pipeline is capable of executing the auxiliary computation and monitoring.  Eventually, the frames are averaged and compressed in order to deliver to ground a single image at the specified cadence. This acquisition scheme has the limitation of a minimum detector integration time of 20~s. 
\subsubsection{Fixed polarisation imaging in visible light} The VL detector acquires frames at a given detector integration time and delivers them at constant rate. This acquisition scheme was created in order to perform acquisitions at a high rate. All frames are acquired by keeping the polarisation angle at the same fixed position.
The CME and Sun-disk monitoring are disabled (see Sect.~\ref{ops:sci:cme} and~\ref{ops:offpoint}). The CR/SEP correction algorithm is disabled, since the frames are not summed. Eventually, the individual frames are compressed and delivered to ground. This acquisition scheme has the limitation of a minimum detector integration time of 1~s and of a maximum number of 64~frames. At 1~s detector integration time and for 64 frames, the total acquisition time is about 11~minutes, including processing time. 
\subsubsection{Temporal noise of visible light} This acquisition scheme is available only in diagnostic mode. The VL detector acquires a given number of frames at a given detector integration time and computes the mean and standard deviation for each sensor pixel. All frames are acquired by keeping the polarisation angle at the same fixed position.
The purpose of this acquisition scheme is to periodically  monitor the detector performance. The two frames, one with the mean and the other with the standard deviation, are compressed and delivered to ground. For this acquisition scheme, no auxiliary computational and monitoring process is active. 

\begin{figure}
\centering
\includegraphics[width=\columnwidth]{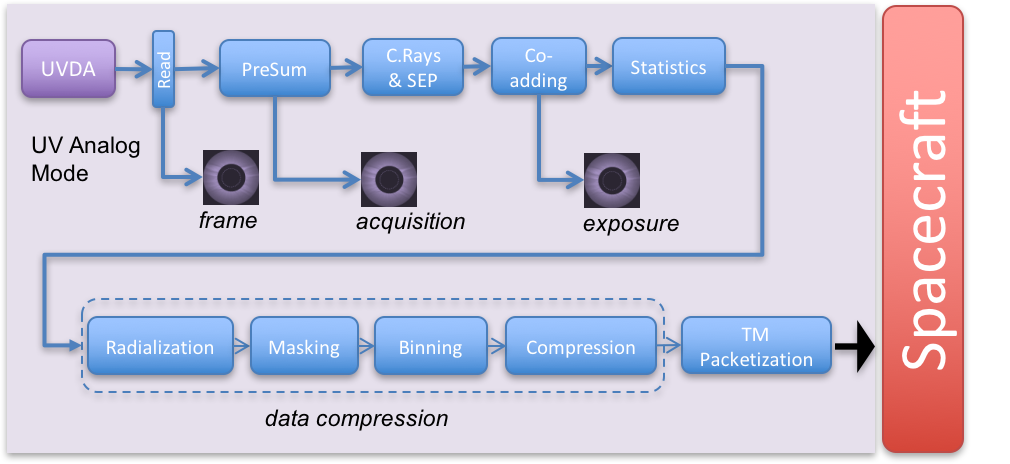}
\caption{The data flow of the UV channel acquiring data in analogue mode: UV-analogue and UV-temporal noise acquisition schemes.}
\label{fig:UV_Analog_flow}
\end{figure}

\subsection {Ultraviolet acquisition schemes}
The UV channel implements different data flows according to the acquisition scheme adopted. The data flow illustrated in Fig.~\ref{fig:UV_Analog_flow} is common to both the UV-analogue and UV-temporal noise schemes.
\subsubsection{Ultraviolet-analogue} The frames are averaged in a two-step process. The first is performed in hardware (pre-sum) and the second via software (co-adding). Eventually, the resulting image is compressed and delivered to ground. This process is repeated in order to produce one image at the requested cadence. The on-board data processing pipeline is capable of executing the auxiliary computation and monitoring, that is, statistics and CR/SEP removal. This acquisition scheme has the limitation of a minimum detector integration time of 1~s. 
\subsubsection{Temporal noise in the UV} In this acquisition scheme, available only in diagnostic mode, the UV detector acquires a given number of frames at a given detector integration time and computes the mean and standard deviation for each sensor pixel. Its purpose is to  periodically monitor detector performances. Eventually, the two frames, one with the mean and the other with the standard deviation, are compressed and delivered to ground. For this acquisition scheme no auxiliary computational and monitoring process is active. 

\begin{figure}
\centering
\includegraphics[width=\columnwidth]{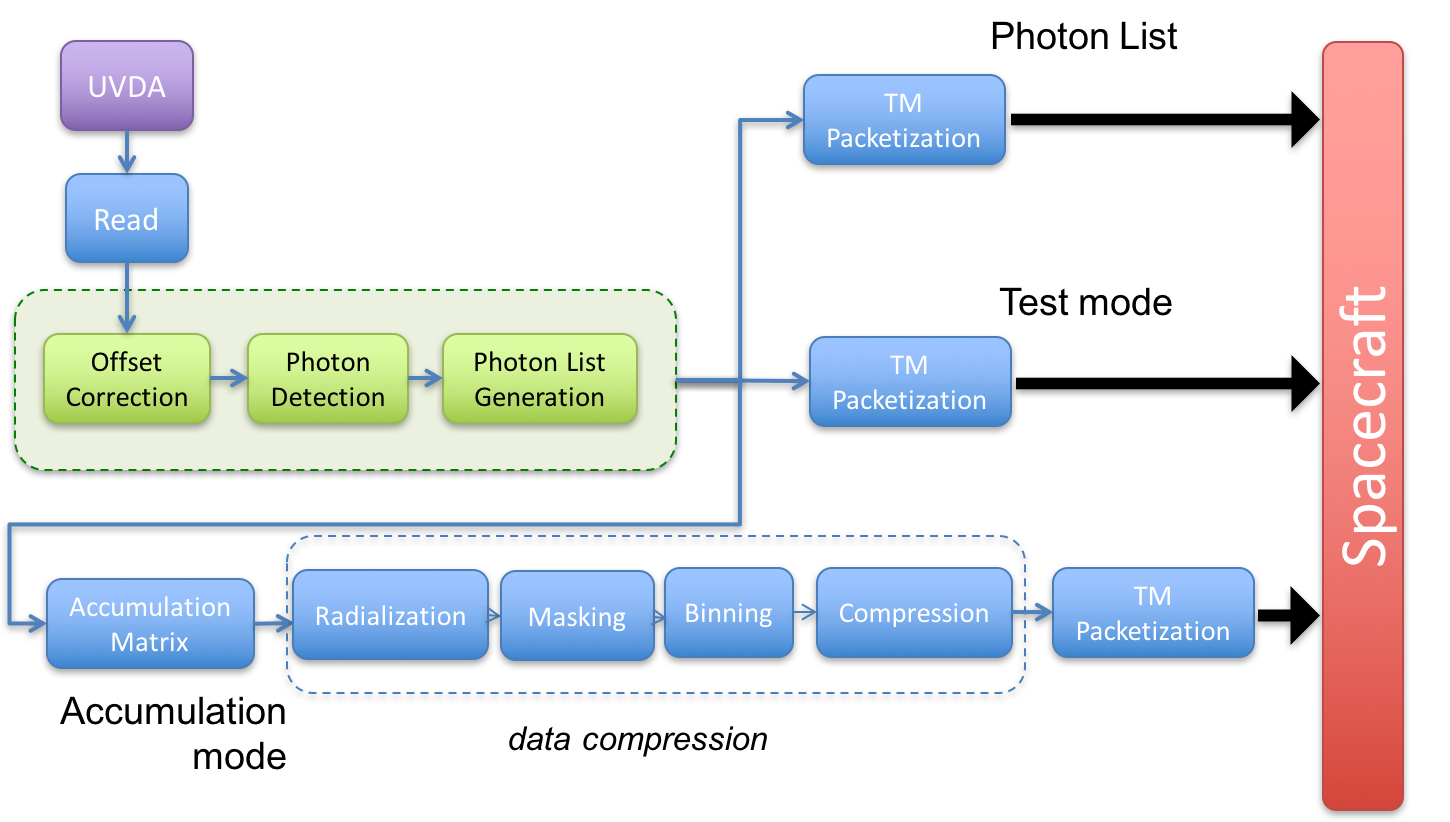}
\caption{The data flow of the UV channel acquiring data in photon counting mode: photon counting acquisition scheme.}
\label{fig:UV_PC_flow}
\end{figure}

\subsection {Ultraviolet photon-counting acquisition schemes}
The photon counting mode data flow is presented in Fig.~\ref{fig:UV_PC_flow}.
\subsubsection{Ultraviolet-photon counting} In this case, the APS is operated continuously, nominally at 97~ms integration time, i.e. the maximum frame rate in order to avoid overlapping of the charge spots generated by each primary photon. The output data are serially acquired by the photon counting unit (PCU) so as to generate a $3\times 3$ pixel event window that dynamically  sweeps the APS matrix. Each window is analysed, searching for the presence of events whose charge content lies within proper limits and satisfies a given set of morphological rules. The PCU algorithm performs three tasks: offset correction, photon detection, and photon list generation. The PCU will determine the centroid coordinates of the identified events and subsequently send a packet containing coordinate pairs and collected photon-charge information to the MPPU. While in photon-counting mode, the UV data can be delivered in three different output formats:
\begin{itemize}
\item Photon list format, if the count rate is below a parametrised threshold.
\item Accumulation, when the count rate is higher than the same parametrised threshold; in this mode the MPPU does not generate the list of events, but it collects the data in a $1024 \times 1024$ matrix.
\item Test mode, which provides the value of each pixel in a 3 x 3 window centred on detected photon events, is used to perform diagnostic and test procedures able to verify the status of the photon counting/photon list operation mode and the UV detector efficiency/functionality.
\end{itemize}
\subsubsection{Ultraviolet-photon counting offset} The purpose of this acquisition scheme is to correct for possible offset variations on the APS pixels electronics chains during the mission. To do so, a set of $N$ ($N=8, 16, 32, 64, 128$) frames is acquired, with the minimum exposure time (97~ms) and with the MCP intensifier switched off. Those frames are then averaged with a pixel-by-pixel arithmetic mean. The averaged frame is sent to ground and at the same time stored onboard to be used in the photon counting acquisition as offset correction matrix. The previous map is stored in a separate memory slot as a backup. No scientific products are delivered to ground in this mode.

\subsection{Data compression}
\label{sec:data_compression}

The Metis instrument generates a significant amount of data, making image compression essential in order to reduce the data volume before transmission to the ground segment. The compression algorithm developed for Metis has low complexity in order to match the limited computational resources available onboard; it can perform both lossless and lossy compression, and supports diversified quality-compression trade-offs in different image regions, since the signal in solar corona images significantly decreases with heliodistance.
Two types of methods were considered in the algorithm design. Prediction-based techniques have been successfully employed for lossless and near-lossless compression; in these methods, a predictor is used to estimate the current sample from the previous ones, so that only the prediction residuals need to be encoded and transmitted. Transform coding techniques have been prevalent for lossy compression \citep[e.g.][]{penna_grs}, also in conjunction with standards such as JPEG 2000. A preliminary analysis was performed by \citet{bemporad2012}, showing that at medium and high bit-rates the predictive approach is best suited to accommodating low-complexity, high-compression efficiency, as well as the efficient handling of regions of interest. Therefore, the algorithm was based on an extension to the recent Consultative Committee for Space Data Systems (CCSDS-123) recommendation, which defines a predictive lossless data compression algorithm, to be applied to three-dimensional image data from payload instruments. The extensions have the objective to adapt the standard to the specific requirements posed by the images of the solar corona. This has resulted in an algorithm design with the following innovative features: a quantisation feedback loop to enable lossy compression; a new entropy coding stage based on the range encoder enabling compression of multi-temporal solar corona image stacks; image reordering through a radialisation process in order to exploit the specific image geometry; ability to encode regions of interest with different quality (including lossless); and variable binning on rectangular or annular image areas. In the following we discuss some of these features in more
detail.

\subsubsection{Quantisation feedback loop and entropy coding} 

The predictor is used in a quantisation feedback loop, whereby the predictor uses as input the previous reconstructed samples generated by a local decoder. Moreover, the Golomb coder employed in CCSDS-123 has been replaced by a range encoder \citep{martinrange} as in \citet{magli_rc}, since the Golomb encoder becomes suboptimal as the bit rate approaches 1 bit/pixel. 

\subsubsection{Multitemporal compression} 
Onboard a coronagraph, several images are usually acquired in temporal sequence, and therefore the multidimensional compression capability of CCSDS-123, initially devised to exploit wavelength correlation in spectral images, can be adapted to take advantage of temporal image correlation. In particular we exploit the similarity of time-consecutive acquisitions, switching the third dimension from wavelength to time; the gain in terms of compression performance is significant with respect to the individual compression of each image.

\subsubsection{Radialisation}
One of the most innovative parts of the compression algorithm consists in a pre-processing routine developed to exploit the radial symmetry of solar corona images. This so-called ``radialisation'' consists in rearranging the positions of the image pixels switching from a Cartesian coordinates system to a polar one, in which the two dimensions are the distance of each pixel from the centre of the image, and the angle with respect to a reference.
Essentially, the radialisation process scans the image from the centre towards its boundary, in circles of increasing radius. In this way, pixels that are expected to have similar values are placed in adjacent positions in the radialised image, minimising discontinuities encountered along the scanning pattern. The radialised image lends itself well to defining regions of interest shaped as annuli, which is a much more natural geometry than rectangular images for a coronagraph. Interestingly, radialisation does not require any interpolation among pixels, which is typical when changing a reference system. Indeed, radialisation can be seen as a permutation of the original pixels according to the new geometry; therefore the lossless property of the compression algorithm can be exactly preserved, since pixel values remain unaltered through the radialisation process. Moreover, radialisation is run just once at initialisation, since the mapping between the input image and the radialised image only depends on the image size and not on the pixel values.
The mapping from Cartesian to polar coordinates is perfectly reversible.
With the selected compression algorithm, errors are controlled individually at the pixel level on the polar image, and the error on each pixel remains the same after inverse mapping to the Cartesian image, since this is a one-to-one mapping.

\subsubsection{Regions of interest with variable quality}

Working with a radialised image allows to define regions of interest in a convenient and very natural way, as an interval of rows in the radialised image corresponds to an annular region in the original image; changing the quality for different row intervals lets the user set a different amount of compression in a selected number of annular regions. As an example, it is possible to tune the parameters so as to leave the inner part of the solar corona coded losslessly, and decrease the quality of reconstruction while moving away from the centre; quality levels within each region of interest are written in a header and used by the decoder to reconstruct each region correctly.

\subsubsection{Variable binning} 

Another option enabled by radialisation consists in applying binning on the radialised as opposed to the original image. Binning reduces the spatial resolution of the image by replacing a sample with its average value over a given window (e.g. $2\times 2$), followed by down-sampling, thereby increasing the signal-to-noise ratio by reducing sensor noise. Binning the radialised image allows for different binning ratios to be chosen over different intervals of rows. For example, this can be used to maintain a very fine spatial resolution in the regions of the image close to the solar corona, while decreasing the spatial resolution far away from the corona.

The combination of the features above provides great flexibility in the compression process, enabling high compression ratios (e.g., 10:1) to be obtained while preserving the scientific quality of the acquired data, through proper allocation of quality levels to different spatial regions of the images.

\subsection{Telemetry}

The packet utilisation standard (PUS) is chosen as infrastructure to communicate with the platform/ground and the ESA CCSDS standard to format the exchanged information. These general directives were adopted and further tailored to address the specific characteristics of the Solar Orbiter mission. Following these guidelines, Metis produces the normal telemetry packets (for housekeeping, events, acknowledgements to the received telecommands and the other foreseen onboard services) and defines the telemetry of its scientific packets. 

Considering the processing onboard and the different acquisition schemes that can be configured to operate the two detectors of the instruments, Metis produces ten different scientific data objects and one additional calibration and ancillary data product. Each data object contains a scientific header, reporting the relevant information concerning the acquisition, in the first telemetry packet. The Metis data products and the relevant processing levels are described in Sect.~\ref{ops:obs_seqs}.

\section{Operations concept and data handling}
\label{ops}

The Solar Orbiter mission profile dictates modes of operations that are more akin to those of an interplanetary probe than to those of a solar space observatory.
In particular, the limited availability of telemetry bandwidth and the one-way propagation delay of up to 16 minutes implies that practically all the operations, with a few exceptions such as those in the commissioning phase, are carried out offline. Instrument telecommands are normally uploaded well in advance for later execution, and all telemetry, including science data, is stored onboard for later retrieval. More specifically, the average telemetry rate available to Metis is 10.5~kbps, while the data storage volume available to Metis is 27.2~Gb ($1\ \text{Gb} = 10^9$~bits) per orbit. 

One of the main consequences of such limitations is that the delay between the acquisition of data onboard and their reception on ground (data latency) can normally be as long as weeks or even months, as described by \citet{sanchez2019}. Advance planning of operations is therefore vital for Metis and for all the instruments onboard Solar Orbiter. This requirement is indeed one of the main drivers of the Metis operation concept. Exceptions are discussed in Sect.~\ref{ops:offpoint} concerning detection of pointing anomalies, in Sect.~\ref{ops:sci:cme} in regards to detection of solar events, and in Sect.~\ref{ops:sw:lldata} with respect to handling of low-latency data.

An additional constraint is given by the fact that Metis can only operate when its occulter is centred on the solar disk, as described in Sect.~\ref{ops:offpoint}.

The following subsections illustrate the basic principle of operations of the Metis coronagraph, including the above-mentioned constraints. The method by which the Metis operations are integrated into the operations of the suite of instrument onboard Solar Orbiter is described in greater detail by \citet{sanchez2019}.

\subsection{Scientific operations}
\label{ops:sci}

During the nominal and extended phases of the Solar Orbiter mission, Metis observations, as well as observations from the other remote-sensing instruments, will be organised along each orbit into three ten-day intervals (remote sensing windows, RSW), centred around perihelion and around maximum and minimum latitudes with respect to the solar equatorial plane \citep{sanchez2019}. In general, Metis observations consist of global maps of the coronal emission obtained simultaneously  both  in the narrow-band UV \ion{H}{i} Lyman-$\alpha$ and in the broadband polarised VL in the 580--640~nm range.  Observations will be obtained at high temporal resolution (down to 1~s) and spatial scale down to about 2000~km in VL. Different spatial resolution and detector exposure time will be used depending on the requirements of the specific science goal, corresponding to specific observing modes of the instrument.
 
When observing during the RSWs along the orbit, the coronagraph FoV, which is in the range from 1.7 to 3.1-3.6~$R_{\odot}$ at closest perihelion (0.28~AU),  will span the corona out to about 9~$R_{\odot}$ (Table~\ref{tab:fov}  reports the FoV limits at different S/C heliodistances), and it will obtain  global maps of the UV \ion{H}{i}~Lyman-${\alpha}$ (121.6~nm) line and the VL polarised emission ($pB$) in the range 580--640~nm (a $pB$ sequence consists of 4 VL images taken at different polarisation angles, e.g. $0\degr$, $45\degr$, $90\degr$, $135\degr$).

These measurements will allow for a complete characterisation of the main plasma components of the corona and the solar wind: electrons and protons. As discussed in Sect.~\ref{sec:sci_objectives} and in greater detail in Sect.~\ref{ops:sci:analysis}, the UV images are used to derive global maps of the outflow velocity of the neutral hydrogen component, by measuring the Doppler dimming of the resonantly scattered component of the \ion{H}{i} Lyman-${\alpha}$ emission line. The coronal electron density needed to apply the Doppler dimming technique, is derived from the VL polarised emission of the corona. The measured expansion rate of the neutral hydrogen component also gives information on the outflow velocity of protons.

\subsubsection{Scientific observing modes}
\label{ops:sci:modes}

\begin{sidewaystable*}
\centering
\caption{Characteristic requirements of the Metis observing modes.}
\begin{tabular}{l|r|c|c|c|c|c|c|c|c|c|c|c|c}
\hline
\hline
\multirow{2}{*}{Mode} & \multirow{2}{*}{} & \multirow{2}{*}{FoV\tablefootmark{a}} & \multicolumn{2}{c|}{\multirow{2}{*}{\specialcell{Binning (pixel)}}} & \multirow{2}{*}{N$_\text{POL}$\tablefootmark{b}} & \multirow{2}{*}{DIT\tablefootmark{c}} & \multirow{2}{*}{T$_\text{ACQ}$\tablefootmark{e}} & \multirow{2}{*}{T$_\text{EXP}$\tablefootmark{g}} & \multirow{2}{*}{T$_\text{CAD}$\tablefootmark{h}} & \multirow{2}{*}{\specialcell{CR\\Removal\tablefootmark{i}}} & \multirow{2}{*}{\specialcell{CME\\Flag\tablefootmark{m}}} & \multirow{2}{*}{\specialcell{Compression\\Ratio\tablefootmark{n}}} & \multirow{2}{*}{\specialcell{Data\\Volume\tablefootmark{o}}}\\
& & & \multicolumn{2}{c|}{} & & & & & & & &\\
\hline
 & & & ($\leq 2.5\degr$) & ($> 2.5\degr$) & & & & & & & & \\
\multirow{2}{*}{WIND} & VL & \multirow{2}{*}{1.7--5~$R_\odot$} & \multirow{2}{*}{$2\times 2$} & $4\times 4$ & 4 & 15--20~s & = DIT & 75--450~s & \multirow{2}{*}{5--30~min} & \multirow{2}{*}{YES} & \multirow{2}{*}{ON} & 21 & \multirow{2}{*}{16.14~Mb}\\
 & UV & & & $2\times 2$ & --- & 1--60~s\tablefootmark{d} & 1--600~s\tablefootmark{f} & 5--30~min & & & & 5 & \\

\hline
\multirow{2}{*}{MAGTOP} & VL & \multirow{2}{*}{1.7--5~$R_\odot$} & \multicolumn{2}{c|}{\multirow{2}{*}{$2\times 2$}} & 4 & 15~s & = DIT & 75--300~s & \multirow{2}{*}{5--20~min} & \multirow{2}{*}{YES} & \multirow{2}{*}{ON} & 13 & \multirow{2}{*}{26.67~Mb}\\
 & UV & & \multicolumn{2}{c|}{} & --- & 1--60~s & 1--600~s & 5--20~min & & & & 3 & \\

\hline
 & & & ($\leq 2.5\degr$) & ($> 2.5\degr$) & & & & & & & & \\
GLOBAL & \specialcell{VL} & \multirow{2}{*}{1.7--3~$R_\odot$} & \multirow{2}{*}{$2\times 2$} & \multirow{2}{*}{$4\times 4$} & 4 & 15--30~s & = DIT & 75--450~s & \multirow{2}{*}{5--30~min} & \multirow{2}{*}{YES} & \multirow{2}{*}{ON} & \multirow{2}{*}{21} & \multirow{2}{*}{13.73~Mb}\\
(0.28 AU) & \specialcell{UV} & & & & --- & 1--60~s & 1--600~s & 5--30~min & & & & & \\

\hline
 & & & ($\leq 2\degr$) & ($> 2\degr$) & & & & & & & & \\
GLOBAL & \specialcell{VL} & \multirow{2}{*}{4--7.5~$R_\odot$} & \multirow{2}{*}{$2\times 2$} & \multirow{2}{*}{$4\times 4$} & 4 & 15--30~s & = DIT & 75--450~s & \multirow{2}{*}{5--30~min} & \multirow{2}{*}{YES} & \multirow{2}{*}{ON} &\multirow{2}{*}{31} & \multirow{2}{*}{9.21~Mb}\\
(0.7 AU) & \specialcell{UV} & & & & --- & 1--60~s & 1--600~s & 5--30~min & & & & & \\

\hline
LT-CONFIG & \specialcell{VL} & \multirow{2}{*}{1.7--3~$R_\odot$} & \multicolumn{2}{c|}{\multirow{2}{*}{$4\times 4$}} & 4 & 15--30~s & = DIT & 300--450~s & \multirow{2}{*}{20--30~min} & \multirow{2}{*}{YES} & \multirow{2}{*}{ON} & \multirow{2}{*}{51} & \multirow{2}{*}{5.70~Mb}\\
(0.28 AU) & \specialcell{UV} & & \multicolumn{2}{c|}{} & --- & 1--60~s & 1--600~s & 20--30~min & & & & & \\

\hline
LT-CONFIG & \specialcell{VL} & \multirow{2}{*}{4--7.5~$R_\odot$} & \multicolumn{2}{c|}{\multirow{2}{*}{$4\times 4$}} & 4 & 15--30~s & = DIT & 300--450~s & \multirow{2}{*}{20--30~min} & \multirow{2}{*}{YES} & \multirow{2}{*}{ON} &\multirow{2}{*}{51} & \multirow{2}{*}{5.70~Mb}\\
(0.7 AU) & \specialcell{UV} & & \multicolumn{2}{c|}{} & --- & 1--60~s & 1--600~s & 20--30~min & & & & & \\

\hline
FLUCTS & \multirow{2}{*}{VL} & \multirow{2}{*}{1.7--3~$R_\odot$} & \multicolumn{2}{c|}{\multirow{2}{*}{$2\times 2$}} & \multirow{2}{*}{1} & \multirow{2}{*}{1~s} & \multirow{2}{*}{= DIT} & \multirow{2}{*}{1~s} & \multirow{2}{*}{1~s} & \multirow{2}{*}{NO} & \multirow{2}{*}{OFF} & \multirow{2}{*}{13} & \multirow{2}{*}{5.34~Mb}\\
(0.28 AU) & & & \multicolumn{2}{c|}{} & & & & & & & & & \\

\hline
TBF & \multirow{2}{*}{VL} & \multirow{2}{*}{1.7--3~$R_\odot$} & \multicolumn{2}{c|}{\multirow{2}{*}{$2\times 2$}} & \multirow{2}{*}{2} & \multirow{2}{*}{20~s} & \multirow{2}{*}{= DIT} & \multirow{2}{*}{20~s} & \multirow{2}{*}{20~s} & \multirow{2}{*}{NO} & \multirow{2}{*}{OFF} & \multirow{2}{*}{13} & \multirow{2}{*}{5.34~Mb}\\
(0.28 AU) & & & \multicolumn{2}{c|}{} & & & & & & & & & \\

\hline
\multirow{2}{*}{CMEOBS} & \specialcell{VL} & \multirow{2}{*}{1.7--9~$R_\odot$} & \multicolumn{2}{c|}{\multirow{2}{*}{$2\times 2$}} & 4 & 15~s & = DIT & 15--75~s & \multirow{2}{*}{1--5~min} & \multirow{2}{*}{YES/NO\tablefootmark{l}} & \multirow{2}{*}{OFF} & 13 & \multirow{2}{*}{26.67~Mb}\\
 & \specialcell{UV} & & \multicolumn{2}{c|}{} & --- & 1--60~s & 1--600~s & 1--5~min & & & & 3 & \\

\hline
\multirow{2}{*}{COMET} & \specialcell{VL} & \multirow{2}{*}{1.7--9~$R_\odot$} & \multicolumn{2}{c|}{\multirow{2}{*}{$2\times 2$}} & 4 & 15--20~s & = DIT & 75--300~s & \multirow{2}{*}{5--20~min} & \multirow{2}{*}{YES} & \multirow{2}{*}{ON} & 13 & \multirow{2}{*}{26.67~Mb}\\
 & \specialcell{UV} & & \multicolumn{2}{c|}{} & --- & 1--60~s & 1--600~s & 5--20~min & & & & 3 & \\

\hline
\multirow{2}{*}{PROBE} & \specialcell{VL} & \multirow{2}{*}{1.7--9~$R_\odot$} & \multicolumn{2}{c|}{\multirow{2}{*}{$2\times 2$}} & 4 & 15--30~s & = DIT & 75--450~s & \multirow{2}{*}{5--30~min} & \multirow{2}{*}{YES} & \multirow{2}{*}{ON} & \multirow{2}{*}{13} & \multirow{2}{*}{22.69~Mb}\\
 & \specialcell{UV} & & \multicolumn{2}{c|}{} & --- & 1--60~s & 1--600~s & 5--30~min & & & & & \\

\hline
\hline
\end{tabular}
\tablefoot{
\tablefoottext{a}{The  FoV spans over a wide range of heliocentric distances owing to the spacecraft eccentric orbit, although only part of this range is accessible at the same time.}
\tablefoottext{b}{Number of polarisation angles used (cycled) during the scientific VL measurement.}
\tablefoottext{c}{Detector Integration Time, representing the actual exposure time of a single read out of the detector frame.}
\tablefoottext{d}{In Photon Counting mode the typical DIT is 97~ms.}
\tablefoottext{e}{Acquisition time, the overall integration time corresponding to a single acquisition. The integration values are set in order to get a limited number of spurious events (cosmic rays) to be removed.}
\tablefoottext{f}{Values valid in Analogue mode only.}
\tablefoottext{g}{Exposure Time, i.e. the overall time corresponding to a single exposure. Single acquisitions are averaged over the exposure time interval, unless a CME flag occurs. Exposure times are determined considering the count-rate estimated for the Sun at its minimum of activity.}
\tablefoottext{h}{Cadence, i.e., time required to get a full set of 4 VL science images or 1 UV science image.}
\tablefoottext{i}{Software procedure to remove CRs and SEPs spurious signals.}
\tablefoottext{l}{Disabled if the number of acquisitions (N$_\text{ACQ}$) is equal to 1; it is enabled if N$_\text{ACQ}>1$.}
\tablefoottext{m}{CME detection algorithm.}
\tablefoottext{n}{Resulting from the combination of the selected binning and masking options with the reduction performed by the compression algorithm, assumed lossless.}
\tablefoottext{o}{For a full set of 4 VL science images plus 1 UV science image. In the case of the ``FLUCTS'' and ``TBF'' modes, 1 VL science image only.}}
\label{tab:wind}
\end{sidewaystable*}

The main instrument scientific observing modes are defined in order to address the science objectives of the Metis investigation (see Sect.~\ref{sec:sci_objectives}):

\paragraph{Solar wind}  Measurement of the electron density and the solar wind outward expansion velocity, in order to identity fast and slow solar wind streams in the global maps according to the values of the outflow velocity of the \ion{H}{i} component. Global maps of the outflow velocity gradient are suitable to locate where and at which rate energy is deposited in corona. 
\paragraph{Coronal magnetic topology}  Wind outflow velocity measurements and relationship with the coronal magnetic field topology. This mode will obtain high-spatial-resolution maps of the outflow velocity of the \ion{H}{i} component along streamer/coronal hole interfaces, above streamers cusps, and inside streamers.
\paragraph{Global corona study} Global corona configuration/evolution measurements before, during, and after CME events. These measurements provide the geometry of the neutral hydrogen and  electron corona, and its evolution, giving information on the timing, mass content, and overall dynamics of CMEs. They are also crucial to determining the directionality of solar plasma eruptions, in particular during the out-of-ecliptic phase, in order to infer their geo-effectiveness and predict their impact on the near-Earth environment. The evolution related to the CME transit can be followed out to the perihelion of the Parker Solar Probe ($\sim9$~$R_{\odot}$).
\paragraph{Long-term evolution of the coronal configuration} These measurements are used to monitor the evolution of the large-scale corona and, during out-of-ecliptic observing windows, to determine the longitudinal distribution and evolution of the electron density in the solar corona, as well as of the mass and energy flux carried away by the solar wind.
\paragraph{Coronal brightness fluctuations} Brightness fluctuation spectra (best if carried out near perihelion). High-spatial-resolution and high-cadence time series of VL frames taken over a range of heliocentric distances from 1.7~$R_{\odot}$ out to 5~$R_{\odot}$, at fixed polarisation angle ($\text{cadence}=1$~s), or changing between two polarisation angles within the same frame in order to reconstruct the total brightness ($tB$) of the corona at a $\text{cadence}=20$~s, to constrain the amplitude of the density fluctuation spectrum.\\
The study at intermediate frequencies ($\text{cadence} = 5$~min) and lower frequencies ($\text{cadence} = 20$~min) will be carried out with the `Coronal magnetic topology' observing mode, with a long temporal baseline (up to more than one day), in order to detect long-period fluctuations.
\paragraph{Coronal mass ejection observation} CME eruption and propagation, prominence eruption, related driven shocks, and solar energetic particles accelerated by CMEs. Measurement of the electron density, \ion{H}{i}~Lyman-${\alpha}$ (121.6~nm) emission and outward expansion velocity gradient at high spatial resolution and temporal cadence (1--5~min). This mode is activated by a proper CME event flag (Sect.~\ref{ops:sci:cme}). Such measurements can also be used to identify the path of the shock front where particles can be accelerated in the outer solar corona. Moreover, combined with radio observations, they contribute to distinguish flare-accelerated SEPs from those associated with CMEs.
\paragraph{Comet observation} Mapping the emission of sungrazing comets. Visible-light and UV measurements are used to monitor the evolution of the cometary emission along its trajectory close to the Sun.
\paragraph{Parker Solar Probe joint science} Coordinated observations with PSP, in order to characterise the properties of the coronal regions crossed by the PSP spacecraft, during its transit close to the Sun.
\paragraph{BepiColombo, Proba-3, ASO-S, and Aditya-L1 joint science} Coordinate observations with BepiColombo and Proba-3, in order to characterise the properties of plasma streams and ejecta directed toward the magnetosphere of Mercury, and provide continuous coverage of the solar wind and CMEs propagation throughout the coronal region from 1.08~$R_{\odot}$ to 5~$R_{\odot}$. Coordinated coronal observations with ASO-S and Aditya-L1 instruments are also foreseen.

Table~\ref{tab:wind} shows the typical requirements of the observing modes described in this section. The data volume to be downloaded per each cadence interval, T$_\text{CAD}$, reported in the last row of Table~\ref{tab:wind}, is estimated taking into account the masking and binning of the detectors, as well as the compression of the images. All the above settings can be fine-tuned to specific observing conditions, for example by increasing the compression rate by using lossy compression (Sect.~\ref{sec:data_compression}).

\subsubsection{Coronal mass ejection flag}
\label{ops:sci:cme}

In addition to the nominal science operations described above, Metis will be able to identify automatically the occurrence of a CME and to switch to the CME observing mode (Sect.~\ref{ops:sci:modes}). To this end, the production of a CME flag by the on-board software will be required \citep{bemporad2014}. The easiest way to generate a CME flag, based on the observed evolution of coronal intensities, is to perform running differences of the acquired images and to identify the occurrence of a limb CME as a sudden increase of the observed brightness in off-limb coronal images. 

The CME flag algorithm uses only the images acquired in the VL channel. This is because the appearance of CMEs in VL images is well known from similar instruments that flew on previous space missions, while no UV coronagraphic images of CMEs have been acquired so far (UVCS/SOHO observations were obtained in limited instantaneous FoVs). In order to identify the occurrence of a CME, the algorithm extracts the total intensity measured over eight angular sectors, $\theta_i$ ($i=1,2,\dots,8$), around the image centre, with an angular size of $45\degr$. The eight intensity values $I_n(\theta_i)$  extracted at any given time from frame $n$ are then compared with the intensities acquired at the previous time from frame $n-1$, in order to compute the relative intensity variations $\Delta I_{n,i} = [I_n(\theta_i)-I_{n-1}(\theta_i)] / I_n(\theta_i)$.

The values $\Delta I_{n,i}$ thus obtained are then compared onboard with two threshold values $T$, defined for non-halo CMEs and  halo CMEs, respectively.  If in at least
one angular sector $\Delta I_{n,i} > T_\text{limb}$ then the major CME flag is raised; while if $\Delta I_{n,i} > T_\text{halo}$ in at least four angular sectors then the halo CME flag is raised. The final values of these thresholds will be optimised during the mission, based on the images obtained with the instrument. Metis will then use the inter instrumental communication (IIC) service to distribute the information, provided by its internal automatic CME-detection algorithm, to the other instruments through a dedicated packet at a cadence of 1~s. Metis will share with the other instruments its actual operational status and will provide, on a coarse scale, the mean value of the VL brightness measured in each of the eight sectors. In particular, the algorithm will be able to provide the time of the initial detection of the
CME, the latitudinal direction of propagation on the plane of the sky (within $45\degr$), and a binary flag indicating whether a halo CME was detected. 
Alternatively, Metis will also be able to exploit the IIC service to monitor flags raised by other instruments, as EUI \citep{Rochus2019} and STIX \citep{krucker2019}, to react to on-disk or near-disk events likely associated to CMEs.  If the capability of reacting to such external monitor flags is enabled, Metis will react as if a CME had been detected by its own internal algorithm, possibly after a time interval set to a suitable value to account for the CME transit time up to the inner edge of the FoV of the  instrument.

\subsection{Commissioning concept and in-flight characterisation}
\label{ops:comm}

The near-Earth commissioning phase (NECP) is the only mission phase when spacecraft and instrument commanding are foreseen to occur in near real-time to support the critical post-launch initial activation and checkout activities. Its duration is expected to be  about 90~days after launch to take advantage of the relatively short distance to Earth in the earliest phase of the mission.

Metis commissioning activities will therefore take place entirely during the NECP. Some of those activities will need to be performed after the release of the Metis sealing cap (described in Sect.~\ref{sec:cap}). The main constraint for the ejection of the Metis cap is the need to wait for a sufficiently low contamination level, expected to be attained not earlier than 20 days after launch. After the cap release and before the subsequent closure of the heat shield door, Metis will be potentially exposed to direct illumination by the Sun (see also Sect.~\ref{ops:offpoint}). This implies the need to eject the sealing cap while the spacecraft is either pointing at the centre of the solar disk or off-pointing by a safe angle greater than 6\degr. Finally, these constraints will be compounded with flight dynamics constraints, such as debris management or thermal/illumination issues. After the cap ejection, the heat shield door of Metis will be closed and a phase of faster outgassing of the Metis cavity lasting at least 5~days will follow. After these 25 days, the need to carry out commissioning activities lasting about 10~days places an additional constraint on the timing of the cap release. 

The commissioning activities will include a version of the full functional test of the instrument and a set of calibration frames obtained with both the VL and the UV channel. Moreover, the instrument boresight will be verified by inspecting the stellar field within the instrument FoV, while the optimal position of the IO is determined by means of a characterisation of the stray light pattern, as described in Sect.~\ref{ops:cal}. Commissioning activities prior the sealing cap release will also provide the opportunity to acquire the only true, in-flight dark frames for the VL camera, as described in Sect.~\ref{ops:cal}. The gap between the heat shield door and the Metis entrance hole may let some scattered light into the instrument, thus preventing the acquisition of proper VL detector dark frames for the entire duration of the mission after the cap release.

Even after the end of the NECP, opportunities will arise during the cruise phase, that is, before the start of the nominal phase, to carry out some performance verification and in-flight characterisation of the instrument. Ideally, a set of the calibration activities at a minimum of two spacecraft--Sun distances should be executed, in order to characterise the instrument response in different thermal/illumination conditions. Already during the cruise phase, the transit within the Metis FoV of radiometric and/or polarimetric stellar standards will also be exploited for calibration purposes.

All these post-NECP activities will however need to be carried out offline, following the planning scheme to be adopted for the rest of the mission.

\subsection{In-flight calibration}
\label{ops:cal}

The nominal Metis operations include engineering activities needed to verify the status and the performance of the instrument as well as to provide calibration data.

A first set of calibration activities concerns the characterisation of the detectors. Since the Metis external door is slightly detached from the heat shield (Sect.~\ref{sec:optical_unit}), some scattered light can still illuminate the interior of the instrument even when it is closed. For this reason, no true visible light detector (VLD) dark frame can be taken during the nominal mission.  However, the VLD performance can be checked by obtaining a set of  frames taken at the shortest allowed integration time ($\sim1$~s). On the other hand, ultraviolet light detector (UVD) dark frames can be obtained with the heat shield door closed, by acquiring frames with no high voltage applied to the intensifier.  Finally, a check on the offset map required to process the UVD frames taken in photon counting mode is also done with the heat shield door closed, and with the high-voltage power supply off. The resulting map, stored in a dedicated memory board for onboard processing, is then transmitted to ground for further checks. This set of detector calibrations are foreseen at least once per orbit, or more frequently (e.g. at different distances from the Sun) if thermal effects turn out to be a significant factor in determining the detector performances.

A second set of calibration activities concerns the characterisation and the minimisation of the instrumental stray light. First of all,  the instrument boresight relative to the nominal spacecraft pointing can be checked by examining the stellar field within the instrument FoV. Once the instrument boresight has been determined, the pattern of stray light is examined and, optionally, the position of the IO is optimised. Further characterisation of the stray-light pattern is done by requiring a given set of spacecraft slews and/or rolls. The former observations are needed to characterise the overall stray-light pattern, while the latter ones allow the measurement of  the asymmetric component of the stray light. Since these activities require spacecraft manoeuvres, they need to be planned at S/C level jointly with other instruments.

Finally, the instrument radiometric (VL and UV) and polarimetric (VL) calibration can be obtained by observing a set of calibration stars transiting within the Metis FoV. The set of potential polarimetric standards is given by \cite{capobianco2014}; these authors in turn draw the list from \cite{axon1976}. The list of UV radiometric standards suitable for Metis is taken from either \cite{Landsman-Simon:93} or \cite{ISSI-13}. During each orbit, there are typically several opportunities for such radiometric calibration activities. These transits might occur outside the nominal remote-sensing windows. 

\subsection{Off-pointing issues}
\label{ops:offpoint}

The Metis coronagraph operates under the assumption that its boresight is centred on the solar disk. A misalignment of the instrument boresight with respect to the centre of the solar disk may produce strong and highly asymmetric stray-light patterns, which can severely hamper  scientific measurements. Prolonged exposure of parts of the instrument to direct sunlight can also result in significant safety risks to the instrument itself.

Since Metis is unable to compensate for Solar Orbiter pointing off the disk centre, its nominal and safe operations depend also on the spacecraft operations. In particular, some coordinated Solar Orbiter science plans may require the spacecraft to re-point to off-centre solar regions or even near the solar limb. In addition, target regions may be followed in their motion across the solar disk \citep{sanchez2019}. Finally, some spacecraft management manoeuvres, involving for instance movements of the high-gain antenna or of the solar panels, can produce spacecraft attitude perturbations that can potentially expose Metis to direct sunlight.

More specifically, the maximum spacecraft offset angle allowing Metis science observations, $\alpha_\text{max}$, is determined by the maximum angle beyond which light from the solar disk hits the inner walls of the entrance cone of the  IEO (Fig.~\ref{fig:metis_raytrace}). Taking the appropriate margins for the nominal uncertainties in spacecraft pointing, we obtain: $\alpha_\text{max}(d) = 57.9\arcmin-15.9\arcmin/d$, where $d$ is the Solar Orbiter distance from the Sun in~astronomical units.

A second angle, more relevant to the safety of the instrument, is the maximum angle, $\beta_\text{max}$, beyond which the solar disk begins to illuminate the detectors. This angle is determined again from purely geometrical consideration as $\beta_\text{max}(d) = 61.1\arcmin-15.9\arcmin/d$.

Direct exposure to sunlight by itself does not normally pose a significant hazard to the health of the VL detector and of the UV detector,  provided that the latter   is off. However, prolonged exposure to direct solar disk light of parts of the instrument can raise the temperature of some components, in particular of the interference filter assembly and of the polarisation module package, beyond critical values.
The conditions of safe thermal operations of the instrument can be characterised by a maximum angle, $\beta_\text{therm}(d)$, and a characteristic timescale, $t_\text{therm}$. More specifically, $\beta_\text{therm}$ is the offset angle beyond which, after a time $t_\text{therm}$, the temperatures in various Metis subsystems rise to levels which might lead to irreversible damage. Conversely, below this limit, Metis temperatures remain within safe ranges indefinitely.

A detailed thermal analysis of the response of the Metis
instrument to a few off-pointing cases (up to 3.5\degr\  from nominal
Sun centre pointing) was carried out, considering both steady-state illumination conditions and transient off-pointing events. The results of this analysis show that $\beta_\text{therm}$ can be set to values up to 1\degr,  with $t_\text{therm}$ being set virtually to infinity, that is,\ the temperature of all the Metis subsystems remains indefinitely within design limits. Above the value of 1\degr, the value of $t_\text{therm}$ is not well constrained by the thermal analysis: the value of 45 minutes for the reference off-pointing angle of 3.5\degr\ considered in the thermal analysis should be regarded as an upper limit if intermediate angles are to be considered. A conservative, safer value would be $t_\text{therm}=0$.

\begin{figure}
\centering
\includegraphics[angle=90,width=1.2\columnwidth]{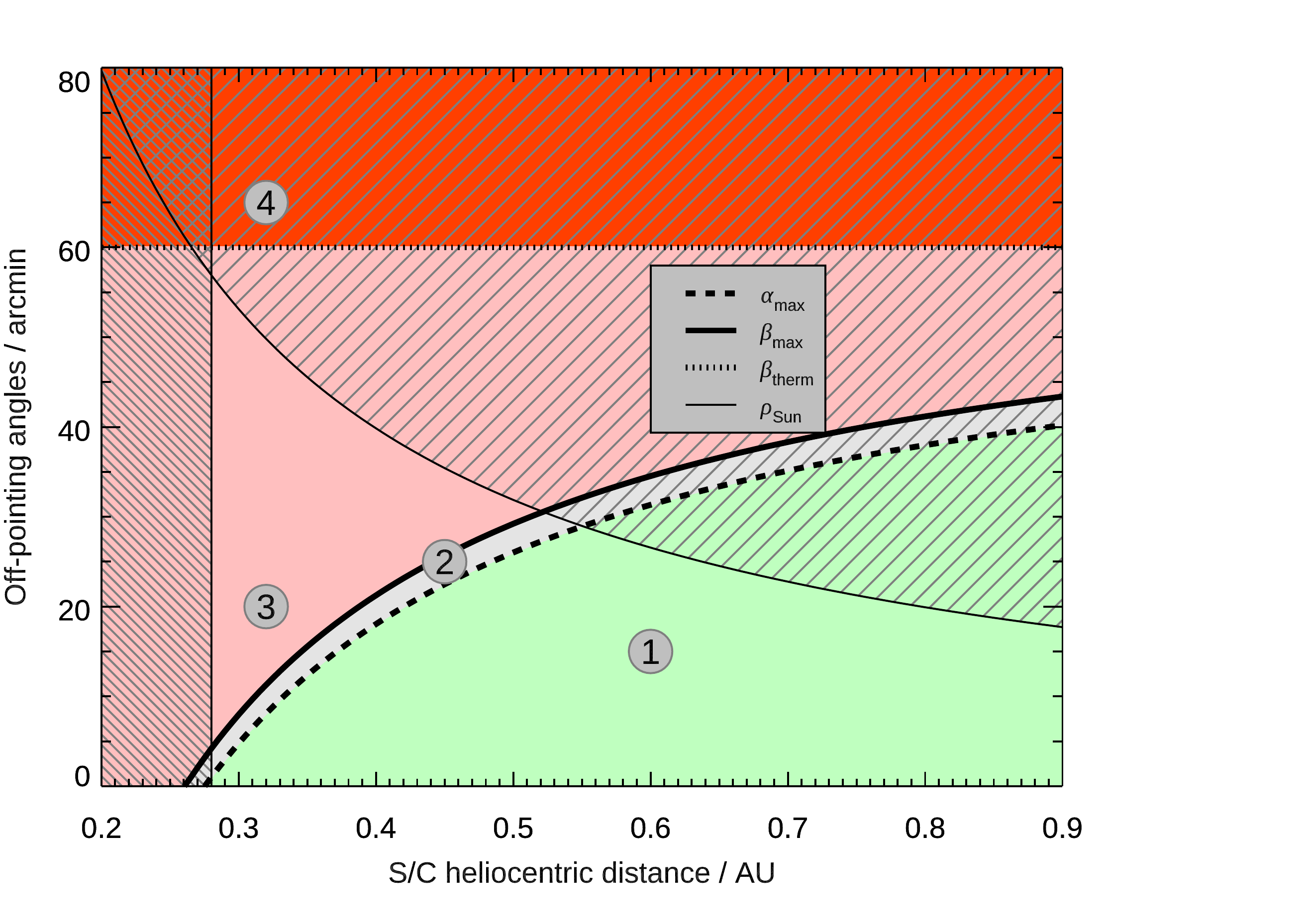}
\caption{Spacecraft off-pointing angles ($\delta$) most relevant for the Metis operations and safety. The Metis heat shield door operations are related to the four areas of the plot: (1) Metis science operations allowed, offset < $\alpha_\text{max}$; (2) Metis external occulter cone is illuminated, $\alpha_\text{max}$ < offset < $\beta_\text{max}$, no harm to instrument, science data to be evaluated on ground; (3) Metis can stand indefinite off-pointing, $\beta_\text{max}$ < offset < $\beta_\text{therm}$; (4) Metis is put on Safe mode and door closed, offset > $\beta_\text{therm}$. The descending solid curve represents $\rho_\text{Sun}$, the angular size of the solar radius $R_\odot$.}
\label{fig:angles}
\end{figure}

In case of excessive spacecraft off-pointing, the safety of the instrument is ensured by the closure of the Metis heat shield door. The closure of the door is managed by the spacecraft and is normally planned in advance by checking whether certain conditions in the planned spacecraft and instrument operations are met (see below). However, the closure of the door can also be triggered by the detection of anomalous spacecraft attitude perturbations. In particular, the Metis on-board software includes a module, the Sun-disk monitor (SDM), which continuously monitors images taken in the VL channel to detect sudden and anomalous changes in the stray-light pattern. If the stray-light asymmetry exceeds a given threshold, Metis issues an emergency flag that causes the spacecraft to command the closure of the heat shield door and the transition of Metis to a safe state. In addition, some temperature sensors within the instrument are also monitored.

In determining the conditions under which the heat shield door must be closed, the relevant parameters are $\beta_\text{therm}(d)$ and $t_\text{therm}$ described above, along with the number of cycles of door opening and closing and the time required for the door to close (currently estimated to be of the order of 10~minutes).

With reference to the regions labelled in Fig.~\ref{fig:angles}, and under the assumption that the number of spacecraft manoeuvres or science operations requiring closure of the Metis door do not exceed -- over the expected mission duration -- the maximum functional number of operations of the door mechanism, the concept for door operations as a function of the spacecraft off-pointing angle, $\delta$, is the following:
\begin{enumerate}
\item $\delta < \beta_\text{max}(d)$ (nominal Sun-centred pointing; regions \#1 and \#2 in Fig.~\ref{fig:angles}): Metis science operations are allowed. In the case of unexpected off-pointing, a flag issued by the SDM internal algorithm causes a request to the spacecraft to close the heat shield door and to put Metis in a safe state. Possible degradation of science data in the case $\alpha_\text{max}(d) < \delta < \beta_\text{max}(d)$ (region \#2) will be evaluated on ground.
\item $\beta_\text{max}(d) < \delta < \beta_\text{therm}(d)$ (region \#3 in Fig.~\ref{fig:angles}): Metis can stay with at least the VL detector on to keep the SDM monitoring module active; no science telemetry is produced. A flag issued by the SDM internal algorithm or by anomalous internal temperature sensor readings causes a request to the spacecraft to close the heat shield door and to put Metis in a safe state.
\item $\delta > \beta_\text{therm}(d)$ (region \#4 in Fig.~\ref{fig:angles}): Any spacecraft manoeuvre entering this region will have the effect of placing Metis into a safe state and to close the heat shield door.
\end{enumerate}

We note that the hatched area in Fig.~\ref{fig:angles} corresponds to the region $\delta >\rho_\text{Sun}$, 
where no nominal spacecraft off-pointings are allowed. 

\subsection{Data products and processing levels}
\label{ops:obs_seqs}

As described in Sect.~\ref{sec:MPPU}, Metis can acquire data from two channels, independently and simultaneously.  Depending on the acquisition scheme adopted and its detailed configuration, Metis can generate (at processing levels 0 to 2 described in Sect.~\ref{ops:sw}) the following different scientific and calibration or ancillary data products, or scientific data objects.

Primary scientific data objects:

\begin{itemize}
\item VL images,
\item UV images,
\item PCU events lists,
\item PCU events accumulation matrices and vectors,
\item PCU test events lists.
\end{itemize}
Secondary scientific data objects:
\begin{itemize}
\item VL temporal standard-deviation matrices,
\item UV temporal standard-deviation matrices,
\item VL cosmic-ray log matrices,
\item UV cosmic-ray log matrices,
\item light curves.
\end{itemize}
Calibration and ancillary data objects:
\begin{itemize}
\item UV PCU offset maps.
\end{itemize}
Metis scientific data objects are then organised into different levels according to their degree of processing, as described in the following section.

\subsection{Planning and data analysis software}

Metis science is planned in coordination with the other instruments of the Solar Orbiter according to the Science Activity Plan \citep{Zouganelis2019}. The Metis contribution to the Solar Orbiter scientific objectives can only be addressed by careful planning of the observations. In order to achieve this goal, the Metis team will develop a planning software package. This software is, in particular, capable of combining data from the instrument archive (ancillary, calibration, and housekeeping data), the low-latency data pipeline (see Sect.~\ref{ops:sw:lldata}) at the Science Operation Centre \citep[SOC; see][]{sanchez2019}, the instrument database, the instrument model, the telemetry corridors and data from other instrument low-latency pipelines, as well as from other missions and earth-based instruments. The software is going to be used by the instrument team to monitor the status of the instrument, plan future observations, prepare the flight control procedures, and check that the expected quantity of telemetry follows the telemetry corridors defined by the SOC for each instrument team.

\subsubsection{Data processing and analysis software}
\label{ops:sw}

Regarding the data processing and analysis software, the in-flight telemetry from the Metis instrument is received by the SOC through the Mission Operations Centre, MOC, and propagated to the instrument team using the generic file transfer system (GFTS) developed by the European Space Astronomy Centre, ESAC. The raw telemetry is archived together with the ancillary data from the instrument integration and calibration phases.

As soon as science telemetry is received through the GFTS, the science data pipeline produces the science data products at different levels of processing. In the first step, the pipeline combines all telemetry packets of each data product, decompresses the data if necessary, and writes the level~0 uncalibrated data products as flexible image transport system (FITS) files. As a second step, the pipeline decodes the data contained in the scientific header and adds all relative data from the archive of the housekeeping data, necessary for the full calibration of the data products. The result is saved in level~1 FITS files that are distributed, if desired, to the scientific community. The final step of the pipeline, calibrates the data products to physical units, converts onboard time to coordinated universal time, UTC, and the instrument coordinate system to the world coordinate system (WCS). These fully calibrated data are distributed as level-2 data products to the scientific community. 

The scientific data pipeline is written in the IDL language and distributed to the community inside the SolarSoft package. If necessary, interested scientist will be able to use this package to calibrate level-1 data products to level-2.

At the time of the writing of this paper, the organisation of level-3 data has not yet been consolidated. It is however foreseen that level-3 data will include additional science products such as $pB$, radial velocity, and density maps derived with the diagnostics techniques described in Sect.~\ref{ops:sci:analysis}.

\subsubsection{Low-latency data pipeline}
\label{ops:sw:lldata}

The Solar Orbiter mission profile shows strong variations in spacecraft-Earth distance causing significant changes in the data downlink capability of the platform, with latencies in data retrieval that can exceed, in some cases, 100~days \citep{sanchez2019}. The payload telemetry is organised by different downlink priorities: the housekeeping data are the first data to be received on ground, while the science data are received last.

\begin{figure}
\centering
\includegraphics[width=\columnwidth]{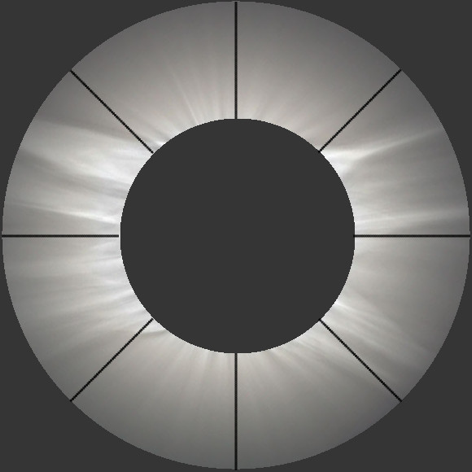}
\caption{Metis FoV with the eight sectors used for the computation of the light curves highlighted.}
\label{fig:lightcurves}
\end{figure}

In order to minimise the effect of the data latency on instrument operation planning and instrument performance checks, an extra priority is introduced to make sure that a subset of recently acquired payload data will be downlinked at each ground contact. This dataset is called low-latency data, and in the case of Metis consists of a set of two VL and one UV images highly compressed in order to comply with the limit of 1 MB/day allowed per instrument.  Furthermore, the VL channel can also provide the light curves, that is, eight temporal series of mean brightness computed averaging the counts of the pixels detected in eight radial sectors of the Metis FoV (Fig.~\ref{fig:lightcurves}). Whenever the production of the light curves is enabled, these data products are downloaded as low-latency data as well. Their main function is to give an overview of the acquired data since the last ground contact, to ensure daily visibility of ongoing operations and to allow changes to future planning if necessary. Metis plans to use this mechanism mainly to monitor the instrument performance and science data quality.
Each instrument has to have a dedicated pipeline installed directly at the ESA SOC, in order to promptly analyse these data and, potentially, to react if needed. The different teams supporting Solar Orbiter adopted a common framework to develop this `quick look' data pipeline in order to simplify the management of ten different processing pipelines at SOC. Each pipeline is developed as a virtual appliance, that is, a virtual-machine image loadable into a system virtualisation software. 

We note that although the low-latency data pipeline performs  steps similar to those of the science data pipeline described in Sect.~\ref{ops:sw}, it is an entirely separate branch of data processing. The data products are immediately available to the Solar Orbiter SOC and to the instrument teams for planning purposes.
 These data could be reprocessed at a later time to incorporate improvements in the calibration and to achieve a higher-level data product.

\section{Diagnostic techniques}
\label{ops:sci:analysis}

The measurements described in the previous sections will allow the investigation, by means of different diagnostic techniques, of a variety of physical parameters and processes occurring in the outer layer of the solar atmosphere.

Metis is primarily designed to study the continuous flow of the expanding corona that generates the solar wind. The outflow speed of the hydrogen component of the solar corona can be inferred by comparing the  Metis UV observations with the \ion{H}{i} line emission expected for a static model corona. In order to synthesize the radiance of the \ion{H}{i} Lyman-$\alpha$ line emitted in the extended corona, among the physical and dynamic parameters describing the line profile formation along the LoS, we account for the electron LoS density $n_{e}$ as a function of the heliocentric distance and heliographic latitude, which can be obtained by analysing the polarised VL data.

\subsection{Electron density diagnostics}

The electron density of the global corona will be derived from the polarised brightness $pB$ data acquired by the Metis VL channel by using the inversion technique developed by \cite{vandeHulst:50}. As a matter of fact, $pB$ is directly related to the coronal electron density, since it depends upon the Thomson scattering of the photospheric white-light radiation by coronal electrons. 

In the spherical symmetry approximation, which is assuming that the coronal electron density is a function of the heliocentric distance only, i.e. $n_{e}=n_{e}(r)$,
\begin{equation}
pB(\rho)=2\frac{B_{\odot}}{1-\frac{u}{3}}\frac{3\sigma_T}{16\pi}\int_{\rho}^{\infty}n_{e}(r)[(1-u)A(r)+uB(r)]\frac{\rho^{2}}{r^{2}}\frac{r\,\mathrm{d}r}{\sqrt{r^{2}-\rho^{2}}},
\label{eq:pb}
\end{equation}
where $\rho$ is the perpendicular distance between the LoS and Sun centre (i.e. the apparent height above the solar limb), $r$ is the radial distance from Sun centre, $B_{\odot}$ is the mean solar brightness, $u=0.63$ is the limb darkening coefficient, $\sigma_T$ is the Thomson scattering cross section for electrons, and $A(r)$ and $B(r)$ are geometric factors, functions of the radial distance only \citep{Minnaert:30}:
\begin{equation}
A(r)=\cos\gamma\sin^{2}\gamma,
\label{eq:geometric_factor_a}
\end{equation}
\begin{equation}
B(r)=-\frac{1}{8}\left[1-3\sin^{2}\gamma\left(\frac{1+3\sin^{2}\gamma}{\sin\gamma}\right)\ln\left(\frac{1+\sin\gamma}{\cos\gamma}\right)\right],
\label{eq:geometric_factor_b}
\end{equation}
where $\sin\gamma=R_{\odot}/r$.

The technique developed by \cite{vandeHulst:50} to obtain the electron density as a function of the heliocentric distance, $n_{e}(r)$, relies on the inversion of the Eq.~(\ref{eq:pb}) describing the relation between the observed Thomson-scattered polarised brightness and the electron density, by expressing the observed values of $pB$ as a polynomial,
\begin{equation}
pB(\rho)=\sum_{i}c_{i}\left(\frac{\rho}{R_{\odot}}\right)^{-d_{i}},
\label{eq:pb_polynomial}
\end{equation}
where the coefficients $c_{i}$ and $d_{i}$ are selected through a $\chi^{2}$ minimisation technique to reproduce the measured quantity to within $\pm5\%$.

It results that the electron density is given by
\begin{equation}
n_{e}(r)=\frac{\sum_{i}a_{i}\left(\frac{r}{R_{\odot}}\right)^{-b_{i}}}{[(1-u)A(r)+uB(r)]},
\label{eq:ne_polynomial}
\end{equation}
where
\begin{equation}
a_{i}=\frac{1}{\sqrt{\pi}KR_{\odot}}\frac{\Gamma\left(\frac{d_{i}+3}{2}\right)}{\Gamma\left(\frac{d_{i}+2}{2}\right)}c_{i}\qquad\textrm{and}
\label{eq:ne_coefficient_a}
\end{equation}
\begin{equation}
b_{i}=d_{i}+1,
\label{eq:ne_coefficient_b}
\end{equation}
where 
\begin{equation}
K=\frac{B_{\odot}}{1-\frac{u}{3}}\frac{3\sigma_{T}}{16\pi}
\end{equation}
and $\Gamma$ is the gamma function.

\subsection{Doppler dimming diagnostics}

The simultaneous detection of the solar corona emission in polarised VL and the \ion{H}{i} Lyman-$\alpha$ line allows the use of the Doppler dimming  technique \citep{noci1987} to measure the radial outflow velocities of neutral hydrogen atoms in the different coronal structures. This diagnostic technique is essentially based on the comparison of the \ion{H}{i} Lyman-$\alpha$ coronal emission, synthesised on the basis of the electron density measurements (static corona),  with the emission observed with Metis.

\begin{figure}
\centering
\includegraphics[width=0.75\columnwidth]{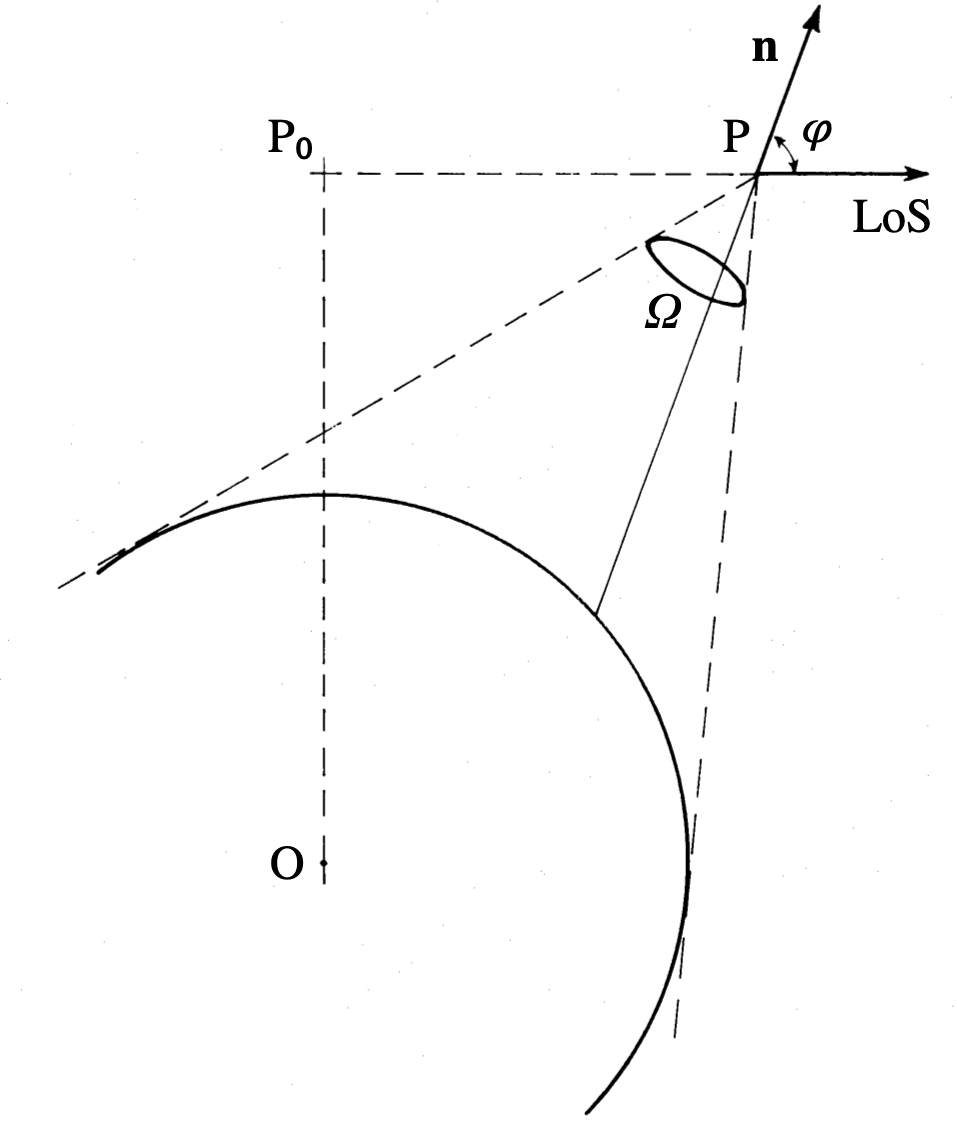}
\caption{Geometry of the scattering process \citep[adapted from][]{noci1987}.}
\label{fig:doppler_dimming}
\end{figure}

Since the \ion{H}{i} 121.6 nm line is dominated by the resonant scattering of the disk radiation, the emission is obtained by computing the radiative component of the line intensity, according to the relation:
\begin{equation}
I_{r}=\frac{1}{4\pi}bh\lambda_{0}B_{12}\int_{LoS}\int_{\Omega}p(\varphi)\,\mathrm{d}\omega\,\Phi(\delta\lambda)\,n_{i}\,\mathrm{d}l,
\label{eq:radiative_component}
\end{equation}
where, the factor $b$ is the branching ratio for radiative de-excitation (equal to one in the case of  \ion{H}{i} Lyman-$\alpha)$, $h$ the Planck constant, 
$\lambda_{0}$ the rest wavelength, $B_{12}$ the Einstein coefficient for absorption, $n_{i}$ the number density of the hydrogen atoms, and $l$ the space coordinate along the LoS \citep[][]{allen1973}.
The function $p(\varphi)$ takes into account the geometry of the scattering process \citep[see][and Fig.~\ref{fig:doppler_dimming}]{noci1987}, $\varphi$ is the angle between the direction of the incident radiation, $\textbf{n}$, and the LoS, $\mathrm{d}\omega$ the infinitesimal solid angle around $\textbf{n}$, and $\Omega$ the solid angle subtended by the solar disk at the point of scattering. 

The factor $\Phi(\delta\lambda)$ represents the Doppler dimming introduced by the presence of coronal outflows \citep{withbroe1982a,noci1987}, and depends on the integral of the product of the intensity of the exciting spectrum, $I_{ex}(\lambda)$, and the normalised coronal absorption profile, $\Psi(\lambda)$, along the direction of the incident radiation  $\textbf{n}$, expressed as a function of wavelength $\lambda$,
\begin{equation}
\Phi(\delta\lambda)=\int_{0}^{+\infty}I_{ex}(\lambda-\delta\lambda)\Psi(\lambda-\lambda_{0})\mathrm{d}\lambda,
\label{eq:doppler_dimming}
\end{equation}
 where $\delta\lambda$ is the shift of the disk spectrum introduced by the outflow velocity $\textbf{w}$ of the coronal absorbing atoms along $\textbf{n}$.
\begin{equation}
\delta\lambda=\frac{\lambda_{0}}{c}\textbf{w}\cdot\textbf{n}.
\label{eq:delta_lambda}
\end{equation}

The coronal absorption profile $\Psi(\lambda-\lambda_{0})$  is assumed to be a Gaussian function with standard deviation $\sigma_{\lambda,n}$. This quantity is related to the kinetic temperature of the coronal hydrogen atoms $T_{k,n}$ along the direction of the incident radiation,
\begin{equation}
\sigma_{\lambda,n}=\frac{\lambda_{0}}{c}\sqrt{\frac{k_{B}T_{k,n}}{m_{p}A}},
\label{eq:standard_deviation}
\end{equation}
where $k_{B}$ is the Boltzmann constant, $m_{p}$ the proton mass, and $A=1$ the atomic mass number.

The coronal number density of ionized hydrogen can be inferred by measuring the electron density from the VL polarised emission detected with Metis. Under the hypothesis of ionisation equilibrium, the ratio between neutral and ionized hydrogen atoms depends on the electron temperature, for example, see \citet{vial2016}. The $T_e$ radial profiles can be selected from the literature. For example, \citet{gibson1999} and \citet{vasquez2003} provide such profiles for an equatorial streamer and polar regions, respectively. These profiles appear to better reproduce the electron temperatures inferred by several observations below 2~$R_{\odot}$, and their reliability at higher altitudes is supported by several studies, as already discussed by \citet{dolei2018}. The kinetic temperature of the coronal hydrogen atoms $T_{k,n}$ can be provided by the extended database that was constructed by \citet{dolei2016} from the analysis of a very large set of UVCS spectrometric data, complemented by the results of the temperature of protons obtained by the in situ instruments.

\begin{figure}
\centering
\includegraphics[width=\columnwidth]{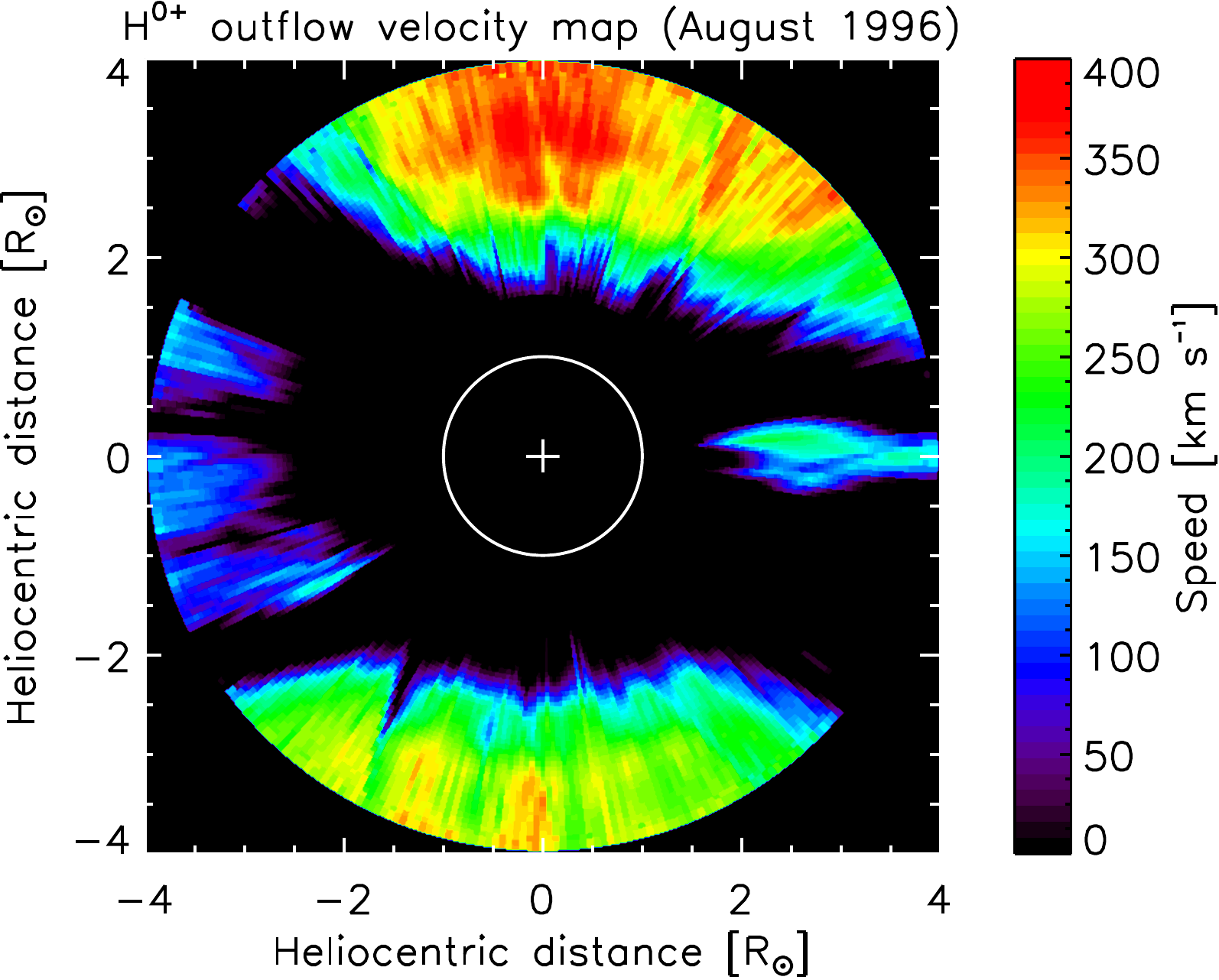}
\caption{Two-dimensional map of the neutral hydrogen outflow velocity in the solar corona out to 4~$R_{\odot}$, inferred from the data obtained with UVCS/SOHO on June 14, 1997 \citep {dolei2018}. The continuous line indicates the solar disk.}
\label{fig:outflow_velocity_map}
\end{figure}

As for the exciting chromospheric Lyman-$\alpha$ line radiation, Metis investigation will be supported by the images of the \ion{He}{ii} line emission at 30.4~nm of the entire solar disk that will be provided almost simultaneously by the Full Sun Imager (FSI) of the  EUI aboard Solar Orbiter \citep{Rochus2019}, after correcting for the contribution of the nearby \ion{Si}{xi} line. These images can be used to reconstruct maps of the chromospheric radiation of 
\ion{H}{i} Lyman-$\alpha$ line over the solar disk through the correlation of the intensity between the two spectral lines, as described by \citet{auchere2005}.
The capability of mapping the coronal outflow velocity using synergistic white-light and UV observations (see Fig.~\ref{fig:outflow_velocity_map}) was recently confirmed by \citet{dolei2018}.

\subsection{Three-dimensional reconstruction of coronal mass ejections}
The close-up observations of the solar corona from different perspectives will represent a step forward in understating the 3D structure of CMEs. In this context, the polarisation-ratio technique \citep{Moran-Davila:04} can be used to perform a 3D reconstruction of the CME and to derive information on the direction and speed of propagation of the ejected plasma. This approach was successively tested by \citet{pagano2015} on CME emission synthesised to simulate the observations obtainable with Metis. Moreover, the combination of polarised VL and UV images can provide complete information on the 2D distributions of the CME plasma density, using the polarisation-ratio technique and temperature, and by comparing expected and measured \ion{H}{i}~Lyman-$\alpha$ intensities as shown by \citet{Susino-Bemporad:16}. In addition, the magnitude of the coronal magnetic fields involved in the CME-driven shock can be derived by applying the Rankine-Hugoniot equations \citep{bemporad2010}.

\subsection{Plasma fluctuation diagnostics}

\begin{figure}
\centering
\includegraphics[width=\columnwidth]{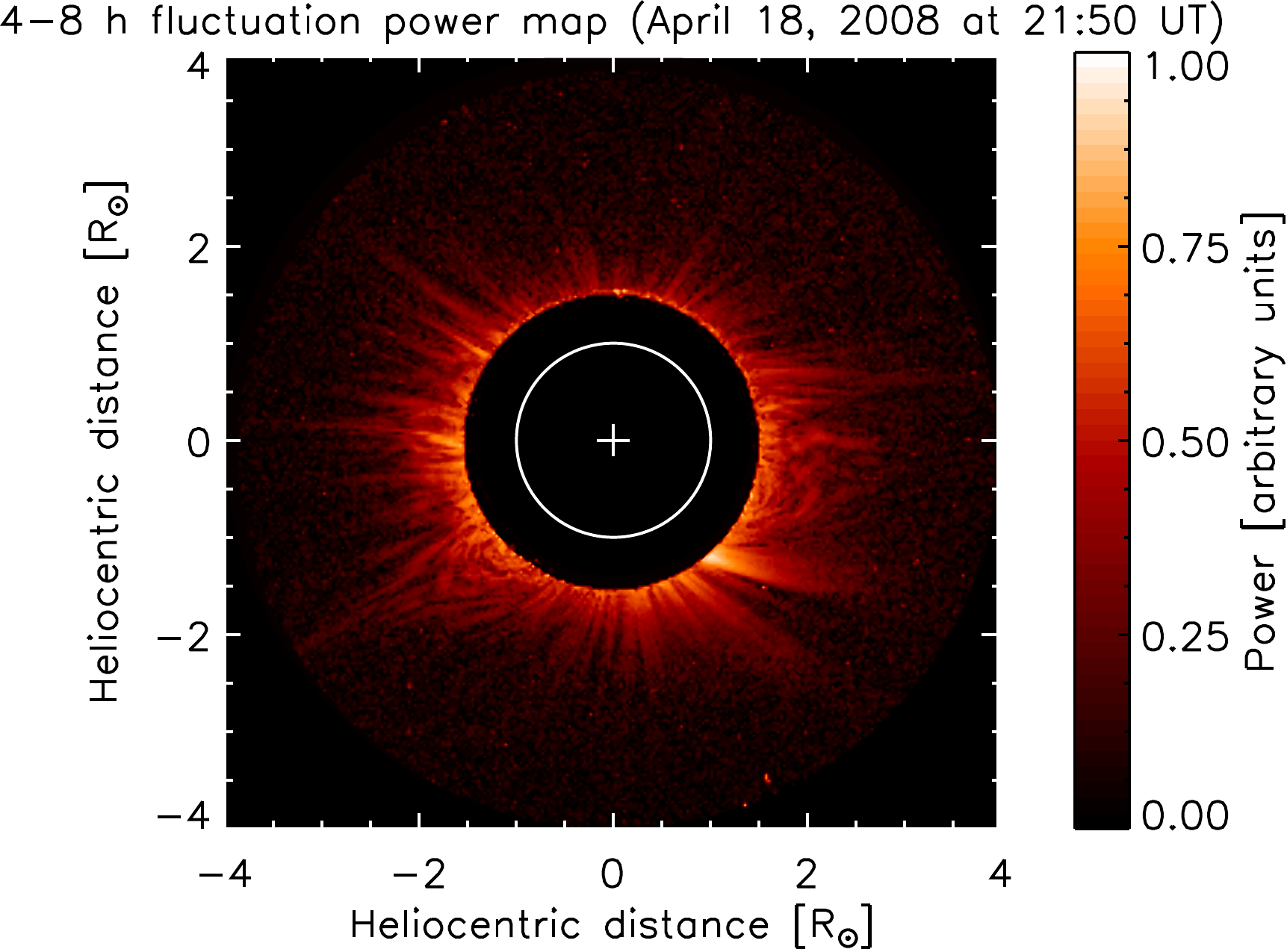}
\caption{Two-dimensional map of the total brightness fluctuation power averaged over the 4--8 hr timescale band on April 18, 2008, at 21:50~UT, inferred from the STEREO A COR-1 observations of the solar corona \citep[][]{telloni2013}. The continuous line indicates the solar disk.\label{fig:fluctuation_power_map}}
\end{figure}

The high temporal and spatial resolution capabilities of Metis allow the investigation of plasma fluctuations in a still unexplored frequency domain and, in addition, the reconstruction of oscillatory patterns on a given timescale in the solar corona (as displayed in Fig.~\ref{fig:fluctuation_power_map}), by applying the Hovm\"oller diagram approach as  shown by \citet{telloni2013}. In essence, by computing the fluctuation power on a certain timescale at multiple locations in the Metis images (pixel by pixel), it is possible to assess the spatial and temporal variability of a field of data, in order to investigate which structures encompass the larger fluctuations. Coronal structures characterised by strong fluctuations can be locations of intense wave--particle interactions. In particular, during periods of lower rotation of the Solar Orbiter spacecraft relative to the solar surface, at minimum perihelion, which has the effect of reducing the frequency contribution of solar rotation, it is possible to map, for the first time, the 2D distributions of both density and velocity fluctuations, thus providing the possibility to discriminate between propagating and local instabilities. In order to map coronal fluctuations, spectral techniques based on the classical Fourier analysis, or more refined methods of analysis such as wavelet decomposition or non-linear analysis techniques based on the structure functions, can be applied to the simultaneous observations in the VL and in the \ion{H}{i}~Lyman-$\alpha$ line obtained with Metis.

\subsection{Turbulence diagnostics}

A series of analysis techniques can provide information on the nature of turbulence in the solar corona, thus definitely assessing the role of coronal fluctuations in the energy deposition processes responsible for the heating and acceleration of the solar wind. The principal techniques applicable to the data are the following: the intermittent analysis, based on a complete set of diagnostic tools \citep[see for instance][]{bruno2014} including the flatness factor on different timescales in order to estimate intermittency \citep{frisch1995}; the Kolmogorov-Smirnov test in order to estimate the degree of intermittency \citep{bi1989}; and the local intermittence measure (LIM) technique to be used to isolate intermittent events \citep{farge1992} at coronal distances where the plasma is not yet reprocessed by dynamical stream--stream interactions and therefore still carries the pristine imprints of the source region.

The study of intermittency in the solar corona is an unexplored field. Intermittency is strictly related to the global scale invariance or self-similarity of the coronal (either density or velocity) fluctuations. The lack of self-similarity affects the shape of the probability density functions (PDFs). Going from larger to smaller scales, PDFs increasingly deviate from a Gaussian distribution showing thicker and thicker tails. Hence, the largest events populating the PDF tails at the smallest scales have a larger probability of occurring with respect to the Gaussian statistics. The increasing divergence from Gaussianity, as smaller and smaller scales are involved, is the typical signature of intermittency \citep{frisch1995}. In other words, the PDFs of a fluctuating field affected by intermittency become more and more peaked when analysing smaller and smaller scales. Since the peakedness of a distribution is measured by its flatness factor $\mathcal{F}$, the behaviour of this parameter at different scales can be used to unravel the presence of intermittency in temporal or spatial coronal fluctuations. Fluctuations can be assumed to be more intermittent as either $\mathcal{F}$ increases faster with changing scales or starts to increase at larger scales.

The flatness factor $\mathcal{F}$ at the scale $\tau$ is defined as
\begin{equation}
\mathcal{F}(\tau)=\frac{\langle\delta\xi_{\tau}^{4}\rangle}{\langle\delta\xi_{\tau}^{2}\rangle^{2}},
\label{eq:flatness_factor}
\end{equation}
where $\delta\xi_{\tau}=\xi(t+\tau)-\xi(t)$ are the field fluctuations at the scale $\tau$, and the brackets indicate an averaging done over the time interval considered.

The local intermittence measure technique is used to localise intermittent events in scale and time. This method, first introduced by \citet{farge1992}, relies on a wavelet decomposition of the signal. The wavelet analysis is indeed a powerful tool for capturing peaks, discontinuities, and sharp variations in the input data. The local intermittence measure at the scale of interest $\tau$ is defined as \citep{farge1992}
\begin{equation}
\mathrm{LIM}(\tau,t)=\frac{\omega(\tau,t)^{2}}{\langle|\omega(\tau,t)|^{2}\rangle}_{t},
\label{eq:lim}
\end{equation}
where $\omega(\tau,t)$ are the coefficients of the wavelet transform. The expression in Eq.~(\ref{eq:lim}) therefore represents the energy content of the field fluctuations at the scale $\tau$ and time $t$ normalised to the average value of the energy at the same scale over the whole data sample.

Finally, the Kolmogorov-Smirnov (K-S) test \citep{telloni2016b} is used to quantify the degree of correlation of the field fluctuations or, in other words, their level of intermittency. The K-S test is indeed a non-parametric and distribution-free test used to compare a data sample with a reference probability distribution, by quantifying the maximum distance $D$ between the empirical and the reference distribution function. In the analysis of the time distribution of the intermittent events, it is possible to test whether or not the sequence of residuals is consistent with a time-varying Poisson process \citep{bi1989}.

Starting from the sequence of time intervals between two successive residuals $\delta t$ (waiting time sequence) as a function of the intermittent event occurrence time $t$, the stochastic variable $h$ can be defined as:
\begin{equation}
h_{i}(\delta t_{i},\delta\tau_{i})=\frac{2\delta t_{i}}{2\delta t_{i}+\delta\tau_{i}},
\label{eq:h}
\end{equation}
where $\delta t_{i}$ and $\delta\tau_{i}$ are the waiting times between an intermittent event at $t_{i}$ and the two (either following or preceding) nearest outliers:
\begin{equation}
\delta t_{i}=\min\left\{t_{i+1}-t_{i},t_{i}-t_{i-1}\right\}
\label{eq:delta_t}
,\end{equation}
\begin{equation}
\delta\tau_{i}=\left\{\begin{array}{ll}
t_{i-1}-t_{i-2} & \mathrm{if}\,\delta t_{i}=t_{i}-t_{i-1}\\
t_{i+2}-t_{i+1} & \mathrm{if}\,\delta t_{i}=t_{i+1}-t_{i}
\end{array}\right.
\label{eq:delta_tau}
.\end{equation}

The stochastic variable $h$ therefore simply represents the suitably normalised  time between intermittent events. Under the null hypothesis that the data sample is drawn from the Poisson reference distribution, that is in the hypothesis that $\delta t_{i}$ and $\delta\tau_{i}$ are independently distributed with exponential probability densities given by $p(\delta t_{i})=2\nu_{i}\exp(-2\nu_{i}\delta t_{i})$ and $p(\delta\tau_{i})=\nu_{i}\exp(-\nu_{i}\delta\tau_{i})$, where $\nu_{i}$ is the instant event rate,  it can be easily shown that the cumulative distribution function (CDF) of $h$, $P(h<H)$, where H is an arbitrary h between 0 and 1, is simply $P(h<H)=H$,
 where $P(h)$ is the probability distribution function of $h$. Hence, if the Poisson hypothesis holds, the stochastic variable $h$ is uniformly distributed in $[0:1]$. The maximum deviation $D$ of the empirical CDF of the outliers from the reference relation $P(h<H)=H$ quantifies the degree of departure of the intermittent events from the Poissonian, that is, non-intermittent, statistics.

By applying the above analytical methodologies to Metis white-light and UV image time series, it will be possible to identify and characterise MHD oscillations in the outer solar corona, and to address the investigation of the turbulent characteristics, if any,  of the coronal plasma, such as its intermittent and fractal nature. In particular, whether or not the solar plasma is turbulent already at coronal heights is of paramount importance with regards to the physical processes invoked to explain the energy deposition of the Alfv\'en waves in the solar corona to heat the plasma and accelerate the solar wind.

To summarise, the instrumental capabilities of Metis (simultaneous observation of the solar corona in UV and VL, high temporal and spatial resolution, out-of-ecliptic and quasi co-rotation observations) together with the diagnostic techniques described above, \cite{vandeHulst:50} $pB$ inversion and Doppler dimming techniques, as well as Fourier and intermittency analyses, will allow the assessment of crucial physical parameters (such as the plasma density and flow) in different coronal structures, and the investigation and characterisation of their spatial--temporal fluctuations, which are likely to characterise wave--particle interactions. This complete set of information will represent a step forward in the understanding of the physical processes at the base of the coronal heating and the solar wind acceleration. Finally, similar diagnostic techniques will be used to derive the basic plasma parameters of eruptive prominences inside CMEs \citep{heinzel2016,jejcic2017,susino2018} and will be applied to Metis CME observations.

\section{Conclusions}
In summary, this paper presents the details of the Metis instrument, which is conceived to investigate and characterise the physical properties and dynamics of the solar corona in the crucial region that links the Sun to the heliosphere. The paper describes the Metis optical and thermo-mechanical design, its subsystems, and the philosophy pursued in the development of the subsystems and the integrated instrument. It reports the test campaigns performed to prove its cleanliness, stability, and performance in the harsh thermal environment that Solar Orbiter will encounter. In addition, it describes the integration, alignment and calibration activities, and relative results. Finally, the operation concept, commanding, data handling, and the diagnostic tools developed to interpret the data in order to obtain the expected scientific results are discussed.

\begin{acknowledgements}
The Metis programme is supported by the Italian Space Agency (ASI) under the contracts to the co-financing National Institute of Astrophysics (INAF): Accordo ASI-INAF N. I-043-10-0 and Accordo ASI-INAF \& Addendum N. I-013-12-0/1 and under the contracts to the industrial partners: ASI-TASI N. I-037-11-0 and ASI-ATI N. 2013-057-I.0. The Metis team thanks Barbara Negri, Enrico Flamini, Marco Castronuovo of the Italian Space Agency and Roberto Della Ceca, Giuseppe Malaguti of the Istituto Nazionale di Astrofisica for their continuous support during the development of the coronagraph. A special thanks to Filippo Marliani of the European Space Agency for his dedication to the program and his high standard of excellence.
\\
The design, development, implementation, and testing of Metis was performed by an industrial consortium constituted by OHB Italia S.p.A. (acting as Prime Contractor towards ASI with the specific responsibility of system engineering, the optical subsystems, the electronics and the basic software) and Thales Alenia Space Italia S.p.A. (Co-Prime Contractor with the specific responsibility of telescope configuration and the thermo-mechanical subsystems, the application software and the instrument AIT). ALTEC has provided logistics and technical support for the INAF Optical Payload Systems facility thanks, in particular, to Alessandro Bellomo and Maurizio Deffacis.
\\
The primary and secondary mirrors were provided as Czech contribution to Metis; the mirror hardware development was possible thanks to the Czech PRODEX Programme.
\\
The UVDA assembly was provided as a German contribution to Metis, thanks to the financial support of DLR (grant 50 OT 1201). The VLDA assembly was provided under Contract 2013-058-I.0 with the Italian Space Agency (ASI). MPS thanks the ASI project responsible for the detectors, Alessandro Gabrielli, for his collaboration and support. MPS thanks the staff of the Metrology Line Source (MLS) at the Physikalisch Technische Bundeshalt (PTB) in Berlin for their collaboration and support during the calibration of the UVDA QM, PFM, and FS units.
\\
For the development, realisation and qualification of the Metis detectors, the Metis team thanks the following personnel at MPS: B. Grauf, M. Kolleck, R. Mueller, S. Frahm, D. Germerott, S. Ramanath, M.-F. Mueller, M. Bergmann, J. Bochmann, S. Meyer, M. Monecke, M. Sperling, S. Meining, M. Muehlhaus, M. Klaproth, C. K Koehler, I. Biswas and T. Keufner.
\end{acknowledgements}

\end{document}